\pdfoutput=1
\documentclass{aa}
\bibpunct{(}{)}{;}{a}{}{,} 
\usepackage[english]{babel}
\usepackage{multirow}
\usepackage{graphicx}
\usepackage{subcaption}
\usepackage{xspace}
\usepackage{xcolor}
\usepackage{booktabs}
\usepackage{txfonts}
\usepackage{lscape}

\setcitestyle{notesep={ }}

\usepackage[hidelinks,colorlinks=true,citecolor=blue,urlcolor=blue,linkcolor=blue]{hyperref}

\newcommand{\vsini}{\ensuremath{v \sin{i}}\xspace}

\newcommand{\ms}{\ensuremath{\mathrm{m\,s^{-1}}}\xspace}
\newcommand{\kms}{\ensuremath{\mathrm{km\,s^{-1}}}\xspace}
\newcommand{\A}{\ensuremath{\mathrm{\AA}}\xspace}

\newcommand{\Halpha}{\ensuremath{\mathrm{H\alpha}}\xspace}
\newcommand{\CaHK}{\ion{Ca}{II}\,H\&K\xspace}

\newcommand{\pEWHalpha}{\ensuremath{\mathrm{pEW}'(\Halpha)}\xspace}

\newcommand{\logLHalphaLbol}{\ensuremath{\log(L_{\Halpha}/L_{\mathrm{bol}})}\xspace}

\newcommand{\Caii}{\ensuremath{\mathrm{\ion{Ca}{II}}}\xspace} 

\newcommand{\Prot}{\ensuremath{P_{\mathrm{rot}}}\xspace}
\newcommand{\Prothalf}{\ensuremath{\frac{1}{2}P_{\mathrm{rot}}}\xspace}
\newcommand{\Protthird}{\ensuremath{\frac{1}{3}P_{\mathrm{rot}}}\xspace}

\newcommand{\days}{\ensuremath{\mathrm{d}}\xspace}

\newcommand{\FAP}{\ensuremath{\mathrm{FAP}}\xspace}

\newcommand{\Msun}{\ensuremath{M_{\odot}}\xspace}

\newcommand{\serval}{\texttt{serval}\xspace}
\newcommand{\raccoon}{\texttt{raccoon}\xspace}

\begin{document}

\title{The CARMENES search for exoplanets around M dwarfs}

\subtitle{
Mapping stellar activity indicators across the M dwarf domain
}

\author{
M.~Lafarga \inst{1,2,3},
I.~Ribas \inst{1,2},
A.~Reiners \inst{4},
A.~Quirrenbach \inst{5},
P.\,J.~Amado \inst{6},
J.\,A.~Caballero\inst{7},
M.~Azzaro \inst{8},
V.\,J.\,S.~B\'ejar \inst{9,10},
M.~Cort\'es-Contreras \inst{7},
S.~Dreizler  \inst{4},
A.\,P.~Hatzes \inst{11},
Th.~Henning \inst{12},
S.\,V.~Jeffers \inst{13},
A.~Kaminski \inst{5},        
M.~K\"urster \inst{12},
D.~Montes \inst{14},
J.\,C.~Morales \inst{1,2},
M.~Oshagh \inst{9,10},
C.~Rodríguez-López \inst{6},
P.~Sch\"ofer \inst{4},
A.~Schweitzer \inst{15},
M.~Zechmeister \inst{4}
}

\institute{
Institut de Ci\`encies de l'Espai (ICE, CSIC), Campus UAB, C/ de Can Magrans s/n, 08193 Cerdanyola del Vall\`es, Spain\label{ice}
\and 
Institut d'Estudis Espacials de Catalunya (IEEC), C/ Gran Capit\`a 2-4, 08034 Barcelona, Spain\label{ieec}
\and 
Department of Physics, University of Warwick, Gibbet Hill Road, Coventry CV4 7AL, United Kingdom
\and 
	 	Institut f\"ur Astrophysik, Georg-August-Universit\"at, Friedrich-Hund-Platz 1, 37077 G\"ottingen, Germany
\and 
Landessternwarte, Zentrum f\"ur Astronomie der Universit\"at Heidelberg, K\"onigstuhl 12, 69117 Heidelberg, Germany
 	\and 
Instituto de Astrof\'isica de Andaluc\'ia (IAA-CSIC), Glorieta de la Astronom\'ia s/n, 18008 Granada, Spain
\and 
Centro de Astrobiolog\'ia (CSIC-INTA), ESAC, Camino Bajo del Castillo s/n, 28692 Villanueva de la Ca\~nada, Madrid, Spain
\and 
Centro Astron\'onomico Hispano Alem\'an, Observatorio de Calar Alto, Sierra de los Filabres, E-04550 G\'ergal, Spain
\and 
Instituto de Astrof\'isica de Canarias, V\'ia L\'actea s/n, 38205 La Laguna, Tenerife, Spain
\and 
Departamento de Astrof\'isica, Universidad de La Laguna, 38026 La Laguna, Tenerife, Spain
\and 
Th\"uringer Landesstenwarte Tautenburg, Sternwarte 5, 07778 Tautenburg, Germany
\and 
Max-Planck-Institut f\"ur Astronomie, K\"onigstuhl 17, 69117 Heidelberg, Germany
\and 
Max-Planck-Institut für Sonnensystemforschung, D-37077, Göttingen, Germany
\and 
Departamento de F\'isica de la Tierra y Astrof\'isica \& IPARCOS-UCM (Instituto de F\'isica de Partículas y del Cosmos de la UCM),
Facultad de Ciencias F\'isicas, Universidad Complutense de Madrid, 28040 Madrid, Spain
\and 
Hamburger Sternwarte, Gojenbergsweg 112, 21029 Hamburg, Germany
}

\titlerunning{Mapping stellar activity indicators across the M dwarf domain}

\authorrunning{Lafarga et al.}

\date{Received, 18 February 2021 / Accepted, 26 May 2021}

\abstract
{Stellar activity poses one of the main obstacles for the detection and characterisation of small exo\-planets around cool stars, as it can induce radial velocity (RV) signals that can hide or mimic the presence of planetary companions.
Several indicators of stellar activity are routinely used to identify activity-related signals in RVs, but not all indicators trace exactly the same activity effects, nor are any of them always effective in all stars.
}
{We evaluate the performance of a set of spectroscopic activity indicators for M dwarf stars with different masses and activity levels with the aim of finding a relation between the indicators and stellar properties.}
{In a sample of 98 M dwarfs observed with CARMENES, we analyse the temporal behaviour of RVs and nine spectroscopic activity indicators: cross-correlation function (CCF) full-width-at-half-maximum (FWHM), CCF contrast, CCF bisector inverse slope (BIS), RV chromatic index (CRX), differential line width (dLW), and indices of the chromospheric lines \Halpha and calcium infrared triplet (IRT).}
{A total of 56 stars of the initial sample show periodic signals related to activity in at least one of these ten parameters. RV is the parameter for which most of the targets show an activity-related signal.
CRX and BIS are effective activity tracers for the most active stars in the sample, especially stars with a relatively high mass, while for less active stars, chromospheric lines perform best. FWHM and dLW show a similar behaviour in all mass and activity regimes, with the highest number of activity detections in the low-mass, high-activity regime.
Most of the targets for which we cannot identify any activity-related signals are stars at the low-mass end of the sample (i.e. with the latest spectral types). These low-mass stars also show the lowest RV scatter, which indicates that ultracool M dwarfs could be better candidates for planet searches than earlier types, which show larger RV jitter.}
{Our results show that the spectroscopic activity indicators analysed behave differently, depending on the mass and activity level of the target star. This underlines the importance of considering different indicators of stellar activity when studying the variability of RV measurements. Therefore, when assessing the origin of an RV signal, it is critical to take into account a 
large set of indicators, or at least the most effective ones considering the characteristics of the star, as failing to do so may lead to false planet claims.}


\maketitle
%

\section{Introduction}

One of the main problems when trying to find and study exo\-planets with Doppler spectroscopy is the intrinsic variability of the host stars.
Different phenomena on the surface of stars, such as solar-like oscillations \citep[e.g.][]{bouchy2002oscillations,butler2003oscillations,kjeldsen2005oscillations,bazot2012oscillations}, surface granulation \citep[e.g.][]{dravins1982lineasymmetry,delmoro2004supergranulation,meunier2017VariabilityGranualtionConvective,cegla2019sunasstar}, stellar magnetic activity features such as cool spots and hot faculae \citep[e.g.][]{saardonahue1997activity,desort2007activity,lagrange2010spot,meunier2010plages}, or long-term magnetic cycles \citep[e.g.][]{santos2010magneticcycle,lovis2011magnetic,suarezmascareno2016photrotation,suarezmascareno2018hadesAct}, introduce signals in radial velocity (RV) measurements that can obfuscate planetary signatures.
Moreover, since stars rotate, stellar activity features often appear as periodic signals modulated with the stellar rotation period (\Prot), which can be mistakenly attributed to the periodic signal caused by a planet \citep[e.g.][]{queloz2001noplanet,desidera2004HD219542Bactivity,bonfils2007gj674,huelamo2008TWHya,boisse2009activityHD189733,figueira2010BD201790activity,hatzes2013aCenBactivity,rajpaul2016aCenBactivity,haywood2014corot7GP,santos2014HD41248,robertson2015gj176}.
Alongside these, long-term magnetic cycles can also result in periodic signals.
Therefore, it is critical to understand how astrophysical variability affects the measurements in order to exclude rotational modulation as the origin of RV variations.

A variety of activity indicators obtained from the same stellar spectra used to extract RVs are employed to probe activity in different ways.
Common examples are measurements of the flux in the core of chromospheric lines, such as the \CaHK lines in solar-like stars \citep{wilson1968CaHK,vaughan1978Sindex,noyes1984rotation,lovis2011magnetic} or the \Halpha line in M dwarfs \citep{west2004mdwarf,schofer2019carmenesActInd}, parametrisations of the cross-correlation function (CCF) profile \citep{queloz2001noplanet,boisse2011disentangling,figueira2013lineprofilevariations,lanza2018lineprofilediagnostics,simola2018ccfskewnormal}, or wavelength-dependent changes in the RVs measured with the so-called chromatic index \citep[CRX,][]{zechmeister2018serval}.
Photometric observations are also commonly used to study activity \citep{boisse2009activityHD189733,cegla2014photRV,haywood2014corot7GP,bastien2014photRV,oshagh2017photRV,hojjatpanah2020photRV}.
The average strength of the indicators is related to the activity level of the stars, and modulations inferred from the indicators' time series measure the inhomogeneity of activity across the stellar surface.

Stellar activity indicators are sensitive to stellar variability and track signals induced in RVs by activity. At the same time, by construction, they are insensitive to planet-induced modulations. Although they could be affected by star-planet interactions enhancing the stellar activity, no firm detection of this phenomenon exists so far \citep[see e.g.][]{cuntz2000spi,shkolnik2003spiHD17}.
Consequently, they are necessary to distinguish between a planetary or stellar origin of periodic signals found in RVs.
Correlations of activity indicators with RVs or the presence of the same periodic signal in both RV and indicators suggest the existence of activity-driven variations in the RVs.
Several techniques have been developed to decorrelate or model the RV activity signal using different activity indicators \citep[e.g.][]{boisse2009activityHD189733,lanza2010activitycorot7,haywood2014corot7GP,rajpaul2015GP,herrero2016starsim,mallon2018gj1214,rosich2020starsim,baroch2020carmenescrxYZCMi,gilbertson2020multivariateGP}.
There are also a number of studies focusing on the temporal variability experienced by common indicators, the correlations between them, and their relation with RV measurements on relatively large sets of cool stars \citep[e.g.][]{suarezmascareno2015rotationchrom,suarezmascareno2017rvrotation,suarezmascareno2018hadesAct,tal-or2018carmenesRVloud,schofer2019carmenesActInd,fuhrmeister2019carmenesPeriodLines}.
However, a general methodology to correct for activity in RV measurements has not yet been established.

Indicators such as asymmetry measurements of the CCF bisector inverse slope (BIS) or the CRX have been observed to show linear correlations with the RVs in M dwarfs, however, this is not always the case for, for example, the full-width-at-half-maximum (FWHM) of the CCF or indicators derived from chromospheric lines \citep[e.g.][]{queloz2001noplanet,bonfils2013harpsMdwarfsample,zechmeister2018serval,tal-or2018carmenesRVloud,gomesdasilva2012actlongtermIndRV}.
The lack of a linear correlation does not mean that these indicators are not correlated with the RVs, but rather, they often show a more complicated relation, which could be due to being out of phase with the RVs \citep[e.g.][]{santos2014HD41248,perger2017hadesGJ3942}.
Usually, indicators display signals at the stellar rotation period or some of its harmonics \citep{boisse2011disentangling,schofer2019carmenesActInd}.
However, differential rotation on the photosphere can result in slight variations in the measured rotation period of the star. 
Complex active region patterns on the stellar surface also alter the frequencies seen in the periodograms, and the limited lifetime of active regions may also impede the detection of signals in long time series \citep[see e.g.][]{schofer2019carmenes4stars}.
Moreover, signals observed in indicators that trace different types of activity, such as activity in the photosphere or the chromosphere, do not necessarily match exactly.
Due to the compound action of all these effects, the behaviour of any given activity indicator can display notable differences between individual stars.
Some stars may show activity-related signals in only some indicators but not in others, while in another set of stars, the `useful' indicators may be different.

The aim of this work is to study the behaviour of a set of spectroscopic activity indicators and evaluate their performance for assessing activity signals.
We want to determine if there is any relation between their performance and the properties of the target stars, with the goal of finding which indicators work best for stars with certain properties.
For this purpose, we used indicators from a sub-sample of the CARMENES M dwarfs and studied their time series.
In Sects. \ref{sec:actindtargets} and \ref{sec:actindused} we present the sub-sample of selected targets, and the observations and parameters analysed.
Sects. \ref{sec:actindanalysis} and \ref{sec:actindresults} describe the methodology we followed to analyse the time series data and the results obtained, respectively.
In Sect. \ref{sec:performance} we evaluate the performance of the indicators and we discuss possible detection biases in Sect. \ref{sec:detbiases}.
We study the relation between the detected activity signals and the RV scatter in Sect. \ref{sec:jitter}, and with the stellar rotational velocity in Sect. \ref{sec:vsini}.
We conclude in Sect. \ref{sec:actindsummary} with a summary of our work.


\section{Stellar sample} \label{sec:actindtargets}

\subsection{CARMENES GTO sample and previous studies on it}

We used observations obtained with the CARMENES instrument \citep[Calar Alto high-Resolution search for M dwarfs with Exo-earths with Near-infrared and optical Echelle Spectrographs,][]{quirrenbach2016carmenes,quirrenbach2018carmenes} as part of its main survey (guaranteed-time observations – GTO program).
CARMENES is installed at the 3.5\,m telescope at Calar Alto Observatory in Almería, Spain, and consists of a pair of cross-dispersed, fibre-fed echelle spectrographs with complementary wavelength coverage, which allow simultaneous observations in the visual and the near-infrared wavelength range.
The visual (VIS) channel covers the spectral range $\lambda$ = 5200--9600\,\A at a resolution of $R=94\,600$, with an average sampling of 2.5 pixels per spectral element, and the near-infrared (NIR) channel covers the range $\lambda$ = 9600--17100\,\A at a resolution of $R=80\,400$, and has an average sampling of 2.8 pixels per spectral element.
The CARMENES survey has been ongoing since 2016. It monitors over 300 M dwarfs across all spectral subtypes with the main goal of detecting orbiting exo\-planets \citep{reiners2018carmenes324}.


\begin{figure*}
\centering
\includegraphics[width=0.8\linewidth]{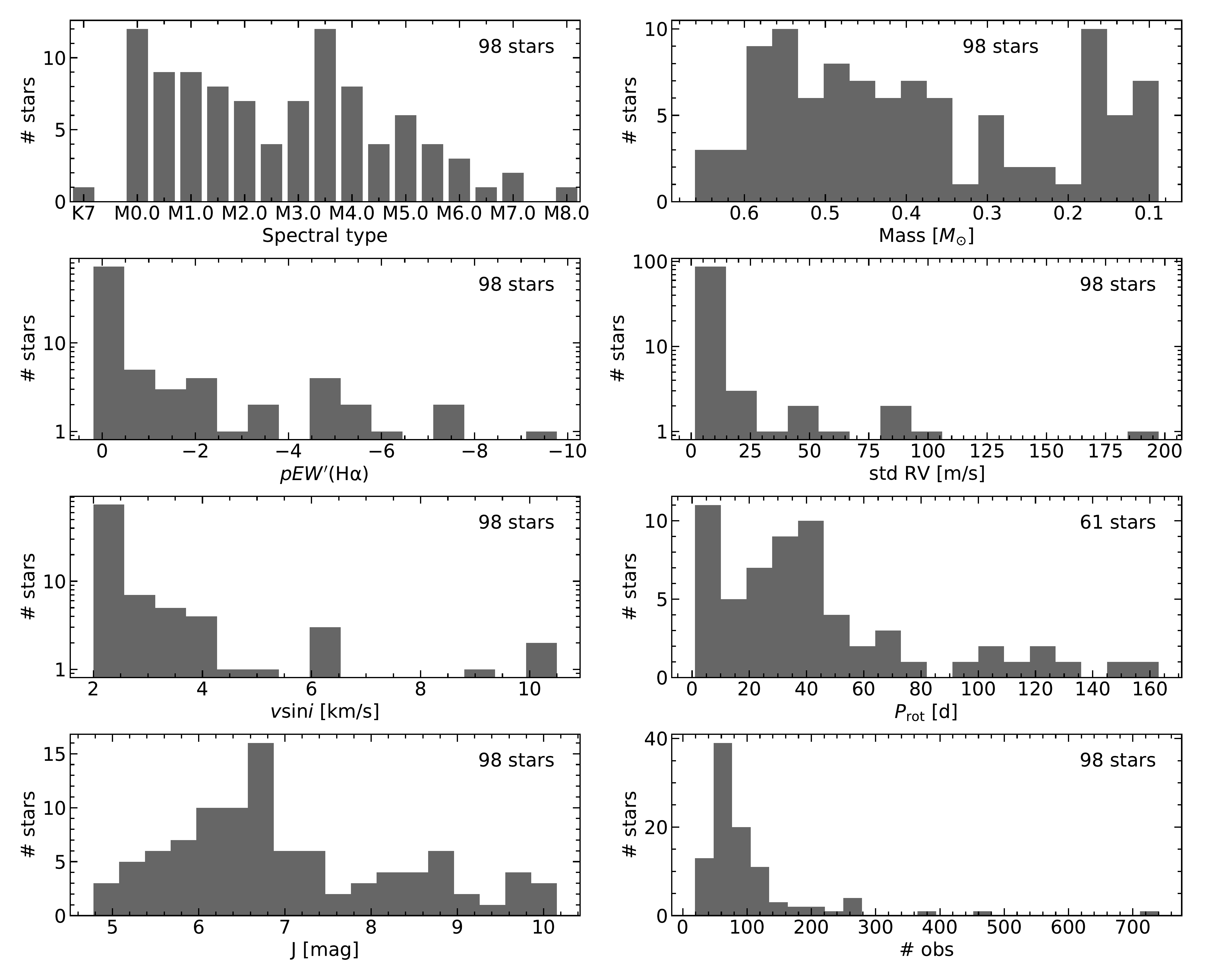}
\caption{Distribution of the spectral type (\emph{top left}), stellar mass (\emph{top right}), mean activity level (\pEWHalpha average measured from the \serval template, \emph{second row left}), RV scatter (measured as the standard deviation of the \serval VIS RVs, \emph{second row right}), rotational velocity (\vsini, \emph{third row left}), rotation period (\Prot, \emph{third row right}), $J$ magnitude (\emph{bottom left}), and number of observations (\#\,obs, \emph{bottom right}) of the 98 selected stars. Only 61 stars have a known \Prot to date. All values compiled from Table \ref{tab:starsactindselected}.}
\label{fig:actindsampleprops}
\end{figure*}


\begin{figure}
\centering
\includegraphics[width=\linewidth]{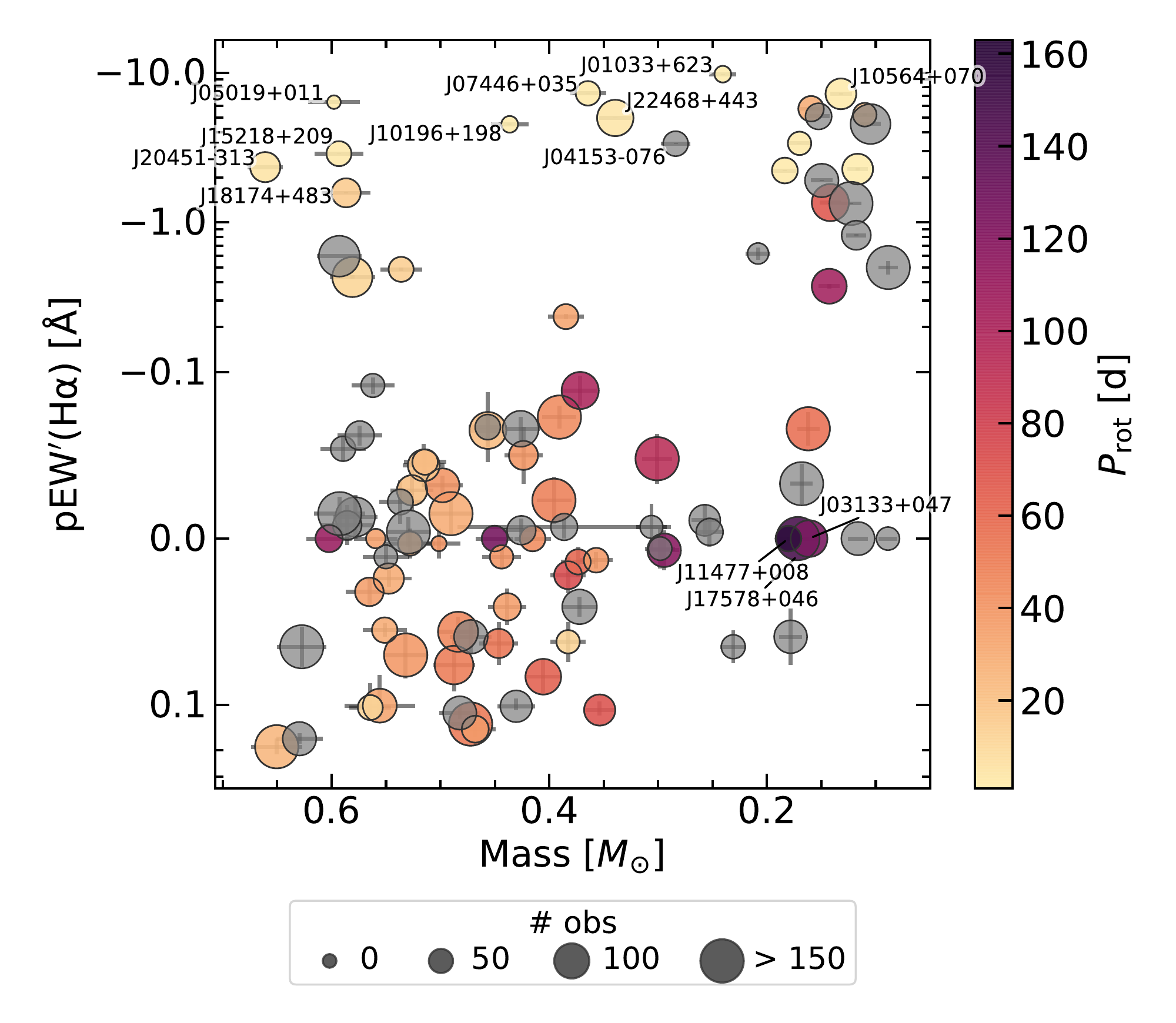}
\caption{Average activity level (measured as the \pEWHalpha of the \serval template, in linear scale between -0.1 and 0.1, and logarithmic scale otherwise) as a function of the stellar mass of the 98 selected stars.
The data points are colour-coded with the target rotation period \Prot (grey points indicate targets with unknown \Prot), and their size is given by the number of CARMENES VIS observations (with all the stars with more than 150 observations having the same size). We have indicated the name of several stars for reference. All values compiled from Table \ref{tab:starsactindselected}.}
\label{fig:actindsummaryobj}
\end{figure}


The CARMENES GTO sample has been previously used to study activity in M dwarfs.
\citet{tal-or2018carmenesRVloud} analysed correlations between RV and activity indicators, focusing on stars with large variability in the RVs (RV scatter\,$\geq10\,\ms$), the so-called `RV-loud' sample.
Of the 31 stars in the RV-loud sample, about a third showed significant anti-correlations between the RVs and the CRX, while for another 20\,\%, there was a marginal detection of an anti-correlation.
For the \Halpha line and the differential line width (dLW, an indicator that measures changes in the width of the absorption lines), the authors excluded the existence of linear correlations with the RVs, although more complex relations could exist \citep[as shown in e.g.][]{zechmeister2018serval}.

\citet{schofer2019carmenesActInd} computed and studied the temporal variability and the correlations of chromospheric indicators and photospheric absorption band indices in almost the whole CARMENES GTO sample, over 300 M dwarfs.
They identified 15 stars (out of the 133 stars with known rotation period longer than 1\,d) with a significant signal related to the rotation period in more than two indicators.
The indicators that were most likely to vary with the stellar period were those measured from the \Halpha line and one of the \Caii infrared triplet (Ca\,IRT) lines, as well as the ones measured from the photospheric titanium oxide bands at 7050 and 8430\,\A.
\citet{fuhrmeister2019carmenesPeriodLines} computed chromospheric indices in a slightly different way than \citet{schofer2019carmenesActInd} and used them to look for activity-related periodicities in 16 early M dwarfs (M0 to M2). Their results agree with those of \citet{schofer2019carmenesActInd}: they find similar periodicities, and of the different chromospheric lines tested, the most useful ones are the \Halpha line and the Ca\,IRT lines.
More recently, \citet{schofer2019carmenes4stars} performed an in-depth study of four stars in which different chromospheric lines and photospheric absorption bands display significant activity-related periodicities.
The analysis of different indicators measured over time indicates changes in the surface feature distribution of the stars, which make the indicators vary with different harmonics of \Prot depending on the observation time.

\subsection{Targets used in this work}

Here, we expand on previous studies by investigating several activity indicators using data from the CARMENES VIS observations from GTO.
To ensure the reliability of our analysis we selected targets having at least 40 observations in the VIS channel, excluding 18 known binaries \citep{cortescontreras2017carmenesinhighres,baroch2018carmenesBinaries}.
A preliminary search for periodic signals related to the stellar rotation period in the GTO data revealed that six stars with less than 40 measurements showed strong signals in the periodogram of RV and/or spectroscopic activity indicators. To increase the number of stars in the sample, we therefore included these stars in our analysis, bringing our final sample to 98 targets.
These six additional stars have 19 to 39 observations (27 on average), and are active stars or have well-defined \Prot. We would expect to find the same kind of activity signals if more observations were available, so considering them in the sample should not bias our results.

Compared to the previous studies mentioned above, our sample includes 8 of the total 31 RV-loud stars, for which correlations between RV and activity indicators (CRX, dLW, and \Halpha index) were previously analysed by \citet{tal-or2018carmenesRVloud}. The rest of RV-loud stars were not considered here because of their low number of observations. Most stars showing large RV scatters, that is, RV-loud stars, were discarded from the GTO survey after about 11 observations.
\citet{schofer2019carmenesActInd} studied chromospheric indicators in 96 of the 98 stars in our sample. The two new stars in our sample not included in \citet{schofer2019carmenesActInd} are J20451-313 (AU~Mic, GJ~803), whose observations started after that work was accepted, and J18198-019 (HD~168442, GJ~710), which has spectral type K7 and therefore is not always considered as part of the GTO sample of M dwarfs.
In this work, instead of focusing on the correlation between the indicators, we perform a time-series analysis, and include, in addition to the indicators mentioned above, three measurements from the CCF profile. Moreover, we use a more up-to-date GTO data release, which in most cases includes more observations per star.

Table \ref{tab:starsactindselected} shows the main properties of the selected stars.
Histograms of the spectral type, stellar mass, average activity level, RV scatter, rotational velocity, rotational period (mostly photometric measurements), apparent brightness, and number of observations are shown in Fig. \ref{fig:actindsampleprops}.
All values are obtained from the latest internal release of Carmencita, the CARMENES input catalogue \citep{alonsofloriano2015carmenesinlowres,caballero2016carmencita}, except for the RV scatter, which we compute from the CARMENES measurements.
Fig. \ref{fig:actindsummaryobj} shows the activity level of our sample as a function of the stellar mass.
The selected stars cover most of the M spectral subtypes, from M0\,V to M8\,V, plus one late-K type dwarf, but most of them have early and mid M spectral subtype.
The mass measurements are computed also from CARMENES observations and other data sets as in \citet{schweitzer2019carmenesMR}.
The sample also covers a wide range of activity levels.
To quantify the average activity level of the stars, we used measurements of the pseudo-equivalent width of the \Halpha line (\pEWHalpha) computed from a coadded stellar template as in \citet{schofer2019carmenesActInd}, instead of averaging the \pEWHalpha measurements of individual observations.
The templates for each star are built with the \serval pipeline \citep{zechmeister2018serval}, which co-adds the different observations of the star.
The average \pEWHalpha goes from $\sim0.2\,\AA$ to $-10\,\AA$, however, the majority of stars are considered \Halpha inactive (71 of them having $\pEWHalpha \geq -0.3\,\AA$, and of these, 42 have $\pEWHalpha\geq 0\,\AA$).
Since the \pEWHalpha values were measured from coadded templates of the same CARMENES observations that we are using in our analysis, they should accurately reflect the activity observed in our data.

A total of 61 stars have a previously measured rotation period, ranging from $\sim$1 to 163~d, which mostly come from photometric measurements \citep{diezalonso2019carmenesRotPhot,newton2016rotation}. Most of these are concentrated towards $\sim$3~d (for the most active stars, all with $\pEWHalpha\leq-2\,\AA$) and $\sim30-40$~d, and only 18 stars have \Prot larger than 50~d. Stars with unknown \Prot are scattered throughout the activity-mass space, but most of them are likely small-amplitude, slow rotators. 


\section{Data: RVs and activity parameters} \label{sec:actindused}

To analyse the presence of activity signals, we used RVs and several activity indicators derived from the CARMENES VIS observations: three parameters derived from the CCF profile (FWHM, contrast, and BIS), CRX, dLW, and indices derived from four lines that show chromospheric emission.
We describe these parameters in the following subsections.

\subsection{RV} 

RVs from the VIS channel were obtained with 
the template-matching pipeline \serval \citep{zechmeister2018serval}, which computes the RV of a set of observations by performing a least-squares fit with a high S/N template built from the observations themselves.
The RVs have been corrected for barycentric motions, secular acceleration, instrumental drifts, and nightly-zero points \citep{trifonov2018carmenes1st, tal-or2019systematichires}.

\subsection{CCF profile: FWHM, contrast, and BIS}

Distortions common to the majority of the photospheric absorption lines are reflected in the CCF profile, and can be studied with different parametrisations of the CCF.
We analysed the FWHM, contrast, and BIS of the CCF, computed with the \raccoon code as explained in \citet{lafarga2020carmenesccf}.
The CCFs of the targets are computed with different weighted binary masks depending on the star spectral subtype and rotational velocity. The masks are created from high S/N stellar templates built from CARMENES observations with \serval.
The resulting CCFs are fitted with a Gaussian, from which we measure the FWHM and contrast.
The BIS is computed directly from the CCF profile as the difference between the average velocity of the top region of the CCF (from 60 to 90\%) and the average velocity of the bottom region (from 10 to 40\%), as in \citet{queloz2001noplanet}.

\subsection{CRX and dLW}

We also used the CRX and the dLW values computed by the \serval pipeline, both parameters as defined in \citet{zechmeister2018serval} \citep[as mentioned above, see][for an analysis of the correlations between these indicators and RV]{tal-or2018carmenesRVloud}.
The CRX measures variations in the RV across the observed wavelength range.
The dLW accounts for differential changes in the line widths of the observed spectrum compared to the spectral template.
Indicators such as dLW, FWHM, or contrast can be affected by instrumental effects such as changes in the instrument profile, sky background (e.g. contamination by light scattered by clouds or the Moon, especially on observations of faint targets), or an artificial broadening of the lines due to barycentric motion during a long exposure.

\subsection{Chromospheric emission lines: H$\alpha$ and Ca\,IRT}

To probe chromospheric activity, we used measurements of the core emission flux of the \Halpha line (6564.60\,\A vacuum wavelength) and the Ca\,IRT lines (with vacuum wavelengths at 8500.35, 8544.44, and 8664.52\,\A).
Initially, we analysed both their $I$ index computed as in \citet{zechmeister2018serval} and pseudo-equivalent width pEW' computed as in \citet{schofer2019carmenesActInd}.
The $I$ index is defined as the ratio of flux around the centre of the line to the flux in reference bandpasses on either sides of the line \citep[e.g.][]{kuerster2003barnard}.
The pseudo-equivalent width pEW' measures the equivalent width of the line with respect to a pseudo-continuum region, normalised by the spectrum of a non-active star in order to remove photospheric contributions \citep[e.g.][]{young1989halpha,montes1995halpha}. Following \citet{schofer2019carmenesActInd}, a non-active star per spectral subtype has been used to compute the pEW' values.
Due to this spectral subtraction, pEW' measurements represent an absolute measure for the (excess) emission, and is most useful when comparing different stars.
Equivalent widths and line indices are closely related, as both quantities measure the integrated flux of the lines.
We observe that, for the four lines under consideration, the two types of measurements show in general a similar behaviour.
Therefore, to simplify our analysis, we chose to only use the $I$ index.
In the following, instead of using, for instance, $I_{\Halpha}$ for the \Halpha index, we refer to this measurement simply as `\Halpha' for the \Halpha line, and as `Ca\,IRT-a', `Ca\,IRT-b', and `Ca\,IRT-c' for the three lines of the Ca\,IRT.


\section{Time-series analysis} \label{sec:actindanalysis}

\subsection{Periodogram analysis}

For each of the selected stars, we computed the generalised Lomb-Scargle (GLS) periodogram \citep{zechmeister2009gls} of the time series of the RVs and the activity indicators mentioned in section \ref{sec:actindused}.
We sampled frequencies from the inverse of the time span of the measurements of each star to 1\,d$^{-1}$, since the shortest known \Prot of the sample stars is $\sim1$\,d.
Within these limits, we used a frequency grid that approximately oversamples each peak with 10 points, to make sure that the grid is fine enough to recover all significant peaks without adding a significant amount of computing time.
To assess the significance of the power of the periodogram peaks, we used the false alarm probability (FAP) computed with the analytical approximation proposed in \citet{baluev2008fap}.

In each time series, we discarded observations with low S/N (average S/N per pixel < 5 around 7456\,\A). 
We also removed outliers (values that deviate from the mean of the corresponding time series by more than 3 times their standard deviation, i.e. 3$\sigma$ clipping), and data points with large error bars (also applying a 3$\sigma$ clipping on the uncertainties), to avoid including in the analysis observations obtained during flare events, which can strongly affect the measured RVs and spectroscopic indicators \citep[see e.g.][]{reiners2009CNLeoflare}. On average, for each star, this process removed less than 5 observations.

All the data sets were corrected for long-term trends (using a linear model) to avoid biases due to long-term magnetic cycles or wide-orbit sub-stellar companions present in the RVs.
Some of the stars in the sample are known to host planets. We did not remove the planetary signals, so the RV periodograms may reflect those signals, alongside those related to activity.
We also averaged all measurements obtained during the same night.


\subsection{Detection of activity-related signals}

Periodograms reflect the convolution of the power spectrum of the true signals present in the data with the window function of the data.
This means that each frequency $f$ can show an alias at $f + n f_W$, where $n$ is an integer, and $f_W$ represents a typical sampling frequency (a strong feature in the window function).
For ground-based observations, this typically results in 1-day (actually, one sidereal day, 0.99726~d) and 1-year aliases.
Harmonics of a specific frequency $f$ can appear for periodic signals that are not strictly sinusoidal and have power at higher harmonics $m f$, where $m$ is a positive integer.
These harmonics can also suffer aliasing effects due to the window function.
Moreover, every peak at a frequency $f$ has a corresponding peak at a frequency $-f$, and both peaks in the positive and negative frequency domain are subject to aliasing.
In general, aliases can appear at frequencies $f_\mathrm{alias} = \lvert m f + n f_W \rvert$ \citep{dawson2010alias,vanderplas2018periodogram,stock2020carmenesYZCet}.


Therefore, we looked for significant signals related to the activity of the star, taking into account the presence of aliases and harmonics of the true periods contained in the data.
Specifically, we looked for significant peaks at the rotation period of the stars, \Prot, and its first two harmonics, $\frac{1}{2}\Prot$ and $\frac{1}{3}\Prot$. 
From previous analyses, we know that, for ground based observations, it is very frequent to find significant peaks at the 1-day alias of \Prot, so we also looked for peaks at the regions around the 1-day aliases of \Prot and its first two harmonics.
For stars with unknown \Prot (37 stars), we looked for coincidences between signals in the RVs and the indicators. In case of a significant peak at the same frequency (we considered the same frequency values within $\sim0.01$ d$^{-1}$) in two or more different indicators, we assumed it to be related to the activity of the star.


\subsection{Examples}

In this section we explain the detection of activity-related signals of five stars as representative examples of the analysis performed.
Figures with the time series data of RV and activity indicators, as well as their periodograms, can be found in Appendix \ref{sec:actindperiodogramexamples}.


\subsubsection{Clear signal at known \Prot: J07446+035 (YZ CMi, GJ 285)}

J07446+035 is a mid-M dwarf (M4.5\,V) with the activity level that counts among the highest in the sample ($\pEWHalpha\sim-7$).
It has a \Prot of 2.78\,d \citep{diezalonso2019carmenesRotPhot} and has been observed 51 times with CARMENES.
This is a clear example of an active star with a very significant activity-related signal present in the RVs as well as most of the activity indicators, which has been previously studied in \citet{zechmeister2018serval,tal-or2018carmenesRVloud,baroch2020carmenescrxYZCMi,schofer2019carmenes4stars}.
In Fig. \ref{fig:actind_tsperiodogramJ07446+035} we show the time series of the RVs and the indicators, the data folded at \Prot, the correlation of the indicators with the RV, and the periodogram of each parameter.

The periodograms of RV, CRX, dLW, and the three CCF parameters all show significant (FAP < 0.1\,\%) periodicities at \Prot, and a smaller but still significant (with FAP < 0.1\,\% and < 10\,\%) signal at 1.56\,d, which corresponds to the 1-day alias of the \Prot signal.
The two highest peaks of the \Halpha periodogram are at \Prot and its daily alias, but they are not significant (FAP > 10\,\%) nor clearly above the rest of the peaks.
The Ca IRT lines do not show any clear peak.
We do not observe any significant power (FAP < 10\,\%) at \Prothalf nor its alias in any of the parameters.
Therefore, for this star, we conclude that only RV, CRX, dLW, FWHM, contrast, and BIS show significant signals at both \Prot and its 1-day alias.
We also note that for \Halpha, there is a signal at \Prot, but with very low significance.


\subsubsection{Multiple signals related to known \Prot: J22468+443 (EV Lac, GJ 873)}\label{sec:periodogramJ22468+443}

J22468+443 is another example of a very active ($\pEWHalpha\sim -5$) mid-M dwarf (M3.5\,V) with a short \Prot \citep[4.38\,d,][]{diezalonso2019carmenesRotPhot}, also previously studied in \citet{tal-or2018carmenesRVloud} and \citet{schofer2019carmenes4stars}.
The latter study finds that the periodicities change over time, indicating significant changes of the surface features.
Here we use all the observations available (111 observations, Fig. \ref{fig:actind_tsperiodogramJ22468+443}).
The periodograms are more complex than for the case of J07446+035, with signals at both \Prot and \Prothalf (and their 1-day aliases).

The periodograms of RV, CRX, BIS, and FWHM show their highest peak at \Prothalf (FAP < 0.1\,\%), and another much less significant (with FAP between $\sim$ 1\,\% and 10\,\%), but still very obvious peak at \Prot.
In contrast, dLW and contrast show their highest peak at \Prot (with FAP < 0.1\,\% and 0.1\,\% FAP < 1\,\%, respectively), and a much less significant peak at \Prothalf (FAP > 10\,\%).
The chromospheric lines have a periodogram with more structure around the peaks. They also display their highest peak at \Prot (with 1\,\% < FAP 10\,\%), and we observe no significant power at \Prothalf.
Additionally, dLW, FWHM, and BIS show another significant peak at \Protthird (0.1\,\% < FAP < 1\,\% for dLW, and FAP $\sim$ 10\,\% for FWHM and BIS), and RV and CRX also seem to have some power at this period, although much less significant.


\subsubsection{Unknown \Prot: J17303+055 (BD+05 3409, GJ 678.1 A)}

In Fig. \ref{fig:actind_tsperiodogramJ17303+055} we show an example of an early M dwarf (M0.0\,V) with low activity level ($\pEWHalpha\sim0$) and unknown \Prot.
RV, dLW, FWHM, contrast, and the chromospheric lines show a significant peak at $\sim33.8$\,d (FAP < 0.1\,\% for FWHM and chromospheric lines, and FAP $\sim$ 1\,\% for the other indicators mentioned). 
There is also a peak at $\sim1.03$\,d, its 1-day alias, with similar significance. We attribute the long-period peak to the true rotation period, because this is an inactive star with a low rotational velocity \cite[$\vsini\leq2\,\kms$,][]{reiners2018carmenes324}.
BIS shows significant signals at 17.6\,d and 1.06\,d (FAP $\sim$ 1\,\%), which are close to \Prothalf and its 1-day alias. CRX does not show any significant peak (no peaks with FAP < 10\,\%).


\subsubsection{Clear but formally non-significant signals: J20451-313 (AU Mic, GJ 803)}

J20451-313 is an active ($\pEWHalpha\sim-2$), early M dwarf (M\,0.5V) with a \Prot of 4.84\,d \citep{messina2011rotation}, for which two transiting planets have been recently announced \citep{plavchan2020aumic,martioli2021aumic}.
We see clear peaks at either \Prot or \Prothalf (and their 1-day aliases) in the RVs and all the indicators (Fig. \ref{fig:actind_tsperiodogramJ20451-313}).
However, for CRX, dLW, \Halpha, and Ca IRT-a, the peaks have low significance, $\FAP>10\,\%$.
Since these peaks are clear, isolated, and above the rest of the peaks present in the periodogram, we still consider those as probably activity-related signals, although in our final analysis, we only consider peaks with FAP < 10\,\%.


\subsubsection{Long \Prot: J03133+047 (CD Cet, GJ 1057)}

Low-activity stars tend to have long rotation periods that are not well constrained, which complicates their identification in periodograms.
J03133+047 is an example of this.
It is an M5.0\,V star with a low activity level ($\pEWHalpha\sim0$) and a long \Prot of $\sim126$\,d \citep{newton2016rotation}.
It hosts a super-Earth orbiting on a 2.29\,d orbit \citep{bauer2020cdcet}.

dLW, FWHM, contrast, \Halpha, and Ca IRT-b show significant (FAP < 0.1\,\%, except 1\,\% < FAP < 10\,\% for Ca IRT-b), wide peaks around \Prot (Fig. \ref{fig:actind_tsperiodogramJ03133+047}).
In this case, we chose the peaks closest to the literature \Prot value as the activity-related signal.
These fall in the range from 137 to 148\,d, depending on the indicator.
The other indicators show some power close to \Prot, but they are less clear and not significant (FAP > 10\,\%).
RV also shows some excess power close to \Prot and \Prothalf. The power at \Prot becomes significant if the 2.29\,d planetary signal is removed from the RV measurements \citep[see][]{bauer2020cdcet}.


\section{Rotation signals in activity indicators} \label{sec:actindresults}

Table \ref{tab:starsactindresults} shows a summary of the results obtained from the periodogram analysis, where we list the indicators for which we found a significant activity-related signal.
We find at least one significant ($\FAP\leq10\,\%$) signal in 56 of the 98 stars investigated. 
In Appendix \ref{sec:actindperiodogramresults}, we show the specific results obtained for each indicator.
Tables \ref{tab:starsactindselected_rv} to \ref{tab:starsactindselected_cairt3} contain, one for each parameter (RV, CRX, dLW, FWHM, contrast, BIS, \Halpha, and Ca IRT-a,b,c), the periods and FAP of the significant peaks that we identified as being related to stellar activity.
Specifically, we show the values of the peaks at \Prot, \Prothalf, and their 1-day aliases (we do not show the values corresponding to \Protthird because we only found significant signals for one star, J22468+443, see Sect. \ref{sec:periodogramJ22468+443}).

\begin{table*}
\caption{Parameters that show an activity-related signal with $\FAP \leq 10\,\%$ (bold-face indicates $\FAP \leq 0.1\,\%$) of the 56 stars for which we find a detection.
We also show the mass and \pEWHalpha values for reference (same as in Table \ref{tab:starsactindselected}).}\label{tab:starsactindresults}
\centering
\begin{tabular}{lccl}
\hline\hline
Karmn & Mass [\Msun] & \pEWHalpha & Detection \\
\hline
J00183+440  &     $0.391\pm 0.016$ &  $-0.0730\pm 0.0070$ &                                                   RV, \textbf{\Halpha}, \textbf{Ca IRT-b}, Ca IRT-c \\
J01025+716  &     $0.488\pm 0.019$ &     $0.076\pm 0.016$ &      \textbf{RV}, CRX, \textbf{dLW}, \textbf{FWHM}, \textbf{\Halpha}, Ca IRT-a, \textbf{Ca IRT-b,c} \\
J01026+623  &     $0.515\pm 0.019$ &    $-0.044\pm 0.013$ &                                              RV, FWHM, BIS, \textbf{\Halpha}, \textbf{Ca IRT-a,b,c} \\
J01033+623  &     $0.241\pm 0.012$ &    $-9.770\pm 0.040$ &                                                                                             RV, CRX \\
J01125-169  &   $0.1418\pm 0.0098$ &    $-1.360\pm 0.020$ &                                                    RV, \textbf{dLW}, \textbf{FWHM}, \textbf{Contr.} \\
J02002+130  &   $0.1497\pm 0.0098$ &    $-1.910\pm 0.025$ &                                \textbf{RV}, CRX, dLW, FWHM, \textbf{Contr.}, BIS, \Halpha, Ca IRT-c \\
J02222+478  &     $0.551\pm 0.020$ &   $0.0550\pm 0.0040$ &                                            RV, \textbf{dLW}, \Halpha, Ca IRT-a, \textbf{Ca IRT-b,c} \\
J02530+168  &  $0.08857\pm 0.0088$ &    $-0.500\pm 0.050$ &                                                        \textbf{dLW}, \textbf{FWHM}, \textbf{Contr.} \\
J03133+047  &     $0.161\pm 0.010$ &                $0.0$ &                                     \textbf{dLW}, \textbf{FWHM}, Contr., \textbf{\Halpha}, Ca IRT-b \\
J03463+262  &     $0.562\pm 0.020$ &  $-0.0920\pm 0.0050$ &                                                                      \textbf{\Halpha}, Ca IRT-a,b,c \\
J04153-076  &     $0.284\pm 0.014$ &    $-3.360\pm 0.028$ &                                                    \textbf{RV}, \textbf{CRX}, \textbf{BIS}, \Halpha \\
J04290+219  &     $0.650\pm 0.024$ &     $0.190\pm 0.023$ &                       \textbf{RV}, \textbf{dLW}, FWHM, BIS, \textbf{\Halpha}, \textbf{Ca IRT-a,b,c} \\
J04376+528  &     $0.578\pm 0.020$ &    $-0.013\pm 0.013$ &                                          RV, \textbf{dLW}, FWHM, BIS, \textbf{Ca IRT-a,b}, Ca IRT-c \\
J04429+189  &     $0.501\pm 0.020$ &   $0.0030\pm 0.0090$ &                                                      \Halpha, Ca IRT-a, \textbf{Ca IRT-b}, Ca IRT-c \\
J04588+498  &     $0.589\pm 0.021$ &  $-0.0540\pm 0.0070$ &                                      \textbf{RV}, dLW, \textbf{BIS}, \Halpha, \textbf{Ca IRT-a,b,c} \\
J05019+011  &     $0.598\pm 0.024$ &    $-6.360\pm 0.019$ &                                                             \textbf{RV}, \textbf{CRX}, \textbf{BIS} \\
J05314-036  &     $0.556\pm 0.033$ &     $0.101\pm 0.019$ &                     \textbf{RV}, CRX, \textbf{dLW}, Contr., \textbf{\Halpha}, \textbf{Ca IRT-a,b,c} \\
J05365+113  &     $0.581\pm 0.021$ &    $-0.432\pm 0.012$ &              \textbf{RV}, dLW, \textbf{FWHM}, \textbf{BIS}, \textbf{\Halpha}, \textbf{Ca IRT-a,b,c} \\
J07274+052  &     $0.301\pm 0.014$ &    $-0.048\pm 0.015$ &                                                                                                  RV \\
J07446+035  &     $0.364\pm 0.017$ &    $-7.280\pm 0.024$ &               \textbf{RV}, \textbf{CRX}, \textbf{dLW}, \textbf{FWHM}, \textbf{Contr.}, \textbf{BIS} \\
J08413+594  &   $0.1228\pm 0.0094$ &    $-1.340\pm 0.013$ &                                                    \textbf{dLW}, \textbf{FWHM}, \Halpha, Ca IRT-b,c \\
J09143+526  &     $0.586\pm 0.025$ &    $-0.008\pm 0.012$ &                                \textbf{RV}, \textbf{dLW}, FWHM, BIS, \Halpha, \textbf{Ca IRT-a,b,c} \\
J09144+526  &     $0.592\pm 0.023$ &    $-0.015\pm 0.010$ &                                  \textbf{RV}, \textbf{dLW}, \textbf{\Halpha}, \textbf{Ca IRT-a,b,c} \\
J09561+627  &     $0.574\pm 0.020$ &  $-0.0620\pm 0.0060$ &                                     \textbf{RV}, dLW, FWHM, \textbf{\Halpha}, \textbf{Ca IRT-a,b,c} \\
J10122-037  &     $0.526\pm 0.020$ &    $-0.029\pm 0.012$ &                                                          RV, \Halpha, Ca IRT-a,b, \textbf{Ca IRT-c} \\
J10196+198  &     $0.436\pm 0.017$ &    $-4.520\pm 0.040$ &                                                                           \textbf{RV}, \textbf{CRX} \\
J10482-113  &   $0.1167\pm 0.0096$ &    $-2.270\pm 0.050$ &                                                                                    RV, \textbf{dLW} \\
J10564+070  &     $0.132\pm 0.010$ &    $-7.220\pm 0.080$ &                                                                      \textbf{RV}, \textbf{CRX}, dLW \\
J10584-107  &     $0.208\pm 0.011$ &    $-0.620\pm 0.060$ &                                                                                    \textbf{RV}, CRX \\
J11026+219  &     $0.536\pm 0.019$ &    $-0.486\pm 0.013$ &                                                                                        RV, CRX, BIS \\
J11302+076  &     $0.439\pm 0.018$ &     $0.041\pm 0.011$ &                                                         \textbf{RV}, \Halpha, \textbf{Ca IRT-a,b,c} \\
J11511+352  &     $0.456\pm 0.018$ &    $-0.065\pm 0.010$ &                                                                                         \textbf{RV} \\
J13299+102  &     $0.490\pm 0.018$ &    $-0.015\pm 0.012$ &                                                                                         \textbf{RV} \\
J15218+209  &     $0.593\pm 0.022$ &    $-2.880\pm 0.018$ &                                                                       \textbf{RV}, CRX, Contr., BIS \\
J16167+672S &     $0.627\pm 0.023$ &     $0.065\pm 0.012$ &                                   \textbf{RV}, \textbf{dLW}, Contr., \Halpha, \textbf{Ca IRT-a,b,c} \\
J16555-083  &   $0.1049\pm 0.0093$ &    $-4.540\pm 0.060$ &                                               RV, \textbf{dLW}, \textbf{FWHM}, \textbf{Contr.}, BIS \\
J16581+257  &     $0.514\pm 0.020$ &  $-0.0460\pm 0.0070$ &                                                      \Halpha, Ca IRT-a, \textbf{Ca IRT-b}, Ca IRT-c \\
J17303+055  &     $0.537\pm 0.019$ &    $-0.022\pm 0.013$ &                RV, dLW, \textbf{FWHM}, Contr., BIS, \textbf{\Halpha}, \textbf{Ca IRT-a,b}, Ca IRT-c \\
J17378+185  &     $0.426\pm 0.017$ &  $-0.0660\pm 0.0070$ &                                RV, CRX, \textbf{dLW}, FWHM, \textbf{\Halpha}, \textbf{Ca IRT-a,b,c} \\
J17578+046  &     $0.172\pm 0.010$ &                $0.0$ &                                                  \textbf{dLW}, \textbf{FWHM}, BIS, \textbf{\Halpha} \\
J18174+483  &     $0.587\pm 0.022$ &    $-1.580\pm 0.012$ &                          RV, CRX, dLW, BIS, \textbf{\Halpha}, Ca IRT-a, \textbf{Ca IRT-b}, Ca IRT-c \\
J18198-019  &     $0.593\pm 0.021$ &               $-0.6$ &                                     \textbf{RV}, \textbf{dLW}, FWHM, \Halpha, \textbf{Ca IRT-a,b,c} \\
J18498-238  &     $0.184\pm 0.011$ &    $-2.220\pm 0.018$ &                                                   RV, CRX, \textbf{dLW}, FWHM, Contr., \textbf{BIS} \\
J18580+059  &     $0.559\pm 0.020$ &                $0.0$ &                                               \textbf{dLW}, \textbf{\Halpha}, \textbf{Ca IRT-a,b,c} \\
J19346+045  &     $0.564\pm 0.019$ &     $0.104\pm 0.017$ &                            \textbf{dLW}, FWHM, Contr., BIS, \textbf{\Halpha}, \textbf{Ca IRT-a,b,c} \\
J20305+654  &     $0.385\pm 0.016$ &    $-0.235\pm 0.010$ &                                                                                             \Halpha \\
J20451-313  &     $0.661\pm 0.016$ &               $-2.3$ &                                                          \textbf{RV}, FWHM, Contr., BIS, Ca IRT-b,c \\
J21164+025  &     $0.430\pm 0.017$ &   $0.1020\pm 0.0060$ &                                              \textbf{FWHM}, \textbf{\Halpha}, \textbf{Ca IRT-a,b,c} \\
J21221+229  &     $0.498\pm 0.019$ &  $-0.0320\pm 0.0060$ &                        \textbf{dLW}, \textbf{FWHM}, Contr., \textbf{\Halpha}, \textbf{Ca IRT-a,b,c} \\
J22021+014  &     $0.548\pm 0.021$ &   $0.0240\pm 0.0050$ &                                                                           RV, \Halpha, Ca IRT-a,b,c \\
J22057+656  &     $0.482\pm 0.019$ &     $0.113\pm 0.011$ &                                                                           \textbf{RV}, \textbf{CRX} \\
J22114+409  &     $0.160\pm 0.010$ &    $-5.740\pm 0.050$ &                                                                           RV, \textbf{FWHM}, Contr. \\
J22115+184  &     $0.565\pm 0.022$ &   $0.0320\pm 0.0090$ &                     RV, \textbf{dLW}, FWHM, \textbf{Contr.}, \Halpha, Ca IRT-a, \textbf{Ca IRT-b,c} \\
J22468+443  &     $0.339\pm 0.015$ &    $-4.980\pm 0.021$ &                            RV, \textbf{CRX}, \textbf{dLW}, FWHM, Contr., BIS, \Halpha, Ca IRT-a,b,c \\
J22565+165  &     $0.532\pm 0.020$ &     $0.070\pm 0.014$ &  \textbf{RV}, \textbf{dLW}, \textbf{FWHM}, \textbf{Contr.}, \textbf{\Halpha}, \textbf{Ca IRT-a,b,c} \\
J23492+024  &     $0.396\pm 0.016$ &    $-0.023\pm 0.014$ &                                                        \textbf{RV}, \textbf{FWHM}, \textbf{\Halpha} \\
\hline
\end{tabular}
\end{table*}


\begin{figure*}
\centering
\begin{subfigure}[b]{0.32\linewidth}
\includegraphics[width=\textwidth]{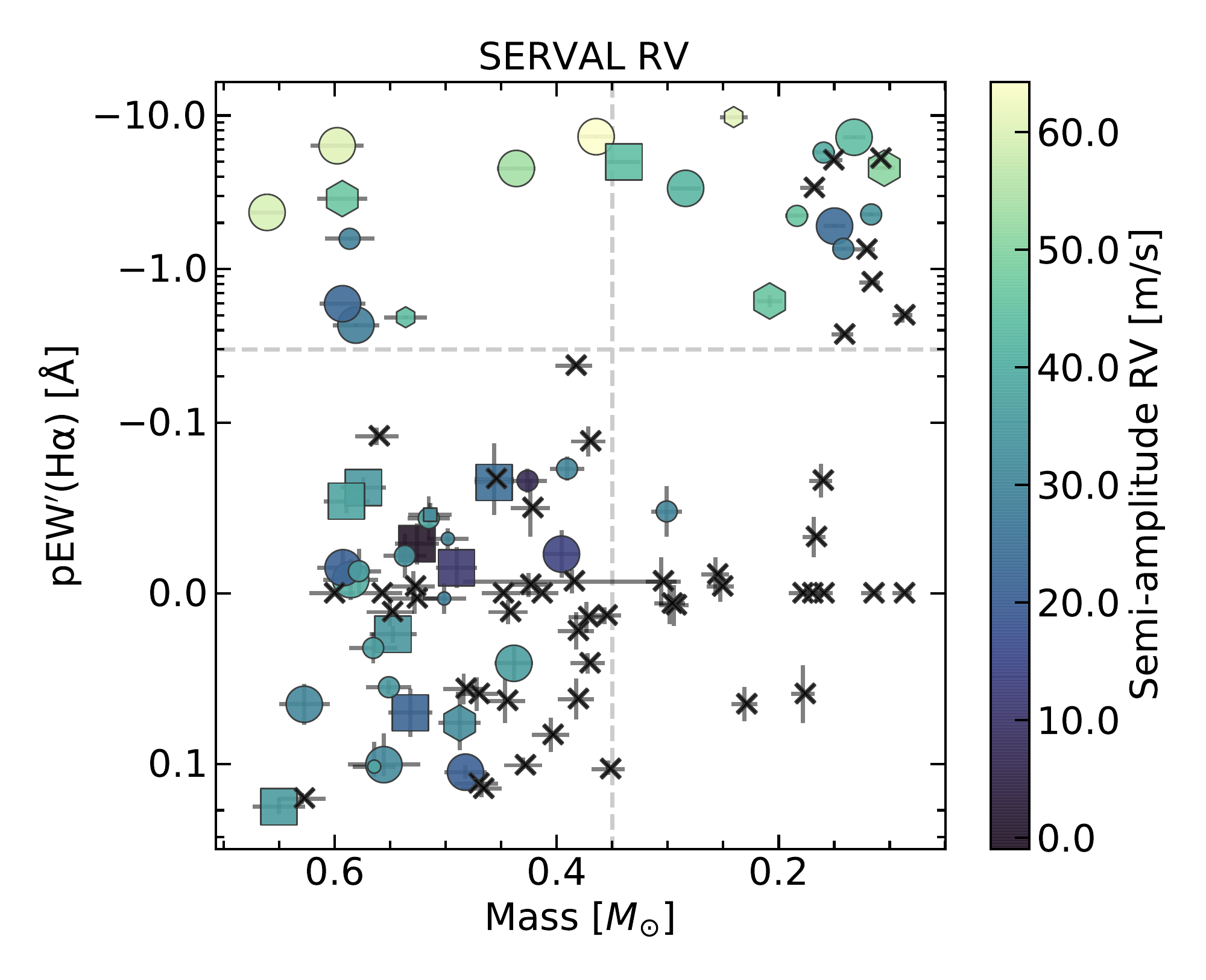}
\end{subfigure}
\,
\begin{subfigure}[b]{0.32\linewidth}
\includegraphics[width=\textwidth]{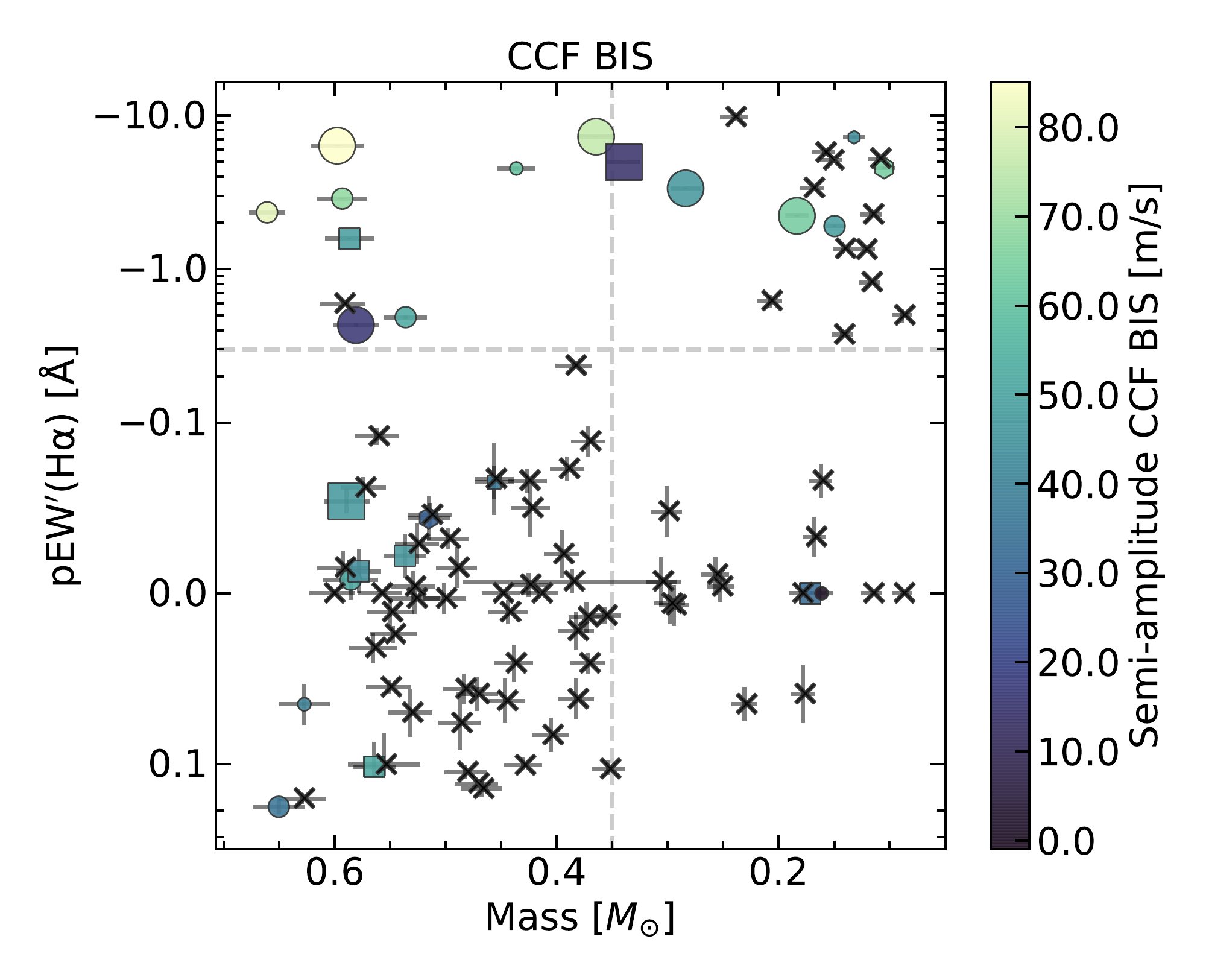}
\end{subfigure}
\,
\begin{subfigure}[b]{0.32\linewidth}
\includegraphics[width=\textwidth]{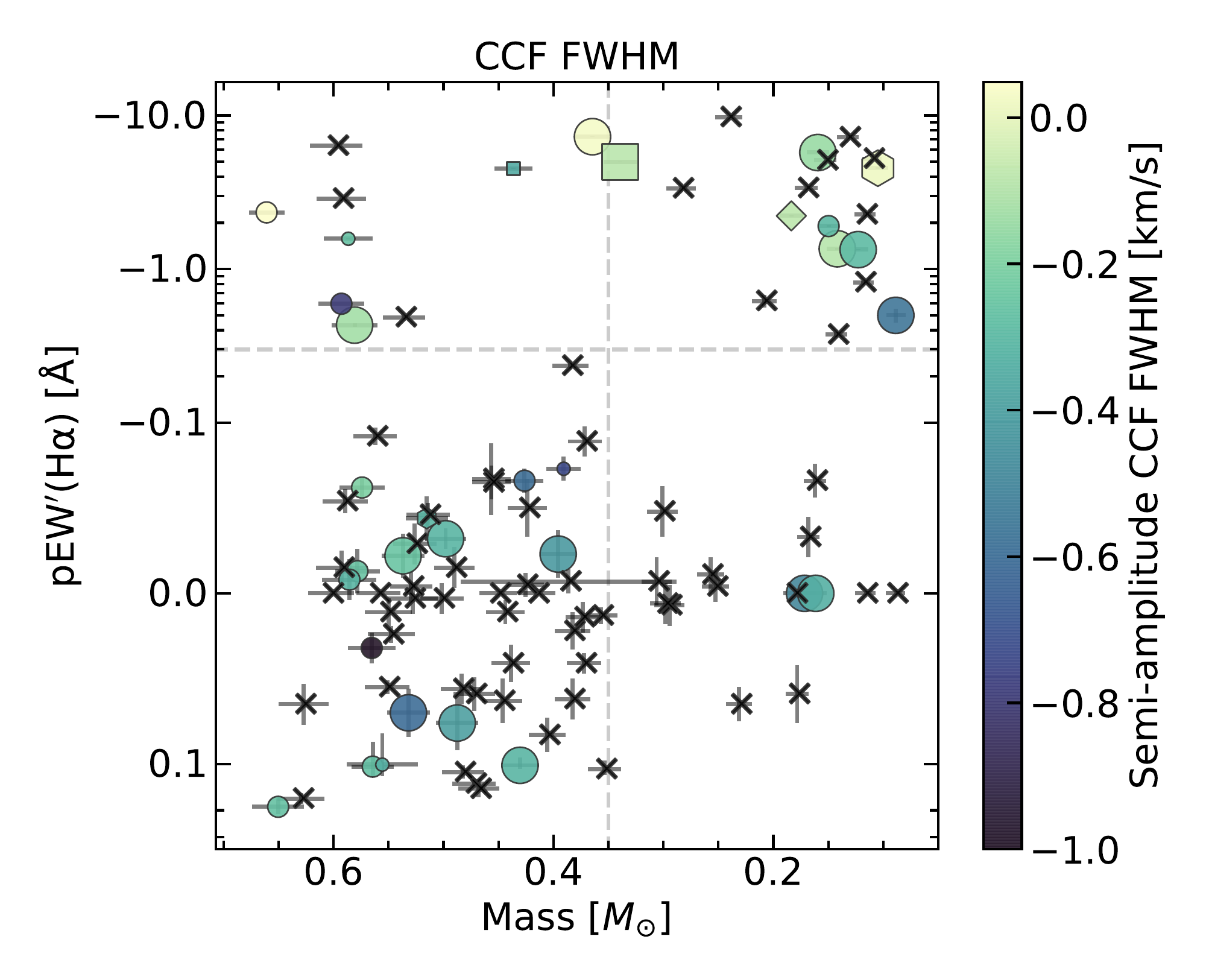}
\end{subfigure}
\\
\begin{subfigure}[b]{0.32\linewidth}
\includegraphics[width=\textwidth]{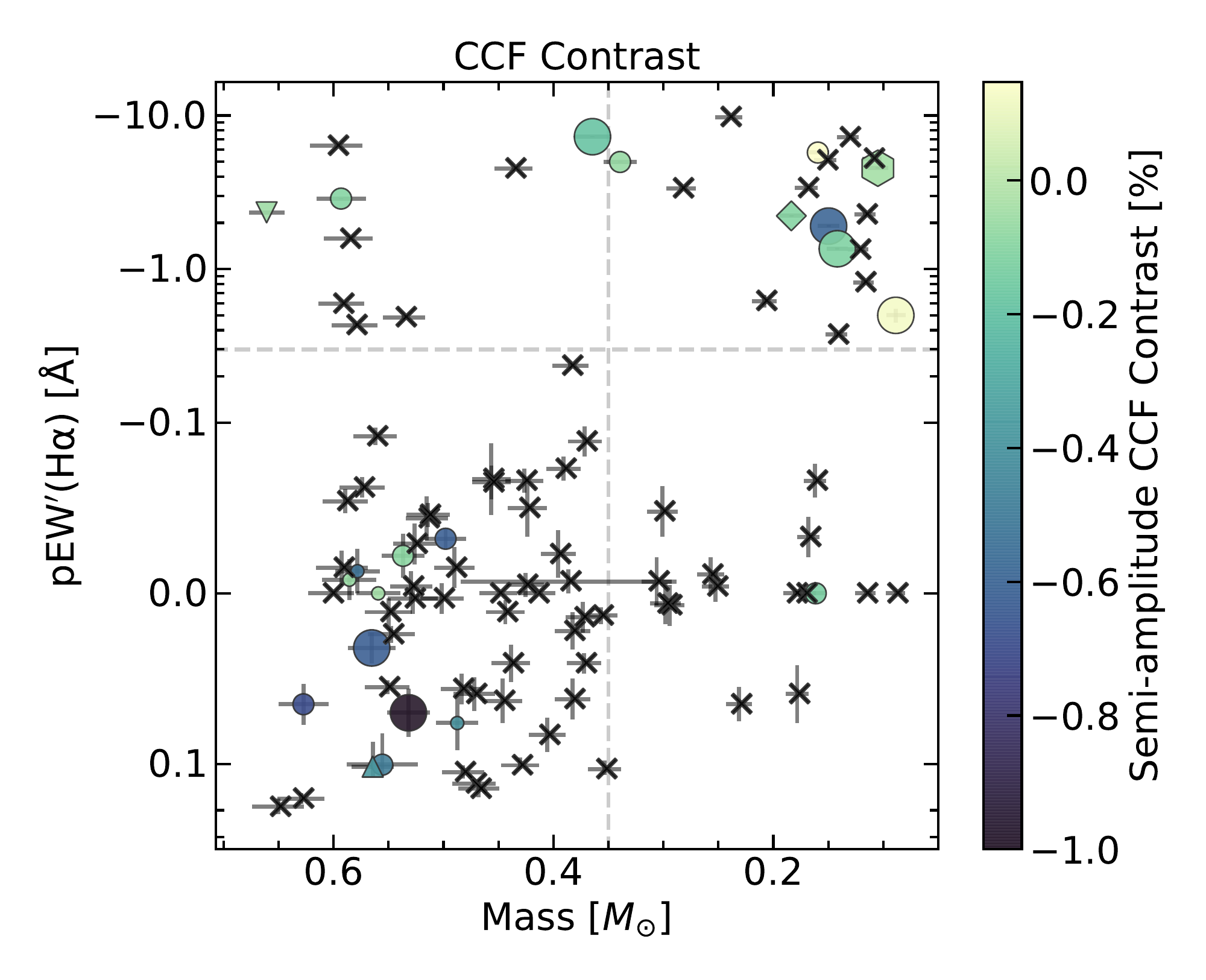}
\end{subfigure}
\,
\begin{subfigure}[b]{0.32\linewidth}
\includegraphics[width=\textwidth]{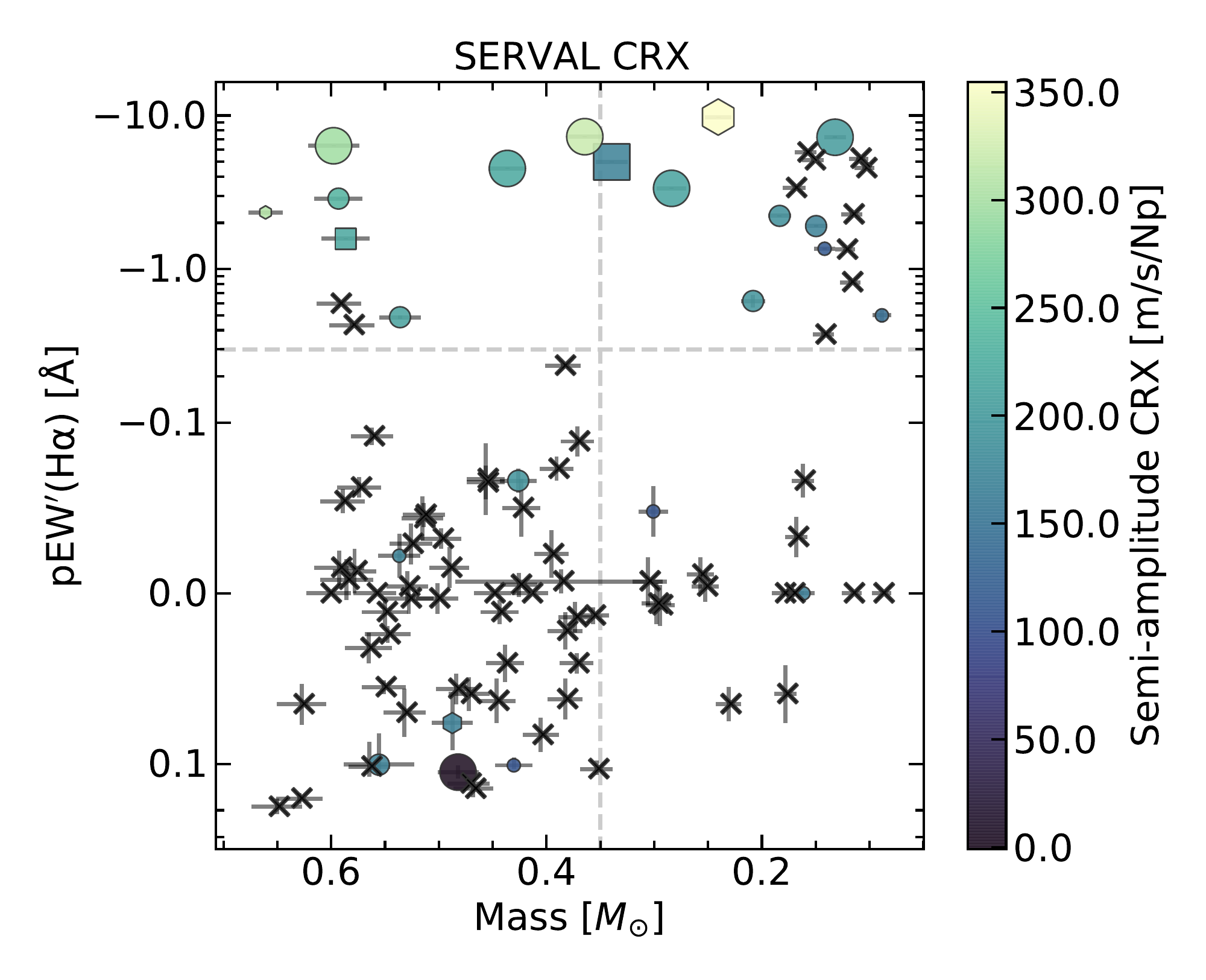}
\end{subfigure}
\,
\begin{subfigure}[b]{0.32\linewidth}
\includegraphics[width=\textwidth]{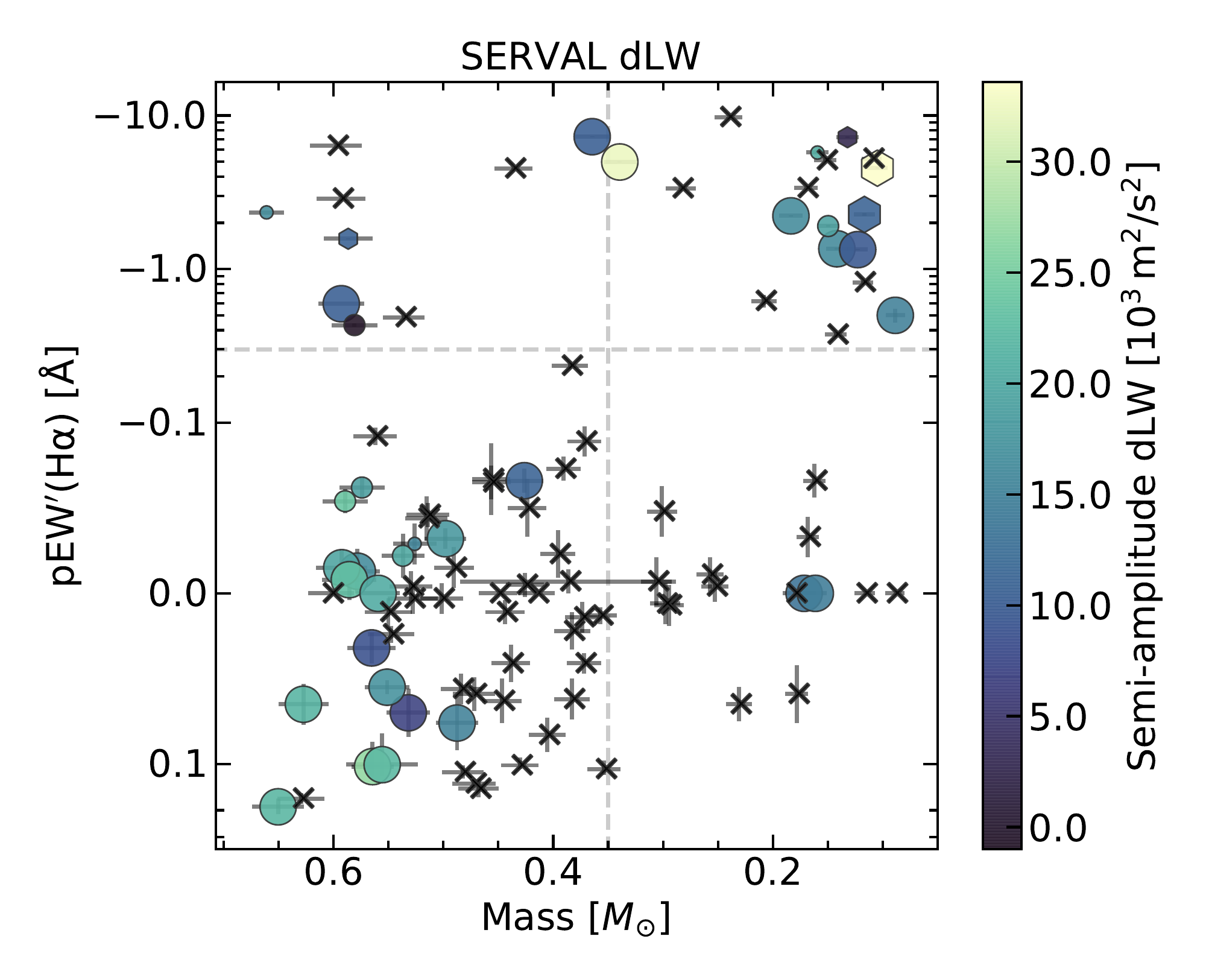}
\end{subfigure}
\\
\begin{subfigure}[b]{0.32\linewidth}
\includegraphics[width=\textwidth]{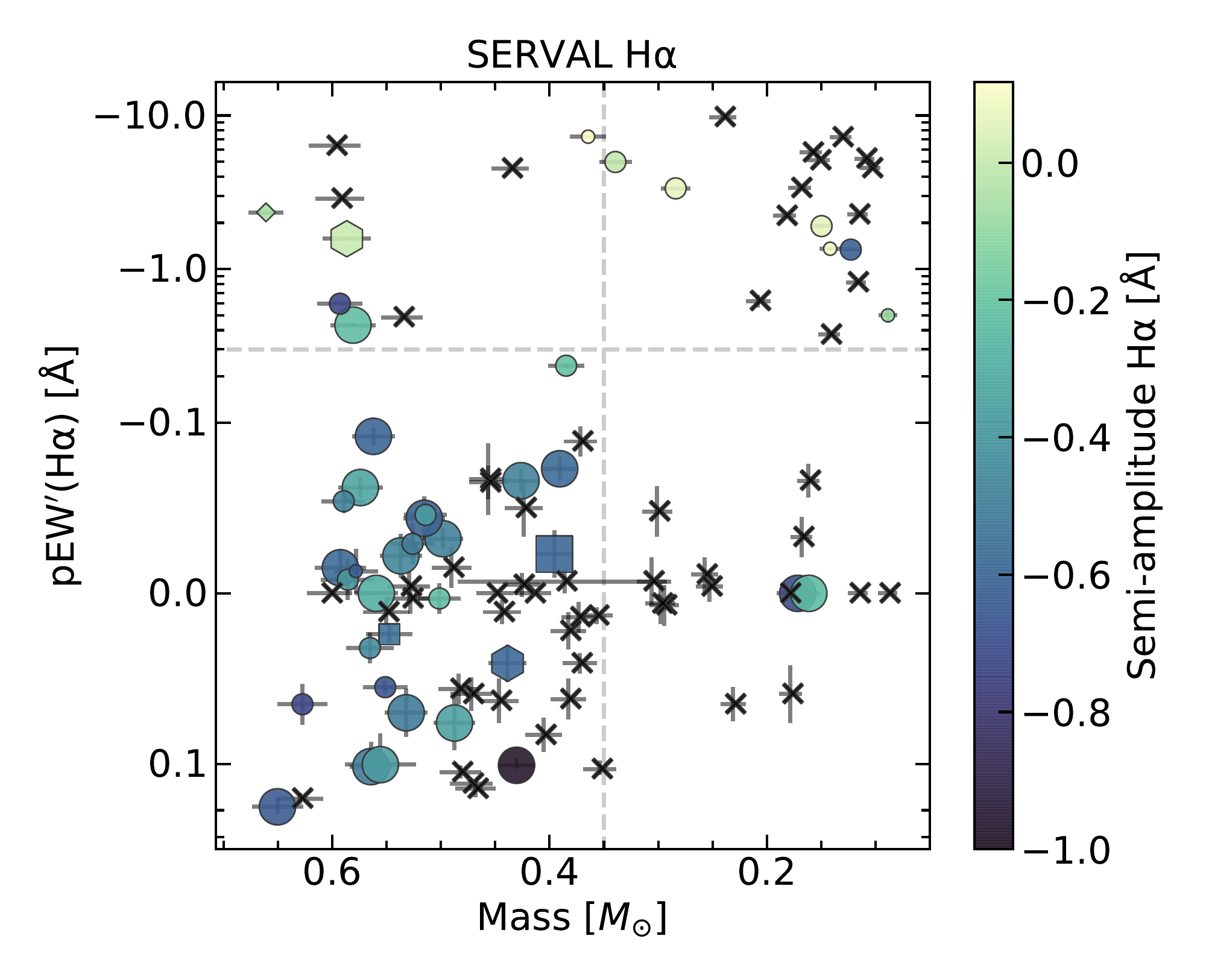}
\end{subfigure}
\,
\begin{subfigure}[b]{0.32\linewidth}
\includegraphics[width=\textwidth]{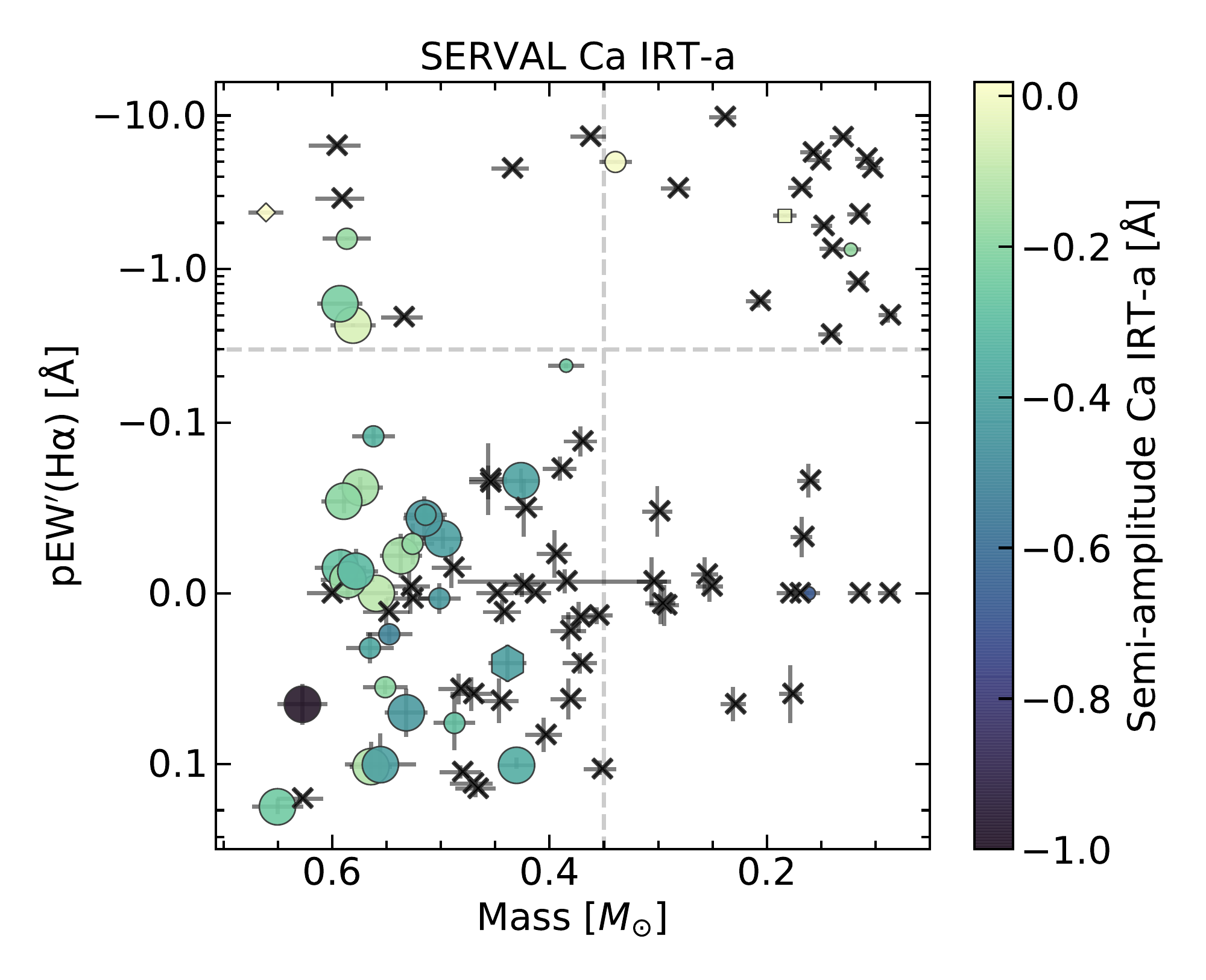}
\end{subfigure}
\,
\begin{subfigure}[b]{0.32\linewidth}
\includegraphics[width=\textwidth]{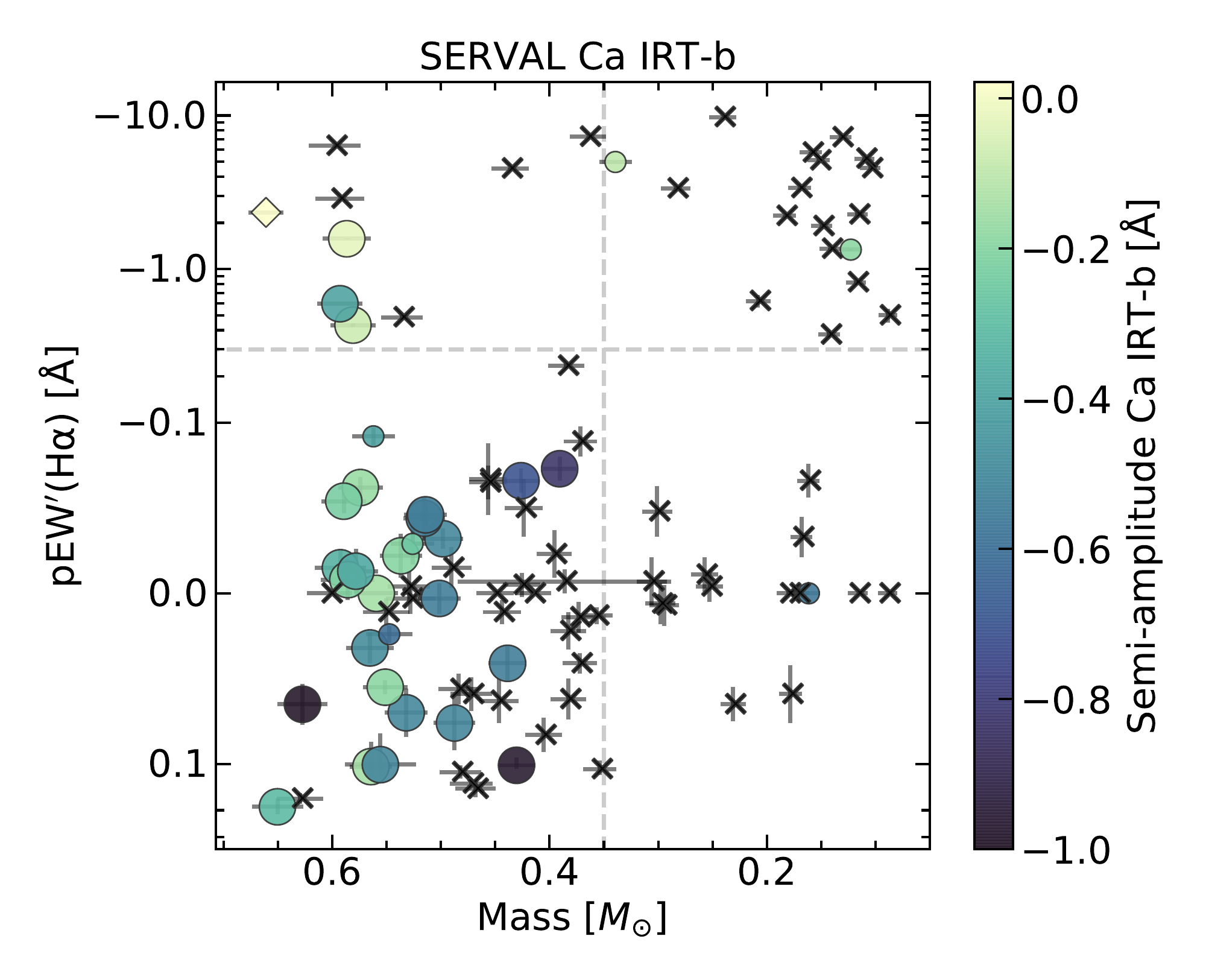}
\end{subfigure}
\\
\begin{subfigure}[b]{0.32\linewidth}
\includegraphics[width=\textwidth]{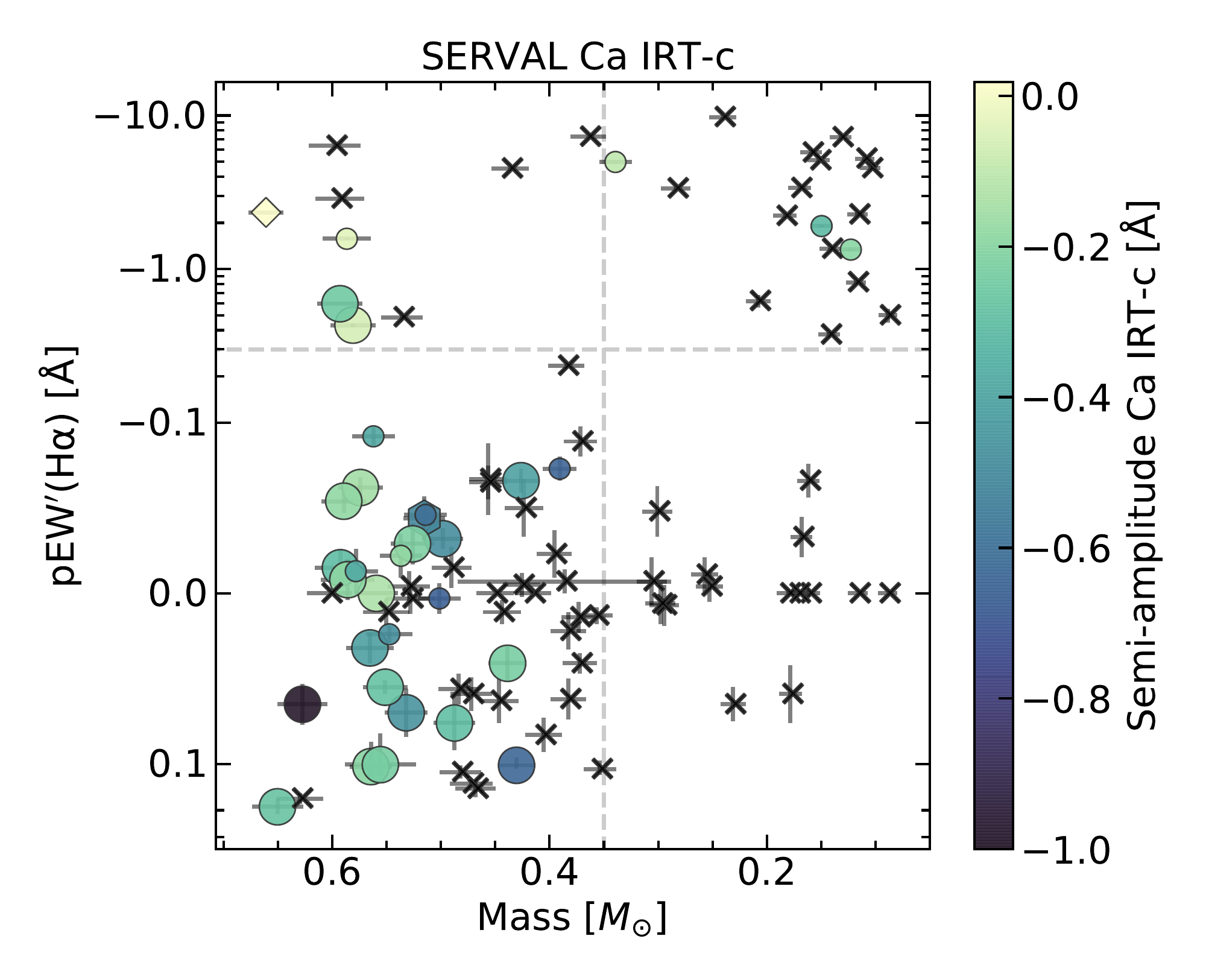}
\end{subfigure}
\\
\begin{subfigure}[b]{0.32\linewidth}
\includegraphics[width=\textwidth]{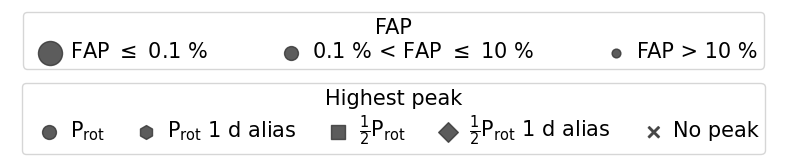}
\end{subfigure}
\caption{Activity detections in the 98 selected stars for the ten parameters analysed (\emph{left} to \emph{right}, \emph{top} to \emph{bottom}: RV, CCF BIS, CCF FWHM, CCF contrast, CRX, dLW, \Halpha, Ca IRT-a, b, and c).
Stars are shown as a function of their average activity level (measured as the \pEWHalpha of the \serval template) and mass.
Different markers represent the activity signal with the smallest FAP identified in the periodogram of each star (circles correspond to \Prot, hexagons to 1-day \Prot alias, squares to \Prothalf, diamonds to 1-day \Prothalf alias, and crosses indicate that no activity-related peak was found).
Symbol sizes indicate the FAP value of the corresponding peak (larger size means smaller FAP), and the points are colour coded as a function of the semi-amplitude of the best-fitting sinusoid to the corresponding period.}
\label{fig:actindsummaryresults}
\end{figure*}

To visualise which of the indicators show a modulation due to activity depending on the properties of the stars, we plot a summary of the results 
in Fig. \ref{fig:actindsummaryresults}. Each panel corresponds to the results obtained for one of the parameters analysed (RVs and the nine activity indicators).
All the panels show the average activity level of the selected targets on the y-axis as a function of their mass on the x-axis.
We used the same quantities as in Fig. \ref{fig:actindsummaryobj}: average \pEWHalpha measured from a template of the CARMENES observations themselves as in \citet{schofer2019carmenesActInd}, and masses from \citet{schweitzer2019carmenesMR}.
In all the panels, each data point corresponds to one of the 98 selected stars.
For many targets, we found more than one peak related to activity (at \Prot, its harmonics, or 1-day aliases).
In the panels, we show the properties of the most significant peak.
Different symbols indicate at which frequency the highest peak was found, and their size corresponds to the FAP of that peak, with larger sizes indicating smaller FAPs.
Finally, the data points are colour-coded according to the semi-amplitude of the sinusoid corresponding to that peak.
This type of representation allow us to see how the different indicators behave depending on the mass and activity level of the stars.


\subsection{RV}

Of the 56 stars for which we find a signal in any parameter, 44 show significant (FAP\,$\leq$\,10\,\%) activity signals in the RVs (25\,\% of them have signals with FAP $\leq$ 0.1\,\%).
In the upper left panel of Fig. \ref{fig:actindsummaryresults}, we see how they are distributed in the activity-mass space.
Most of the stars with a relatively high mass ($M\gtrsim0.3 - 0.4\,\Msun$) and with high activity levels ($\pEWHalpha\lesssim -0.3$, stars near the top left corner of the panel) show significant (FAP\,$\leq$\,0.1\,\%) activity signals at \Prot.
The only exceptions are J22468+443, for which the most significant peak is at \Prothalf (see Sect. \ref{sec:periodogramJ22468+443}), and J15218+209, for which the highest peak is found at the 1-day alias of \Prot.
Two other targets show slightly less significant signals ($0.1\,\%<\FAP\leq10\,\%$) at \Prot and its 1-day alias (J18174+483 and J11026+219, respectively).
Some of the less massive ($M\lesssim 0.3 - 0.4\,\Msun$) but active stars ($\pEWHalpha\lesssim -0.3$, top right corner of the panel), also show clear signals (with FAPs from $\leq0.1$ to 10\,\%) at either \Prot or its 1-day alias, but for seven stars (all with $M\leq 0.2\,\Msun$ and spectral type later than M5.0\,V), we do not observe any significant signal in the periodogram.
Six of those seven are some of the faintest stars in the sample ($J \gtrsim 8\,\mathrm{mag}$), and another one (J23419+441), has a long \Prot, of $\sim 106\,\days$ \citep{diezalonso2019carmenesRotPhot}.
Regarding the stars with low activity level ($\pEWHalpha\gtrsim -0.3$), less than half of the high-mass sub-sample ($M\gtrsim 0.3 - 0.4\,\Msun$, bottom left corner of the panel) show clear signals at either \Prot or \Prothalf.
In the low-mass regime ($M\lesssim 0.3 - 0.4\,\Msun$, bottom right corner of the panel), almost none of the stars show any signal.

\subsection{CRX and BIS}

Both CRX and CCF BIS show a similar behaviour (middle panels in the first and second rows of Fig. \ref{fig:actindsummaryresults}).
For these indicators, most of the detections correspond to the most active stars ($\pEWHalpha\lesssim -0.3$), especially in the high-mass regime ($M\gtrsim0.3 - 0.4\,\Msun$).
In the active and low-mass regime ($\pEWHalpha\lesssim -0.3$ and $M \lesssim 0.3 - 0.4\,\Msun$), the situation is similar to the RVs: over half of the targets (including some of the faintest ones) do not show any significant signal.
There are some detections for the stars in the low-activity regime ($\pEWHalpha\gtrsim -0.3$), but they are in general much less significant.

\subsection{dLW, FWHM, and contrast}

CCF FWHM and dLW (right panels in the first and second rows of Fig. \ref{fig:actindsummaryresults}) show some significant detections across all the activity levels and masses.
In the high-activity regime ($\pEWHalpha\leq-0.3$), they show significant detections, mostly at \Prot and its 1-day alias, for some of the lowest-mass ($M<0.3\,\Msun$) stars (six with $\FAP\leq0.1\,\%$), for which CRX and BIS do not show signals.
About half of the more massive stars show signals (four with $\FAP\leq0.1\,\%$).
For the low-activity, high-mass stars, we find significant detections, mostly at \Prot, in less than half of the targets, in general common to the ones with RV detections.
The low-activity, low-mass stars, do not show any signals, as for the RVs, except for two targets, 
J03133+047 (CD~Cet) and J17578+046 (Barnard's~star).
These two stars also show significant peaks in the \Halpha index, but not in the Ca IRT lines.
CCF contrast (second row, left panel of Fig. \ref{fig:actindsummaryresults}) shows a behaviour similar to CCF FWHM, but with slightly 
lower number of detections and with lower significance.

\subsection{Chromospheric lines}

In the chromospheric indicators (bottom panels of Fig. \ref{fig:actindsummaryresults}), we find significant peaks especially in the high-mass, low-activity regime, where most of the stars with detections coincide with the ones that have detections in the RV.
About half of the high-activity, high-mass stars also show some activity-related peaks.
In the high-activity, low-mass regime, one to four active stars, depending on the chromospheric line, show peaks with very low significance ($\FAP>10\,\%$), and, except for the two targets mentioned above for dLW and FWHM (J03133+047 and J17578+046), none of the low-activity, low-mass stars show any signals.
In general, the four lines show similar results. Differences in the detections could be due to the Ca IRT lines containing a photospheric component, or to telluric contamination, which affects the targets to varying degrees, depending on their absolute RV \citep{schofer2019carmenesActInd}. 


\section{Performance of activity indicators as a function of stellar activity and mass}\label{sec:performance}

\begin{figure*}
\centering
\includegraphics[width=0.8\linewidth]{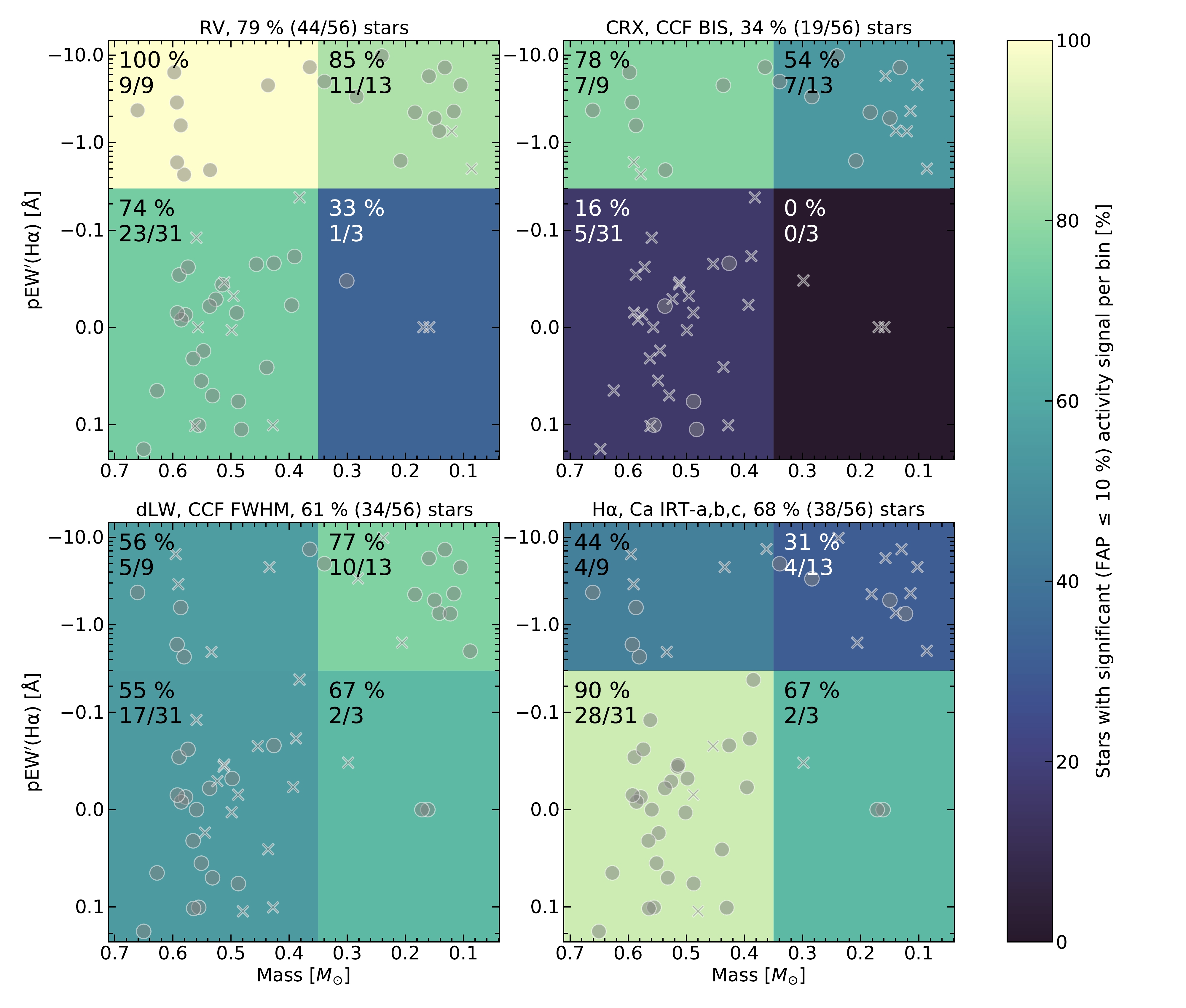}
\caption{Number of stars with activity detections in RV (\emph{top left}), CRX and BIS (\emph{top right}), dLW and FWHM (\emph{bottom left}), and chromospheric lines \Halpha and Ca IRT-a,b,c (\emph{bottom right}).
The stars are divided into four bins, depending on their average activity level and mass. Axes are the same as in Fig. \ref{fig:actindsummaryresults}.
The colours of each bin indicate the number of stars (in percentage) for which we found an activity-related signal with $\FAP\leq10\,\%$.
The text in each bin also shows that percentage, together with the absolute number of stars that have such a detection.
The title of each panel shows the same numbers, but for all the stars (i.e. for the four bins together).
Grey data points indicate the position of the 56 stars considered in the activity-mass space, with large circles representing the stars with a detection in the specific indicator, and crosses, stars with no detection.}
\label{fig:boxes4p0.1}
\end{figure*}

\begin{figure*}
\centering
\includegraphics[width=0.8\linewidth]{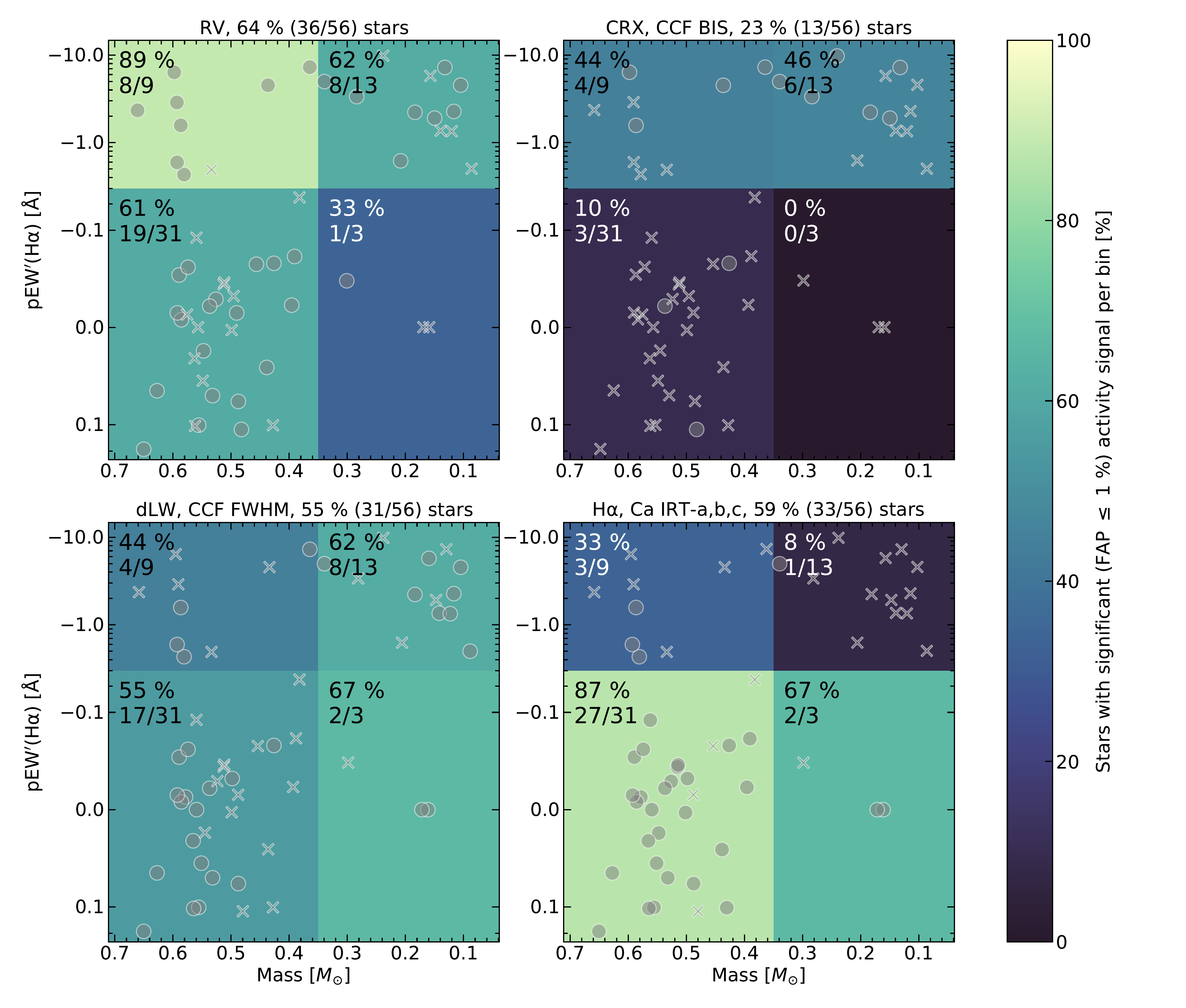}
\label{fig:boxes4p0.01}
\includegraphics[width=0.8\linewidth]{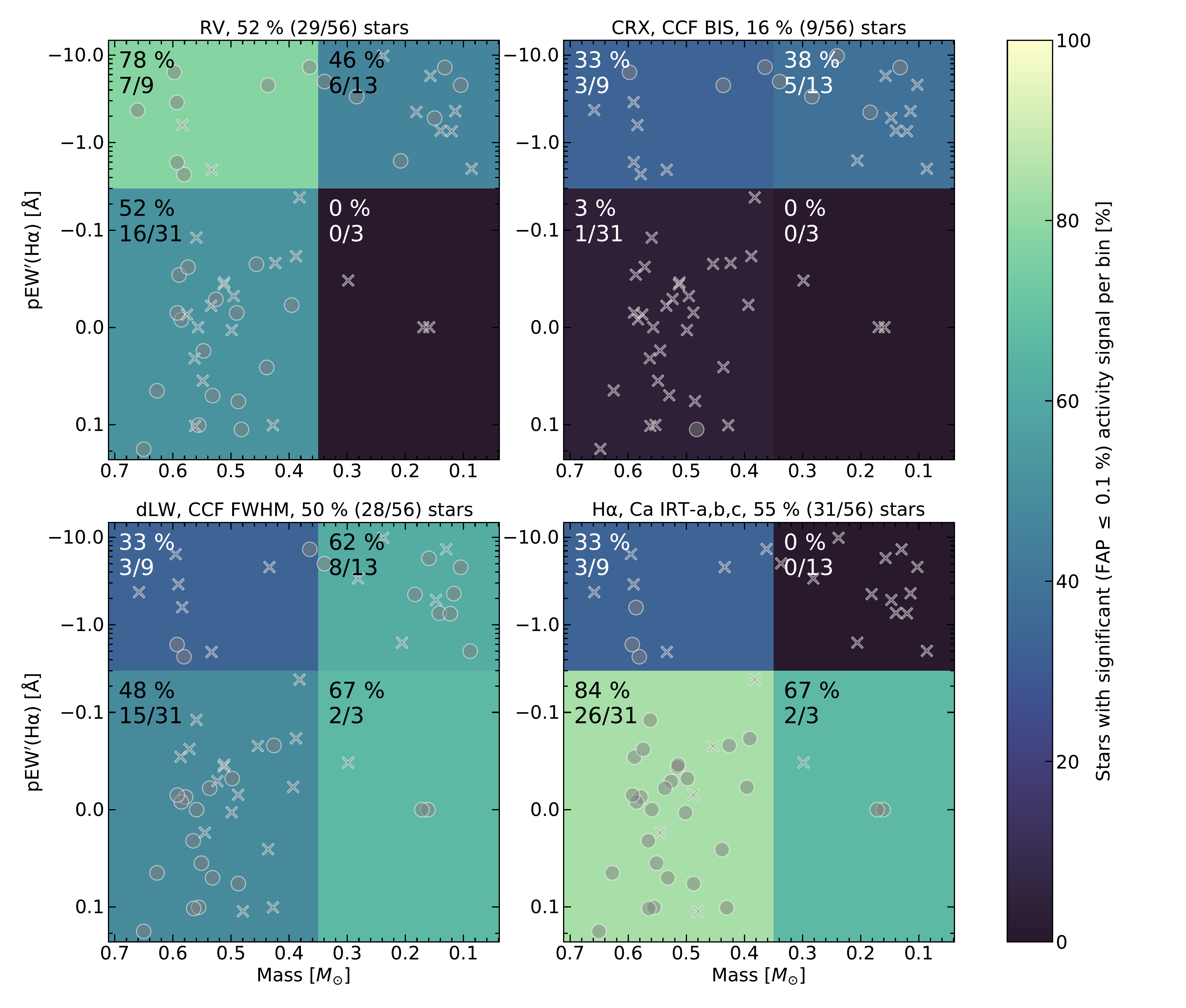}
\caption{Same as Fig. \ref{fig:boxes4p0.1}, but for signals with $\FAP\leq1\,\%$ (\emph{top}) and $\FAP\leq0.1\,\%$ (\emph{bottom}).}
\label{fig:boxes4p0.01_0.001}
\end{figure*}

Given their similar behaviour discussed before, we can group the indicators into 3 categories: (1) CRX and BIS, which trace chromaticity and average line asymmetry, respectively, (2) dLW and FWHM, which trace changes in the width of the stellar absorption lines, and (3) \Halpha and Ca IRT lines, which are proxies of chromospheric emission.
We do not include the contrast because its results were similar to FWHM and dLW, but less significant in general.
We also grouped the 56 stars for which we find at least one activity-related signal into four subsets, depending on their activity level (high, $\pEWHalpha\leq-0.3$, 22 stars, or low, $\pEWHalpha>-0.3$, 34 stars), and their mass (high, $M\geq0.35\,\Msun$, 40 stars, or low, $M<0.35\,\Msun$, 16 stars).
In the four panels of Fig. \ref{fig:boxes4p0.1}, we show the performance of these three groups of indicators, together with the detections in the RVs, for each of the four subsets of stars.
In each panel, the bins correspond to the four subsets of stars, and the colour code and numbers indicate how many of them show a significant (FAP\,$\leq$\,10\,\%) activity signal.
As in Fig. \ref{fig:actindsummaryresults}, we selected the activity-related peak with smallest FAP.
Fig. \ref{fig:boxes4p0.01_0.001} shows the same as Fig. \ref{fig:boxes4p0.1}, but for signals with FAP\,$\leq$\,1\,\% and FAP\,$\leq$\,0.1\,\%.

\subsection{RV}

We find significant ($\FAP\leq10\,\%$) signals in the RVs for most of the stars (almost 80\,\% of them, especially in the high-activity regime), except for low-mass, low-activity, where only one of the three stars in the bin (the target with the highest mass) shows a signal.
The total number of detections decrease if we only consider signals with FAP $\leq1\,\%$ (64\,\% of the stars show a detection) or  FAP $\leq0.1\,\%$ (52\,\% of the stars show a detection), by about the same fraction for each bin, that is to say, we still find more detections for the most active, highest mass stars, and significantly less for the least active, lowest mass ones.
Globally, RV variations are the best tracers of stellar activity signals, which implies that nearly always activity signals can be found in RVs, but not always in the activity indicators.


\subsection{CRX and BIS}

As noted above, CRX and BIS are effective activity tracers for the stars showing the highest activity levels, especially for the ones with the highest mass, where 78\,\% (seven of nine) of the targets in the high-mass high-activity subset show a signal with $\FAP\leq10\,\%$.
If we restrict the FAP to values $\leq1\,\%$, the number of stars with signals decreases, especially for the high-activity high-mass subset, where only the four most active targets (44\,\%) show a signal with that significance (this number goes down to three stars, 33\,\% for signals with FAP $\leq0.1\,\%$).
Of the low-activity stars, only five out of 34 (16\,\%) show clear signals, all of them in the high-mass regime.
For more restrictive FAPs, the number of stars with signals approaches zero: three stars (10\,\%) show a signal with $\FAP\leq1\,\%$, and only one star (3\,\%) has a signal with $\FAP\leq0.1\,\%$.

Both CRX and BIS seem to trace the same type of activity variability, as we observe the same behaviour in the periodogram analysis and similar correlations with RV.
CRX traces RV changes with wavelength, while BIS is sensitive to changes in the average asymmetry in the absorption lines.
Both indicators are sensitive to the brightness distribution over the stellar disk, that is, to the presence of active regions. Active regions such as spots create non-uniform weighting of the blueshifted and redshifted wings of the absorption lines, creating asymmetries, and have a larger effect at shorter wavelengths, that is, have a colour-dependence. Therefore, both BIS and CRX (and RV) show similar variations.
Furthermore, since BIS quantifies the RV difference between the upper and the lower part of the CCF, it should be sensitive to differences between lines formed at different photospheric heights, lines with different depths.
Opacity is higher at shorter wavelengths, so lines in the blue part of the spectrum are formed at larger photospheric heights than those in the red. This means that there should be a dependence on the line depth with wavelength, so that the BIS would be tracing a chromatic effect as the CRX does.

\subsection{Chromospheric lines}

Chromospheric lines show the opposite behaviour as CRX and BIS.
We find significant ($\FAP\,\leq10\,\%$) signals in most (almost 90\,\%, 30 targets) of the stars with low activity levels, and in less than half of the most active ones (eight stars out of 22, 36\,\%).
The signals in most of the low activity stars are all highly significant, since all but two of the total 30 have $\FAP\,\leq0.1\,\%$.
In the high-activity regime, the number of detections decreases with more restrictive FAPs, where only four stars show signals with FAP $\leq$ 1\,\%, and of these, only three with FAP $\leq$ 0.1\,\%.
Regarding the mass, we detect a periodic signal in the chromospheric lines more often in the most massive stars, similarly to RV, and CRX and BIS.

\subsection{dLW and FWHM}

We find that dLW and FWHM perform better for the high-activity, low-mass stars, with signals with $\FAP\leq10\,\%$ in 77\,\% of them (and of these, still 62\% have  $\FAP\leq1\,\%$ and $\FAP\leq0.1\,\%$).
None of the other indicators (except RV) has that many significant signals in this regime.
dLW and FWHM show significant signals for about half of the targets in the three other subsets: 56\,\% for the high-activity high-mass stars, 55\,\% for the low-activity high-mass stars, and 67\,\% for the low-activity low-mass ones.
These numbers slightly decrease when considering FAP $\leq$ 1\,\% or $\FAP\leq0.1\,\%$, but the proportion between subsets remains more or less constant.

\subsection{Overview}

About half of the stars in the sample show clear activity signals in the RVs, especially the most active ones.
Of the different indicators analysed, we find that they behave differently depending on the mass and activity level of the target star.
CRX and BIS work best for the most active stars, while the chromospheric lines \Halpha and Ca~IRT are the most effective in tracing activity in targets with lower activity levels.
dLW and FWHM behave similarly in all mass and activity regimes, but are especially useful for the most active and least massive stars.

We find that 50 to 80\,\% (30 to 60\,\%) of the stars with a high activity level 
show signals with FAP\,$\leq10\,\%$ (FAP\,$\leq$\,1\,\% and \,$\leq$\,0.1\,\%) in photospheric indicators (CRX, BIS, dLW, and FWHM), depending on the mass regime.
For the most massive stars, CRX and BIS work best, with signals in about 80\,\% of the targets.
The other photospheric indicators (dLW and FWHM) and the chromospheric lines (\Halpha and Ca IRT) show signals with FAP\,$\leq10\,\%$ in about 50\,\% of these stars (30\,\% to 40\,\% for FAP\,$\leq$\,1\,\% and \,$\leq$\,0.1\,\%).
For the least massive stars, the most effective indicators are dLW and FWHM, with FAP\,$\leq10\,\%$ signals in over 75\,\% of the targets (and about 60\,\% for FAP\,$\leq$\,1\,\% and \,$\leq$\,0.1\,\%), followed by CRX and BIS, with signals in about 50\,\% of them for FAP\,$\leq10\,\%$ and FAP\,$\leq1\,\%$ (40\,\% of stars for FAP\,$\leq0.1\,\%$).
Only a third of these stars show signals in the chromospheric lines, all of them with large FAPs ($\sim$\,10\,\%).

Regarding stars with low activity levels, most of them ($\sim90\,\%$ in the high-mass, and almost 70\,\% in low-mass regime) show very significant activity signals (FAP\,$\leq$\,0.1\,\%) in chromospheric indicators.
About half of them ($\sim50\,\%$ high-mass, and 70\,\% low-mass stars) also show signals in dLW and FWHM (most of them very significant, with FAP\,$\leq$\,0.1\,\%).
Finally, CRX and BIS indicators are not effective in tracing activity in these low activity stars, with detections in about $\sim15\,\%$ of the most massive stars (most of them with relatively large FAPs, 10\,\% to 1\,\%), and none in the low-mass regime.


These results clearly show the need to use different indicators to assess the presence of activity in RV measurements depending on the characteristics of the target star.
Failing to use the adequate indicators may result in RV signals classified as `planet detections' that are actually false positives.
It follows that it is possible to rule out the existence of planets that have only been cross-checked with a limited set of indicators \citep[e.g.][]{feng2020planets}.
To show a specific example, the unrefereed manuscript \citet{tuomi2019planets} claim the existence of two planets in the very active ($\pEWHalpha\sim-7.2$), late-type (M6.0\,V) star J10564+070 (CN~Leo, GJ~406) based on RV data obtained with the HARPS \citep[23 observations,][]{mayor2003harps} and HIRES \citep[41 observations,][]{vogt1994hires} high-resolution spectrographs.
One of the claimed planet candidates is a hot super-Earth in a $\sim2.68\,\days$ period orbit.
The analysis of different spectroscopic activity indicators, CCF BIS and FWHM from the HARPS data, \CaHK $S$-index from both HARPS and HIRES, and photometric time series from ASAS \citep{pojmanski1997asas}, results in no significant periodicities related to the planet orbital period. Therefore, the authors conclude that the 2.68\,d signal has no stellar origin.
However, we have just seen that in active and low-mass stars such as this one, the chromospheric lines tend to not show activity signals, and other indicators such as CRX and BIS, or dLW and FWHM work better.
In this case, CARMENES observations (see Fig. \ref{fig:actind_tsperiodogramJ10564+070}) show a clear and strong ($\FAP<10^{-13}$) signal at 2.71\,d in the CRX data, which was not analysed in \citet{tuomi2019planets} (we note here that HARPS observes in a different wavelength range than CARMENES, and that indicators such as CRX or BIS could depend on the wavelength band used). This clearly points at a stellar origin for this signal.
There are weaker activity-related periodicities in dLW and BIS, and none in the chromospheric lines.
Moreover, \citet{diezalonso2019carmenesRotPhot} identified the rotation period of the star to be $\sim 2.7$\,d from photometric time series.
Overall, the findings presented here strengthen the notion that caution should be exercised when vetting RV planet candidates, and that a complete activity indicator analysis needs to be performed.


\section{Detection biases} \label{sec:detbiases}

We selected stars with on average more than 40 observations and spanning several nights so that a peridogram analysis of their activity indicators could reveal information related to rotation, which has a time scale of days.
Despite that, there are still several reasons explaining why we do not observe signals in some of the stars.

\begin{figure}
\centering
\includegraphics[width=\linewidth]{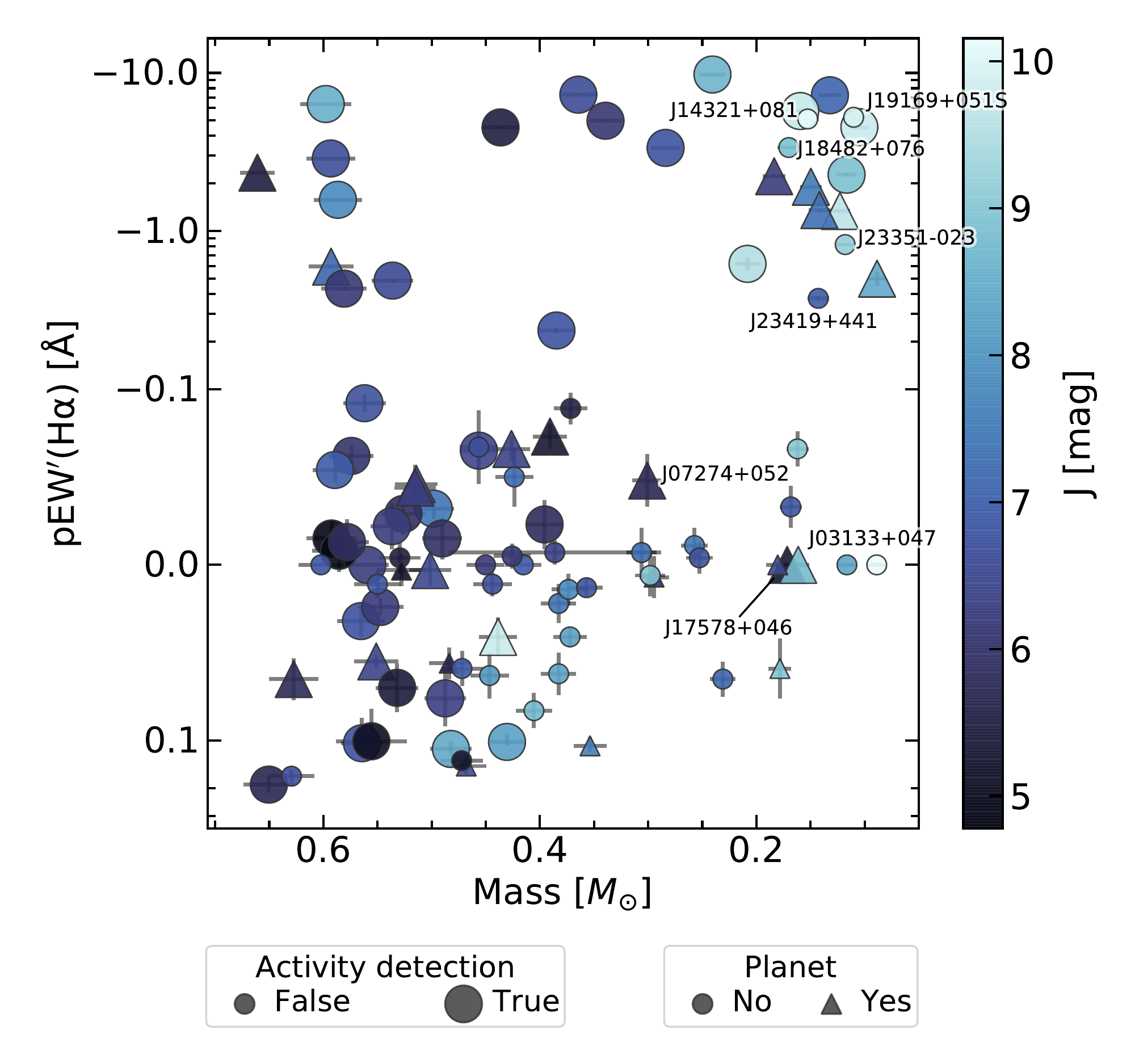}
\caption{Average activity level (\pEWHalpha) as a function of the mass of the 98 sample stars, colour-coded with the $J$ magnitude.
Large data points indicate that a signal (FAP\,$\leq$\,10\,\%) related to the activity of the star was found in at least one parameter, while small data points indicate non-detections.
Stars with no known planetary companions are indicated with circles, while those with confirmed planetary companions are indicated with triangles.
}
\label{fig:m_act_detall}
\end{figure}

We find activity-related signals in 56 of the 98 selected stars.
In Fig. \ref{fig:m_act_detall} we show the average activity of the stars and their mass, as a function of their brightness, and we indicate for which targets we found a detection (with $\FAP\leq10\,\%$) in any of the indicators.
In general, stars for which we find no detections have low activity levels, and are distributed across the entire mass range.
Of the 71 stars in the low-activity regime ($\pEWHalpha\gtrsim -0.3$), 53\,\% (37~stars) show no detection in any parameter.
This number increases if we focus only on the low-mass, low-activity regime ($M\lesssim0.3\,\Msun$, $\sim13$\,stars), where only three stars show a detection: J07274+052 (Luyten's star), J17578+046 (Barnard's star), and J03133+047 (CD Cet).
These are well-studied stars with planetary companions \citep[see, respectively,][]{astudillo-defru2017planetsMdwarf, ribas2018barnard, bauer2020cdcet} and long rotation periods ($\Prot\gtrsim100\,\days$). Since the main aim of the CARMENES survey is to find and characterise exo\-planets, these targets have been observed more frequently than other similar stars, which could explain why we have a detection. 
Of the most active stars ($\pEWHalpha<-0.3$, 27 stars), only 19\,\% (five stars) show no detections (J14321+081, J18482+076, J19169+051S, J23351-023, and J23419+441).
They are low-mass ($M\leq0.17\,\Msun$) stars, and four of them are at the faintest end of the sample ($J\geq8.8\,\mathrm{mag}$), which implies less precise measurements. The fifth star (J23419+441) is brighter ($J\sim6.9$) but shows a long \Prot ($\sim 106\,\days$). These five targets have between 47 and 99 observations.

Clearly, for the least active stars, signals induced by activity can be below the noise floor of our measurements, making them impossible to detect.
Also, stars with lower activity levels show longer rotation periods.
In general, the observations of each star in the sample cover relatively long time spans, of $\sim1000$\,d, and most of the targets have been observed between 50 and 130 times.
However, in stars with long \Prot, this may not always sufficiently sample the rotational modulation so that a significant signal appears in the periodogram.
As mentioned above, another observational constraint is the apparent brightness of the stars, which results in observations with lower S/N.
Therefore, activity signals could be hidden in the noise.
The faintness of some stars at the low-mass end of the sample could be impeding activity detections, and the combination of a relatively small number of observations and long rotation periods could be a problem when looking for activity signals in the least active targets.

Aside from the overall S/N of the spectra, our ability to detect activity-related signals in different types of stars could also be limited by other stellar properties.
CRX measurements may vary depending on the stellar temperature, since for the coolest stars, the RV information is contained in a shorter wavelength range than earlier type M dwarfs, making stellar colour more relevant than brightness.
The shape of photospheric absorption lines, which depends on spectral type, can also differently affect the indicators derived from the line profiles.

The average activity level of the stars may vary with time as a result of a long-period magnetic cycle.
Some of the stars could have been observed at a low-activity phase of the cycle, which would decrease the level of the activity signals, making them more difficult to detect.
Also, some stars have groups of observations separated by several months, and we could be combining epochs with very different activity levels, which could also reduce the significance of the signals.

The inclination of the stellar rotation axis with respect to our line of sight also plays a role in the total amplitude of the activity signals.
In stars with nearly pole-on inclination, visible active regions do not appear and disappear as the star rotates.
This results in smaller-amplitude modulations, which again makes them more difficult to detect.
For perfectly pole-on configurations, in principle \Prot cannot be measured.
However, pole-on configurations are unlikely, and statistically should only occur for a small fraction of the targets.
The distribution of active regions also affects the amplitude of activity modulations.
Similarly to the case of a pole-on configuration, stars with a homogeneous distribution of active regions on their surface may display signals with smaller amplitudes.
We also do not expect a clear modulation in the case of large polar spots common in fast-rotators, for stars with edge-on inclinations \citep[e.g.][]{schuessler1992polarspots,piskunov1994polarspots,yadav2015polarspots}.
Complex activity patterns on the stellar surface may induce signals that are not exactly at \Prot, but at its harmonics \citep[e.g.][]{boisse2011disentangling}, although we tried to account for that.
Stars can also have differential rotation, which can make the very definition of \Prot fuzzy.
Also, active regions could have short life-spans. The fact that we are able to identify activity-related peaks in the periodograms of several sample stars could indicate persistent active region patters on these stars. 
Changes in the activity surface patterns on a time scale shorter than the total time span of the observations could hamper the detection of a coherent signal related to \Prot. 


\section{RV jitter-activity relation} \label{sec:jitter}

\begin{figure*}
\centering
\includegraphics[width=0.42\linewidth]{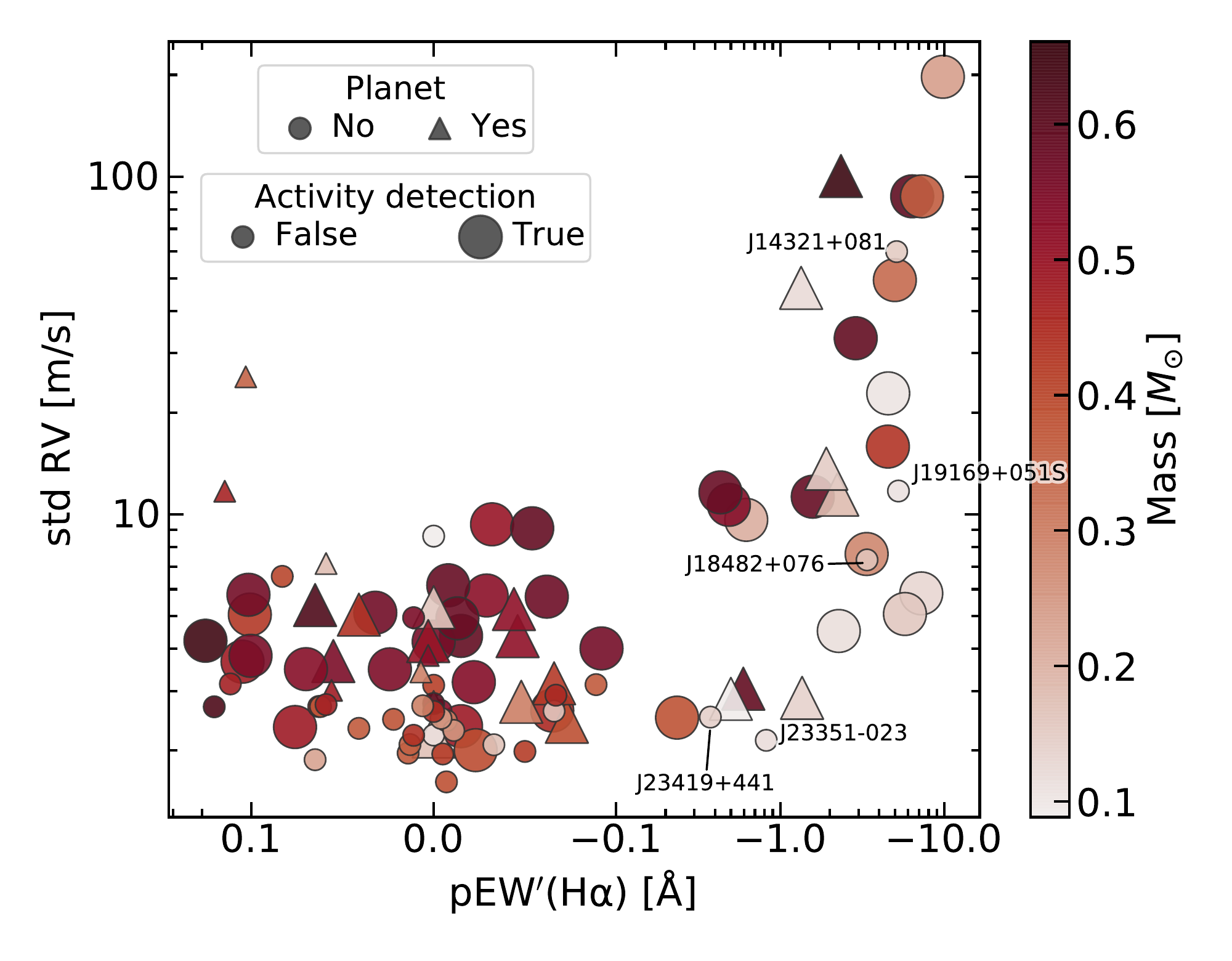}
\,
\includegraphics[width=0.42\linewidth]{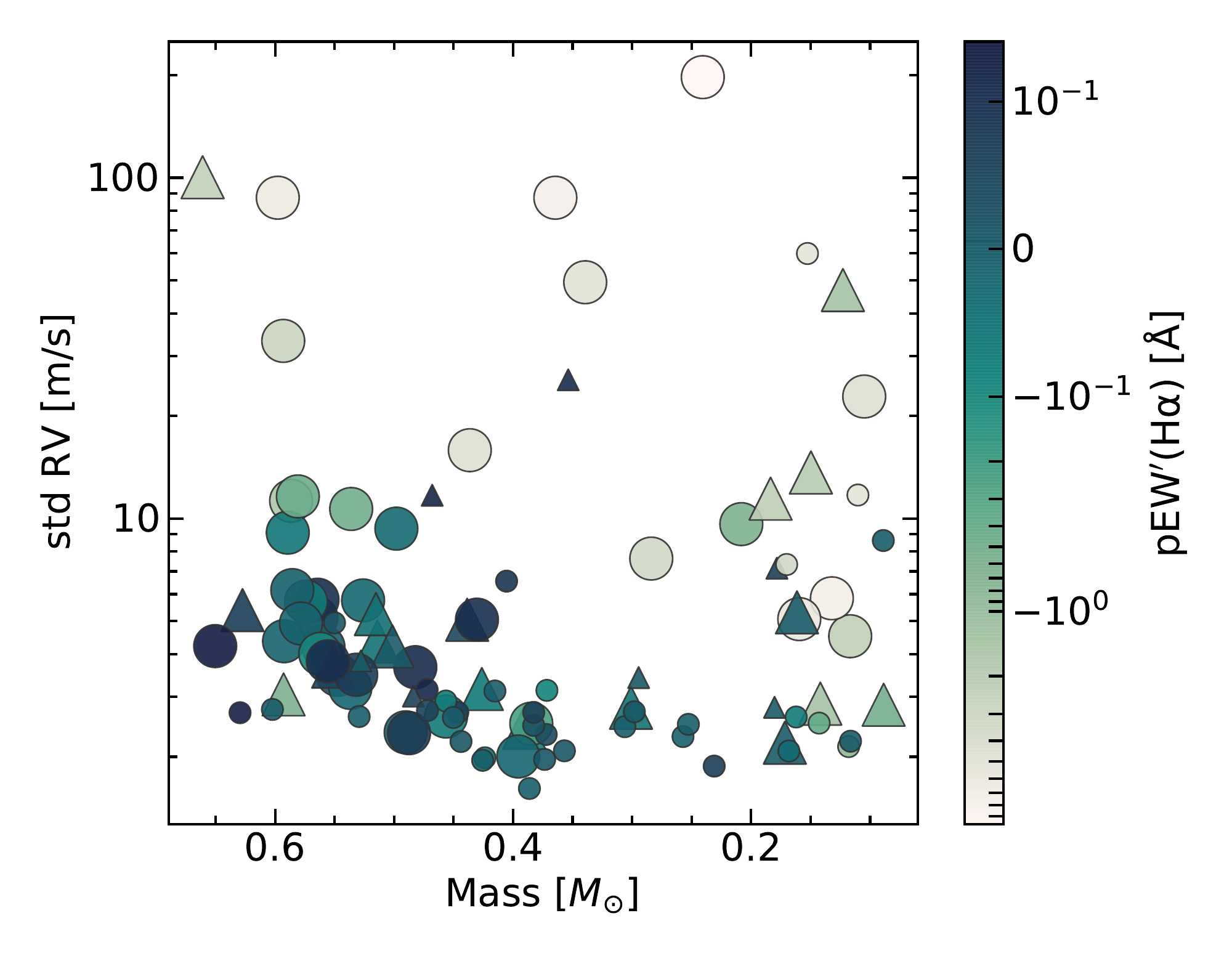}
\caption{\emph{Left}: RV scatter (std RV) as a function of the average activity level (\pEWHalpha) of the 98 stars in the sample, colour-coded with the stellar mass.
Stars with no known planetary companions are indicated with circles, while those with confirmed planetary companions are indicated with triangles.
Large data points correspond to the 56 stars for which we find an activity-related signal in any of the indicators, and small data points, stars with no detection.
\emph{Right}: RV scatter as a function of the stellar mass of the 98 stars in the sample, colour-coded with the average activity level (\pEWHalpha). Same symbols and sizes as in the left panel.}
\label{fig:rvjitter_act_mass}
\end{figure*}

Next, we analyse the RV scatter of the 98 stars in the sample.
Fig.~\ref{fig:rvjitter_act_mass} shows the relation between the RV scatter (measured as the standard deviation of the RV, std RV), the average activity level (measured as the average \pEWHalpha), and the stellar mass of the 98 stars in the sample.
We also indicate the presence of planetary companions, and whether we detected any kind of activity-related signal in any of the parameters studied.
We did not remove RV variations induced by planetary companions, so the RV scatter of stars identified as having planets is only an upper limit to the RV variation caused by activity.

\subsection{High-activity stars}

In the high-activity regime ($\pEWHalpha\lesssim-0.3$, 27 stars), the RV scatter increases with the average activity level of the stars, from values $\sim2\,\ms$ to over $100\,\ms$.
Of these 27 stars, 11 have RV scatter $\lesssim10\,\ms$.
They correspond mostly to low-mass stars (ten of them, i.e. 90\,\%, have masses $<0.4\,\Msun$). We note that, according to our definition of active stars (\Halpha in emission), most low-mass stars are active.
In the other 16 targets with std\,RV\,$>10\,\ms$, the mass distribution is approximately equally divided between low masses (nine stars, 6\,\%, with $M<0.4\,\Msun$) and high masses (seven stars, 4\,\%, with $M\geq0.4\,\Msun$).

For most of these 27 stars, we were able to identify an activity-related signal in at least one of the indicators. As mentioned above (Sect. \ref{sec:detbiases}), there are five targets with no detections, which have low mass, are faint, or have a long \Prot, which could impact on the possibility of detecting a signal.
From Fig. \ref{fig:actindsummaryresults} (top left), we also see that the RV semi-amplitudes of the stars for which we find activity signals are, in general, large ($\gtrsim10\,\ms$).

\subsection{Low-activity stars}

In the low-activity regime ($\pEWHalpha\gtrsim-0.3$, 71 stars), most of the stars have low RV scatter, $\leq10\,\ms$. There are two outliers, J11417+427 (Ross~1003, GJ~1148) and J22096-046 (BD-05~5715, GJ~849), with larger RV scatter, which contain the modulation due to large planetary companions \citep[see e.g.][, respectively]{trifonov2018carmenes1st,butler2006gj849}.
Here, 26 of the stars have $M<0.4\,\Msun$ and 45 stars $M\geq0.4\,\Msun$.
As for the high-activity regime, we find that lower RV scatter values correspond in general to the lowest mass stars, while most of the highest mass stars show larger RV scatter.
Specifically, of the 33 targets with RV scatter $\leq3\,\ms$, 20 ($\sim60\,\%$) have $M<0.4\,\Msun$, while in the 38 stars with RV scatter $>3\,\ms$, only six ($\sim16\,\%$) have $M<0.4\,\Msun$.
This is clearly observed in the right panel of Fig.~\ref{fig:rvjitter_act_mass}, where we see that the RV scatter decreases with the stellar mass.
This difference in RV scatter between higher- and lower-mass M dwarfs could hint at varying manifestations of stellar variability, such as different surface granulation, active regions, or dynamo mechanisms. 

Regarding activity detections, we identify signals in almost half (34 of the 71 stars, 48\,\%) of these low-activity stars.
Most of the targets with detections have RV scatter $>3\,\ms$ (26 of the total of 34 stars, $\sim76\,\%$).
In the group of stars with RV scatter $\leq3\,\ms$, only eight of the 33 targets ($\sim24\,\%$) show reliable detections (J01025+716, J17578+046, J13299+102, J23492+024, J07274+052, J11511+352, and J00183+440).
These eight targets all have a large number of observations, $\geq50$ (seven of them with $>110$ observations), while the majority of the remaining targets with $\mathrm{std\,RV}\leq3\,\ms$ and no detections have been observed significantly less (17 of the other 25 stars, $\sim70\,\%$, have $<60$ observations).
All of the seven targets with detections show clear activity signals in their RVs (except for J17578+046, one of the stars with a planet, and J20305+654).
For three of them (J07274+052, J11511+352, and J13299+102), the only activity signal found is in the RVs.

\subsection{Overview}

In summary, our sample stars with low activity levels ($\pEWHalpha>-0.3$) show RV scatter from $\sim2$ to 10\,\ms, while for active stars, the scatter can reach values one order of magnitude larger.
The RV scatter in the active stars increases with the average activity level, and most active targets show clear periodic signals.
Nevertheless, we identify activity signals in almost half of the inactive stars, especially in the ones with the largest RV scatter ($\mathrm{std\,RV>3\,\ms}$).


\begin{figure*}
\centering
\includegraphics[width=0.42\linewidth]{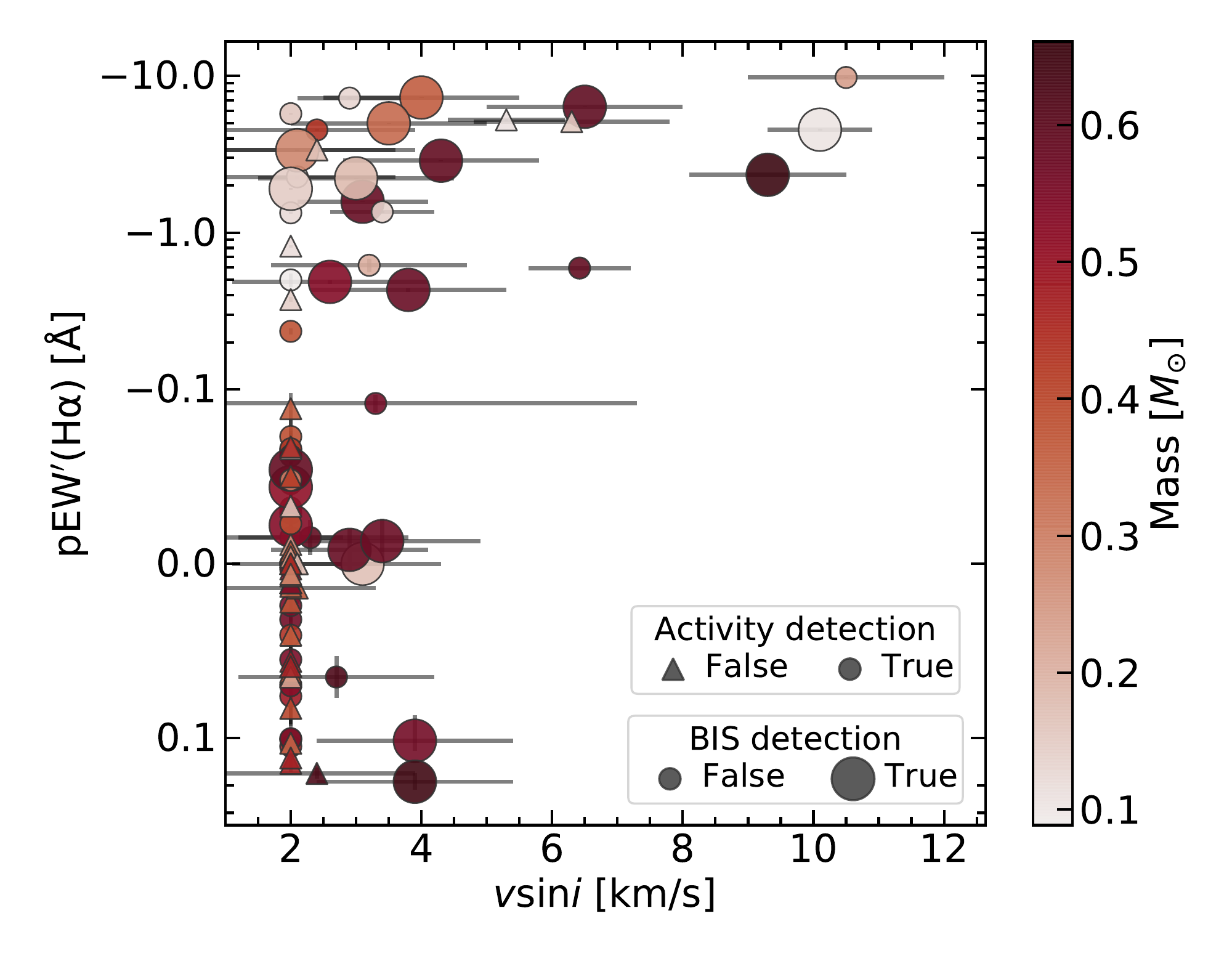}
\,
\includegraphics[width=0.42\linewidth]{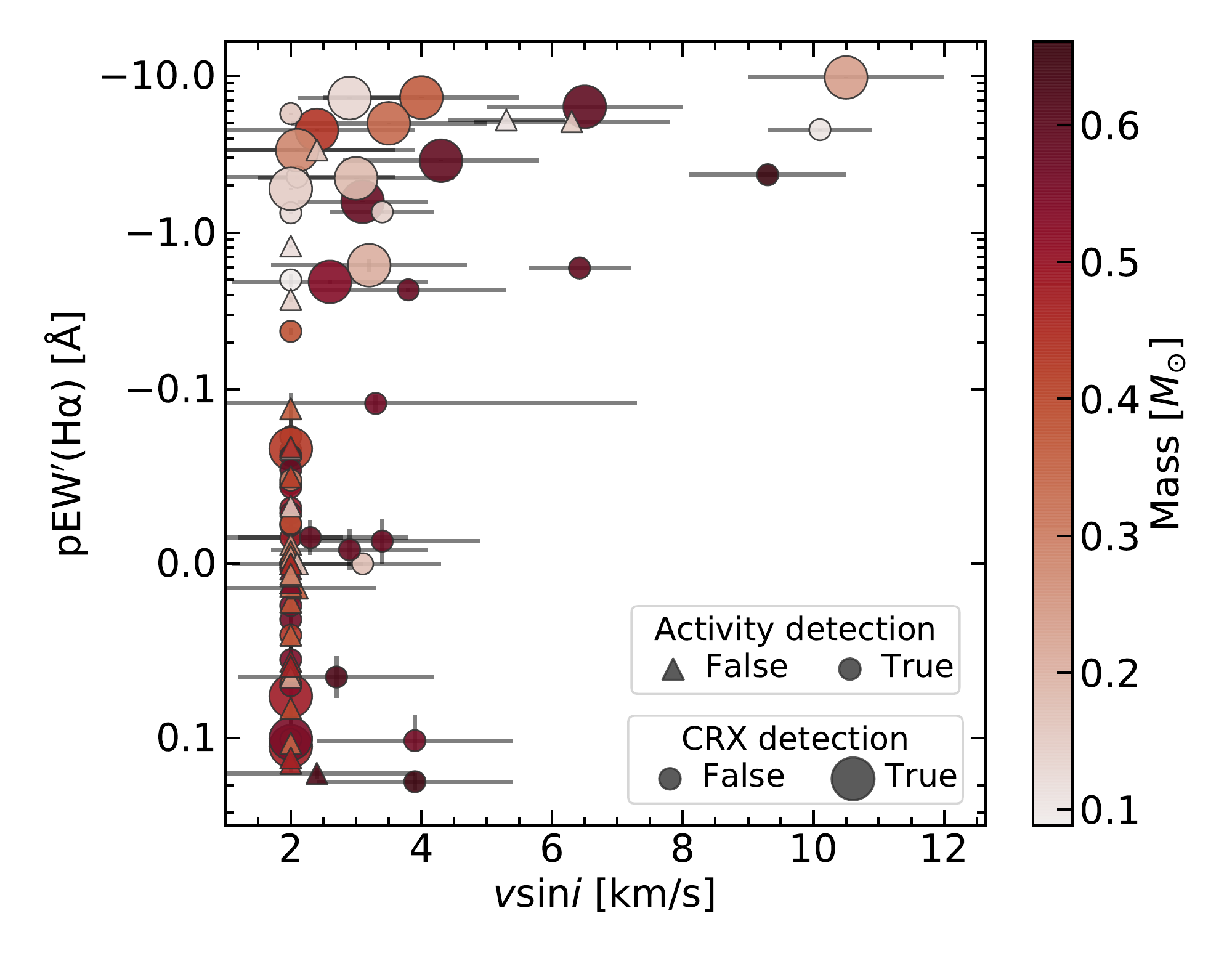}
\caption{\emph{Left}: Average activity level (\pEWHalpha) as a function of the rotational velocity (\vsini) of the 98 stars in the sample, colour-coded with the stellar mass.
Stars with activity-related signals ($\FAP\leq10\,\%$) in any of the parameters are shown with circles, and stars with no detections with triangles.
For the targets with detections, large circles indicate a detection ($\FAP\leq10\,\%$) in the BIS.
\emph{Right}: Same as the \emph{left panel}, but for the CRX instead of the BIS.}
\label{fig:vsini_act_mass}
\end{figure*}

In both active and inactive stars, the RV jitter shows a gradient with stellar mass.
Both low- and high-mass stars show a range of RV scatters, with more active stars showing in general larger scatter values, but we observe that the minimum RV scatter decreases with stellar mass.
Despite having in general higher activity levels, stars with lower RV scatters tend to be those of lower masses, while targets with a larger scatter correspond to the most massive ones. This could indicate different manifestations of activity between M dwarfs with high- and low-mass.
A relevant implication of these results is that, since lower-mass stars (in general, late-type M dwarfs) show the lowest RV jitter, they may be better targets for planet searches than more massive (earlier-type) M dwarfs, independently of their activity level.

\section{\vsini-activity relation and BIS performance}\label{sec:vsini}

Measurements derived from the CCF bisector have been found to show significant variations related to activity in stars with large \vsini, but do not provide significant information for slow rotators \citep{saardonahue1997activity,desort2007activity,bonfils2007gj674}.
The \vsini values of the sample stars increase with the average activity level \pEWHalpha in the high-activity regime.
The results presented here show that activity modulation in BIS increases with the average \pEWHalpha (Fig. \ref{fig:actindsummaryresults}).
Given this behaviour of BIS with \pEWHalpha, we then could expect some correspondence between the detections in BIS and \vsini.

Fig. \ref{fig:vsini_act_mass} (left) shows the average activity level \pEWHalpha of the 98 sample stars as a function of their \vsini, where we indicate the ones that show an activity signal in any of the parameters, and also the ones with a detection in BIS.
About 60\,\% of the targets with $\vsini>2\,\kms$ and an activity detection in any parameter, show a significant detection in BIS (16 out of 26 stars), while for the slower rotators, this number is of only about 13\,\% (four out of 30 stars).
This agrees with the fact that bisector measurements are usually found to be useful in fast rotators, and not in slowly-rotating stars.

If we consider the targets with detections in CRX, the other indicator that we find to behave in a way similar to BIS, we obtain comparable results (Fig. \ref{fig:vsini_act_mass}, right panel). About 45\,\% of the stars with $\vsini>2\,\kms$ and detection in at least one of the parameters, show a significant detection in CRX (12 out of 26 stars), and about 16\,\% of the slowest rotators show a signal (five out of 30 stars).

If we focus on the low-activity regime ($\pEWHalpha>-0.3$), most of the targets have very low \vsini, $\leq2\,\kms$, but there are $\sim10$ stars with larger \vsini values, up to $\sim4\,\kms$.
Half of these outliers show significant signals in BIS (although all with low FAP, see Fig. \ref{fig:actindsummaryresults}), but none in the CRX.
This could indicate that BIS is more sensitive to activity variations in stars with a relatively large \vsini than CRX, at least in the low-activity regime.


\section{Summary} \label{sec:actindsummary}

We performed a search for activity-related periodicities in a sample of 98 M dwarfs observed with CARMENES.
We carried out a periodogram analysis on the time series of their RVs and nine activity indicators: CCF FWHM, contrast and BIS, CRX derived from the RVs, dLW derived from the template matching method used to compute the RVs, and indices measured from four chromospheric lines, \Halpha and the Ca IRT.

Of the initial sample of 98 stars, we find that 56 of them show a signal that we attribute to activity (i.e. a signal related to \Prot) in at least one of the ten parameters analysed.
Most of these 56 stars show an activity signal in RV, which is the most effective tracer of activity (i.e. it shows an activity-related signal for most of the stars).
Aside from RV, CRX and BIS are the most effective for the active stars in our sample (stars with $\pEWHalpha\lesssim-0.3$), but they are not as sensitive in the low-activity regime ($\pEWHalpha\gtrsim-0.3$).
Low-activity stars tend to be slow rotators, which agrees with the fact that BIS is usually not found to trace activity in such stars.
For the chromospheric lines, we observe the opposite behaviour.
Low-activity stars show a larger fraction of targets with periodic signals in these indicators than the most active ones.
dLW and FWHM are similarly effective for all stars, but especially in the high-activity, low-mass ($M\lesssim0.35\,\Msun$) regime, performing better than any other indicator.

This implies that none of the activity indicators can efficiently trace stellar activity for all stars.
Activity in the most active stars tends to be detected in photospheric indicators.
CRX and BIS work best for the active, higher-mass ($M\gtrsim0.35\,\Msun$) stars, followed by dLW and FWHM, and chromospheric lines.
In active, low-mass stars, the most efficient indicators are dLW and FWHM, followed by CRX and BIS. Chromospheric lines are only effective in a small subset of such stars.
Most low-activity stars show activity signals in chromopheric indicators, and about half of them also show signals in dLW and FWHM. CRX and BIS are not good activity tracers in these stars.
From these results it is clear that, when assessing the presence of stellar activity in RV measurements, it is key to take into account the adequate activity indicators, depending on the characteristics of the target star.
Even more importantly, an RV signal should not be assumed to be of planetary nature on the basis of only a small set of indicators, as false positives can abound.

Stars for which we detect no signature of activity are located mainly in the low-activity regime, where they account for about half of the sample.
In the high-activity regime, stars with no detection are a minority and correspond to faint, low-mass targets.
Our inability to detect any activity signals in these stars could be due to the weakness of these signals compared to our measurement precision (in the case of the stars with low activity levels), to complex or changing active region patterns giving rise to incoherent signals, or to observational limitations (observations not covering the rotation phase well enough so that a significant peak appears in the periodogram, or low S/N).

As expected, the RV jitter increases with the average activity level of the stars, especially in the high-activity regime, where the RV scatter (std RV) can reach values as high as $\sim100\,\ms$.
In the low-activity regime, the scatter is usually lower than $\sim10\,\ms$.
In general, lower RV scatter correspond to stars with no detections of signals related to activity.
We observe that, independent of the activity level, stars with lower mass are the ones with lower RV scatter, which could be due to different manifestations of stellar activity.
This also means that late-type M dwarfs could be better candidates for planetary searches than their earlier-type siblings.

Overall, this study highlights the fact that, depending on the properties of the star considered, different indicators of stellar activity will behave differently.
We find that, in a sample of M dwarfs with a relatively large range of masses and activity levels, each type of indicator performs best (is most effective in tracing activity signals) in a specific range of mass and activity.
Therefore, analyses such as the one presented here can be used as a guide for studies of activity signals in spectroscopic observations.
Nevertheless, the effectiveness of the indicators depends on several factors, from observational constraints to the specific properties of the stellar active regions, so thorough studies of all available indicators should be conducted in conjunction with RV searches for planets.

\begin{acknowledgements}
We thank the anonymous referee for the very helpful review of the article.
CARMENES is an instrument for the Centro Astron\'omico Hispano-Alem\'an de
Calar Alto (CAHA, Almer\'{\i}a, Spain).
CARMENES is funded by the German Max-Planck-Gesellschaft (MPG),
the Spanish Consejo Superior de Investigaciones Cient\'{\i}ficas (CSIC),
the European Union through FEDER/ERF FICTS-2011-02 funds,
and the members of the CARMENES Consortium
(Max-Planck-Institut f\"ur Astronomie,
Instituto de Astrof\'{\i}sica de Andaluc\'{\i}a,
Landessternwarte K\"onigstuhl,
Institut de Ci\`encies de l'Espai,
Institut f\"ur Astrophysik G\"ottingen,
Universidad Complutense de Madrid,
Th\"uringer Landessternwarte Tautenburg,
Instituto de Astrof\'{\i}sica de Canarias,
Hamburger Sternwarte,
Centro de Astrobiolog\'{\i}a and
Centro Astron\'omico Hispano-Alem\'an),
with additional contributions by the Spanish Ministry of Economy,
the German Science Foundation through the Major Research Instrumentation
Programme and DFG Research Unit FOR2544 ``Blue Planets around Red Stars'',
the Klaus Tschira Stiftung,
the states of Baden-W\"urttemberg and Niedersachsen,
and by the Junta de Andaluc\'{\i}a.
Based on data from the CARMENES data archive at CAB (INTA-CSIC).
We acknowledge support from the Spanish Ministry of Science and Innovation and the European Regional Development Fund through grant PGC2018-098153-B-C33, the Generalitat de Catalunya/CERCA programme, PID2019-109522GB-C51/2/3/4 (CAB, IAA, IAC, UCM), and UKRI FLF grant MR/S035214/1. SVJ acknowledges the support of the DFG priority program SPP 1992 ``Exploring the Diversity of Extrasolar Planets (JE 701/5-1).''
\end{acknowledgements}

%
%

\bibliographystyle{aa} 
\bibliography{indicators.bib} 

\begin{appendix}

\section{Targets}


\longtab[1]{
{\tiny
\begin{landscape}
\begin{longtable}{lllcccccccccc}
\caption{Properties of the 98 selected stars.}\\
\hline\hline
Karmn & Name & Sp. & Mass\tablefootmark{(b)} & $J$\tablefootmark{(c)} & \vsini\tablefootmark{(d)} & \Prot\tablefootmark{(e)} & \pEWHalpha\tablefootmark{(f)} & \logLHalphaLbol\tablefootmark{(g)} & Planet & \# obs & Span & std RV \\
&      & type\tablefootmark{(a)} & [\Msun] & [mag] & [\kms] & [d] & & & & (\# nights) & [d] & [\ms] \\
\hline
\noalign{\smallskip}
\endfirsthead
\caption[]{Continued.}\\
\hline\hline
Karmn & Name & Sp. & Mass & $J$ & \vsini & \Prot & \pEWHalpha & \logLHalphaLbol & Planet & \# obs & Span & std RV \\
&  & type & [\Msun] & [mag] & [\kms] & [d] & & & & (\# nights) & [d] & [\ms] \\
\hline
\noalign{\smallskip}
\endhead
\hline
\endfoot
J00067-075  &                     GJ 1002 &               M5.5 V &   $0.1164\pm 0.0092$ &   $8.320\pm 0.019$ &                 < $2.0$ &                        ... &                $0.0$ &                ... &   No &    91 (91) &  1188 &    2.22 \\
J00183+440  &              GX And, GJ 15A &         M1.0 V$^{*}$ &     $0.391\pm 0.016$ &     $5.30\pm 0.26$ &                 < $2.0$ &          $45.00\pm 4.40^{*}$ &  $-0.0730\pm 0.0070$ &                ... &  Yes &  150 (118) &  1441 &    2.43 \\
J00184+440  &              GQ And, GJ 15B &               M3.5 V &     $0.168\pm 0.010$ &   $6.790\pm 0.024$ &                 < $2.0$ &                        ... &    $-0.033\pm 0.012$ &                ... &   No &  150 (112) &  1446 &    2.08 \\
J01025+716  &             Ross 318, GJ 48 &               M3.0 V &     $0.488\pm 0.019$ &   $6.300\pm 0.034$ &                 < $2.0$ &              $52.00\pm 2.60$ &     $0.076\pm 0.016$ &                ... &   No &  120 (119) &   719 &    2.35 \\
J01026+623  &            BD+61 195, GJ 49 &         M1.5 V$^{*}$ &     $0.515\pm 0.019$ &   $6.230\pm 0.021$ &                 < $2.0$ &              $20.00\pm 0.40$ &    $-0.044\pm 0.013$ &                ... &  Yes &    83 (81) &   780 &    5.24 \\
J01033+623  &             V388 Cas, GJ 51 &         M5.0 V$^{*}$ &     $0.241\pm 0.012$ &   $8.610\pm 0.027$ &         $10.00\pm 1.50$ &             $1.020\pm 0.010$ &    $-9.770\pm 0.040$ &  $-3.5340\pm 0.0017$ &   No &    26 (26) &  1213 &  197.35 \\
J01125-169  &             YZ Cet, GJ 54.1 &               M4.5 V &   $0.1418\pm 0.0098$ &   $7.260\pm 0.020$ &      $3.40\pm 0.80^{*}$ &              $69.00\pm 2.40$ &    $-1.360\pm 0.020$ &  $-4.3400\pm 0.0070$ &  Yes &   111 (84) &  1081 &    2.86 \\
\end{longtable}
\tablefoot{
The values are taken from the latest version of the Carmencita database available at the time. We also show the number of CARMENES VIS observations (before performing any $\sigma$ clipping or discarding any observations due to low S/N), the number of different nights covered by the observations, their time span, and their RV scatter, measured as the standard deviation (std) of the corrected \serval RVs (instrumental drift and nightly average corrected, averaged same-night observations, and linear trend removed).\\
\textbf{References.}
\tablefoottext{a}{\citet{hawley1996PMSU}, except $^{*}$\citet{alonsofloriano2015carmenesinlowres}, $^{\dag}$\citet{gray2006NStars}, $^{\ddag}$\citet{gray2003NStars}, $^{\S}$\citet{lepine2013mdwarf}, $^{\P}$\citet{benneke2017K2-18}, $^{\#}$\citet{newton2014Mdwarfs}, $^{\textbar}$\citet{riaz2006mdwarf}, $^{**}$\citet{kirkpatrick1991spectra}.}
\tablefoottext{b}{\citet{schweitzer2019carmenesMR}, except J20451-313 computed using \citet{mann2019MLMet}.}
\tablefoottext{c}{\citet{skrutskie20062mass}.}
\tablefoottext{d}{\citet{reiners2018carmenes324}, except $^{*}$\citet{fouque2018spiroucatalogue}, $^{\dag}$\citet{delfosse1998rotationM}, $^{\ddag}$\citet{martinez-rodriguez2014thesis}, $^{\\S}$\citet{lopez-santiago2010activity}, $^{\P}$\citet{torres2006youngstars}.}
\tablefoottext{e}{\citet{diezalonso2019carmenesRotPhot}, except $^{*}$\citet{suarezmascareno2018hadesAct}, $^{\dag}$\citet{newton2016rotation}, $^{\ddag}$\citet{suarezmascareno2017rvrotation}, $^{\S}$\citet{morin2008mdwarfmag}, $^{\P}$\citet{morin2010magnectictopologies}, $^{\#}$\citet{suarezmascareno2015rotationchrom}, $^{\textbar}$\citet{messina2011rotation}, $^{**}$\citet{watson2006vsi}.}
\tablefoottext{f}{\citet{schofer2019carmenesActInd}.}
\tablefoottext{g}{\citet{schofer2019carmenesActInd}.}
}
\end{landscape}
}
}


\section{Periodogram analysis examples}\label{sec:actindperiodogramexamples}

\begin{figure*}
\centering
\includegraphics[width=0.82\linewidth]{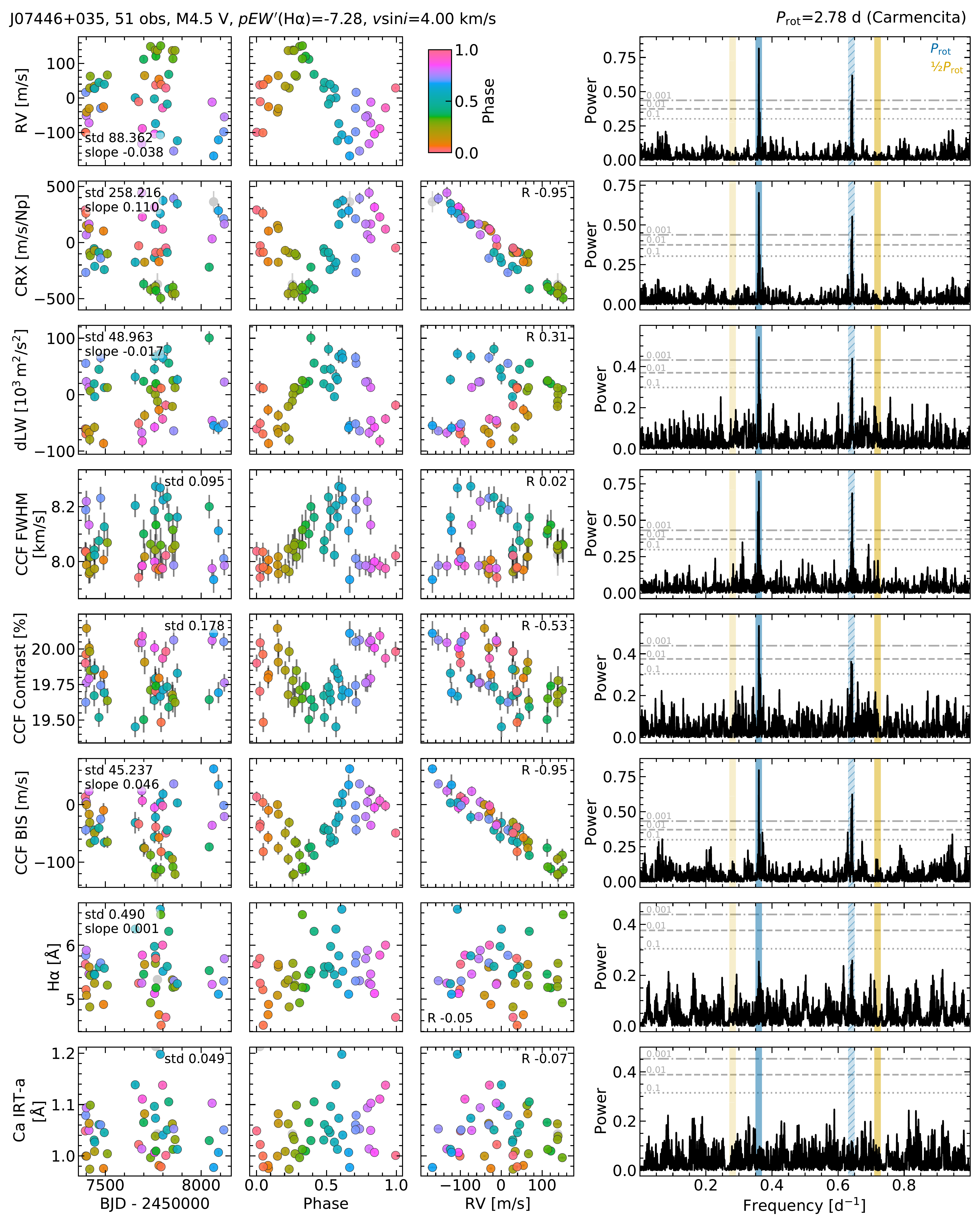}
\caption{RV and indicator time series (\emph{left}), data folded at \Prot (\emph{middle left}), correlation with the RV (\emph{middle right}), and periodogram (\emph{right}) of J07446+035 (YZ~CMi, GJ~285). The parameters are, from \emph{top} to \emph{bottom}: corrected \serval RV, CRX, dLW, CCF FWHM, CCF contrast, CCF BIS, \Halpha index, and Ca IRT-a index. We only show the first IRT line because the other two show very similar time series and periodograms with no significant signal.
Each time series is corrected for a linear trend, and the periodograms are computed on nightly-averaged observations.
Data points are colour-coded with the rotation phase, and outliers not considered are marked in grey (some outliers may be outside the data range shown).
The text in the time series panels shows the standard deviation of the data (std) and the slope of the linear trend that is subtracted from the data (only slopes with absolute values $\geq0.001$).
The text in the correlation panels indicates the Pearson correlation coefficient of the data (R).
In the periodogram panels, horizontal grey lines indicate the FAP level at 10\,\% (dotted), 1\,\% (dashed), and 0.1\,\% (dotted dashed), and coloured shaded regions indicate the location of \Prot (blue),  \Prothalf (yellow), and their 1-day aliases (lighter hatched regions).}
\label{fig:actind_tsperiodogramJ07446+035}
\end{figure*}

\begin{figure*}
\centering
\includegraphics[width=0.82\linewidth]{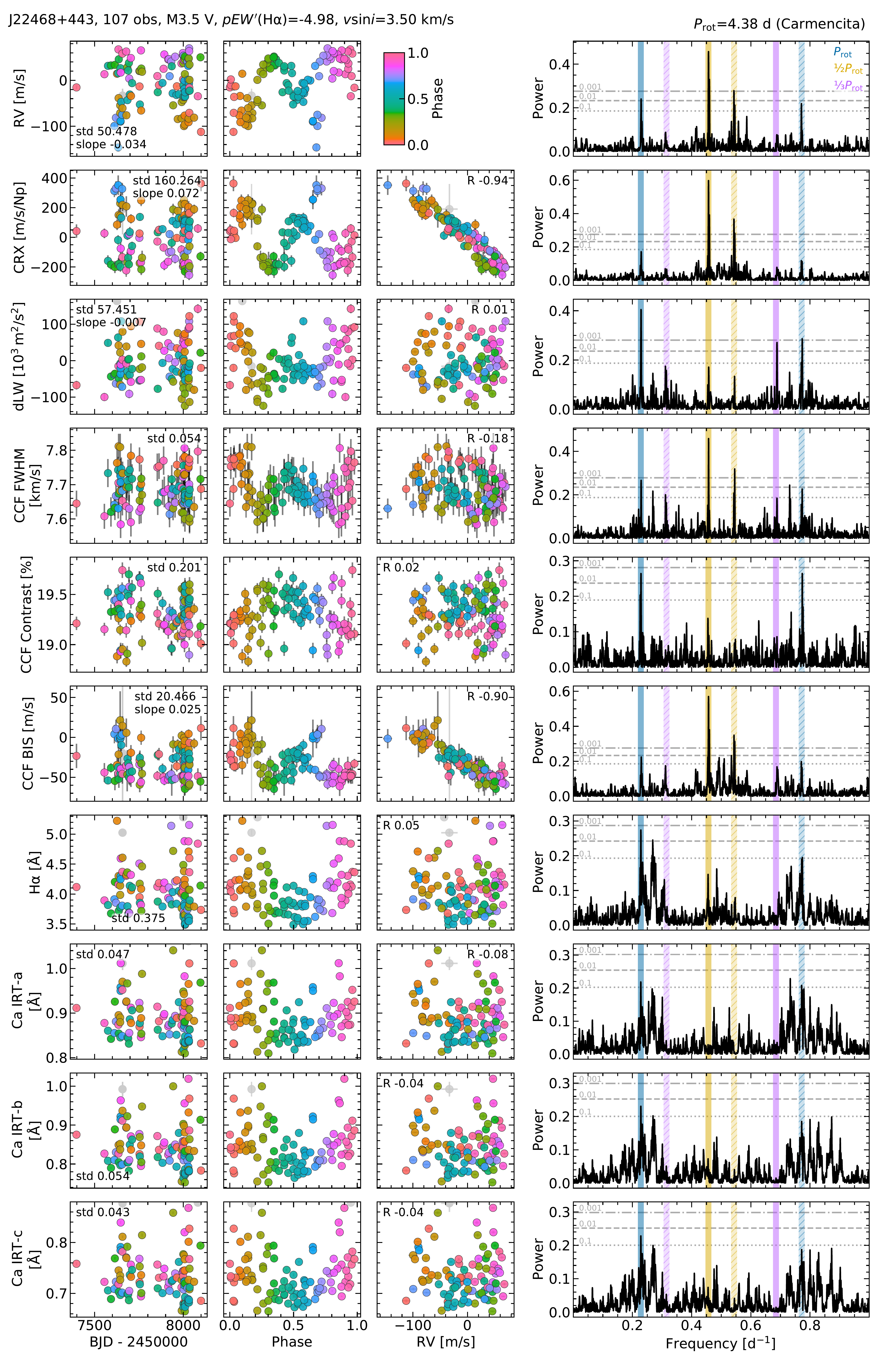}
\caption{Same as Fig. \ref{fig:actind_tsperiodogramJ07446+035}, but for J22468+443 (EV~Lac, GJ~873).
We now display also the Ca IRT-b and -c data because they show significant signals.
In the periodogram panels, we add a purple shaded region around $\frac{1}{3}\Prot$ (and its corresponding 1-day alias indicated by a hatched purple region).}
\label{fig:actind_tsperiodogramJ22468+443}
\end{figure*}

\begin{figure*}
\centering
\includegraphics[width=0.82\linewidth]{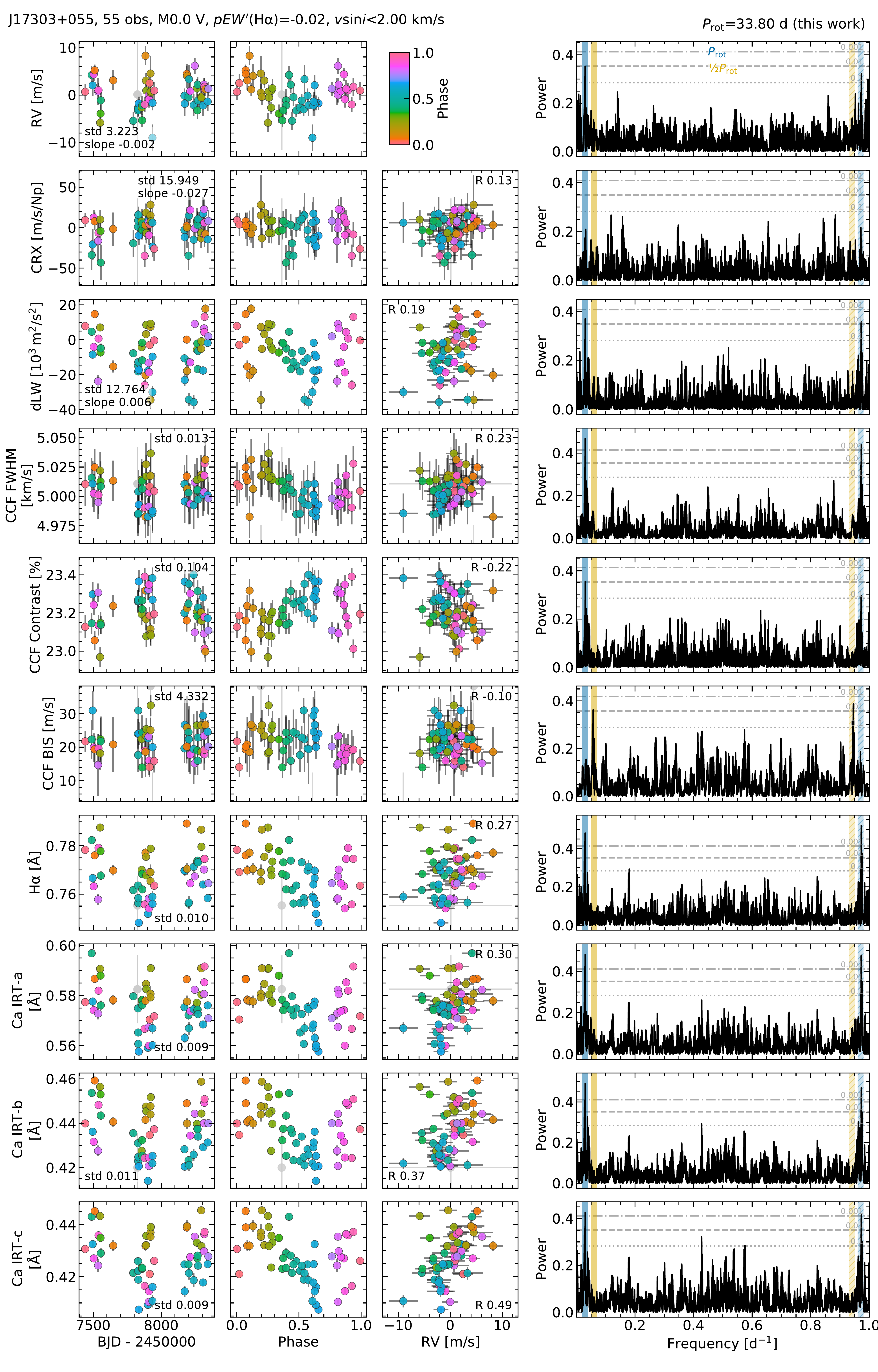}
\caption{Same as Fig. \ref{fig:actind_tsperiodogramJ07446+035}, but for J17303+055 (BD+05~3409, GJ~678.1~A).
}
\label{fig:actind_tsperiodogramJ17303+055}
\end{figure*}

\begin{figure*}
\centering
\includegraphics[width=0.82\linewidth]{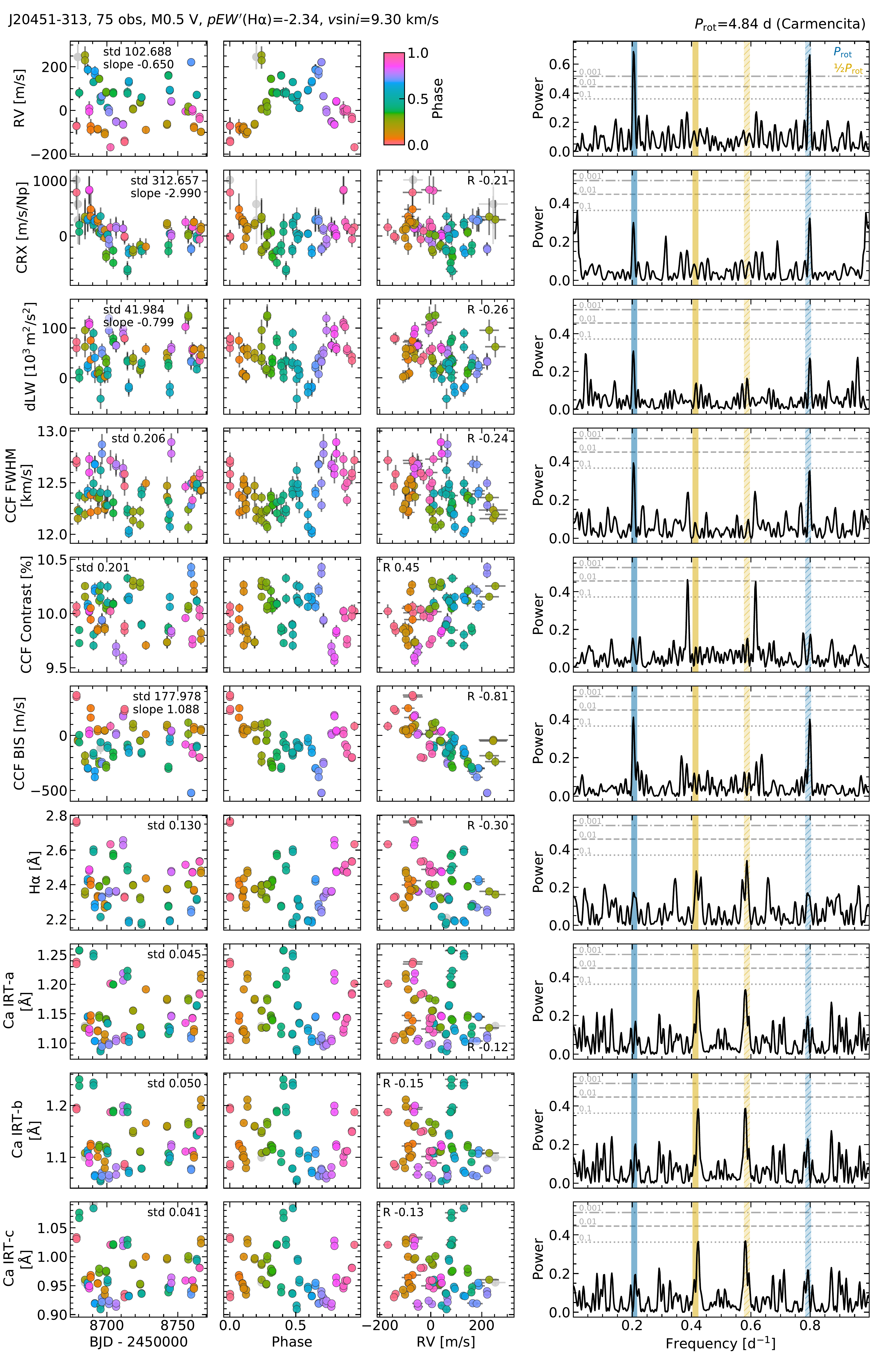}
\caption{Same as Fig. \ref{fig:actind_tsperiodogramJ07446+035}, but for J20451-313 (AU Mic, GJ~803).}
\label{fig:actind_tsperiodogramJ20451-313}
\end{figure*}

\begin{figure*}
\centering
\includegraphics[width=0.82\linewidth]{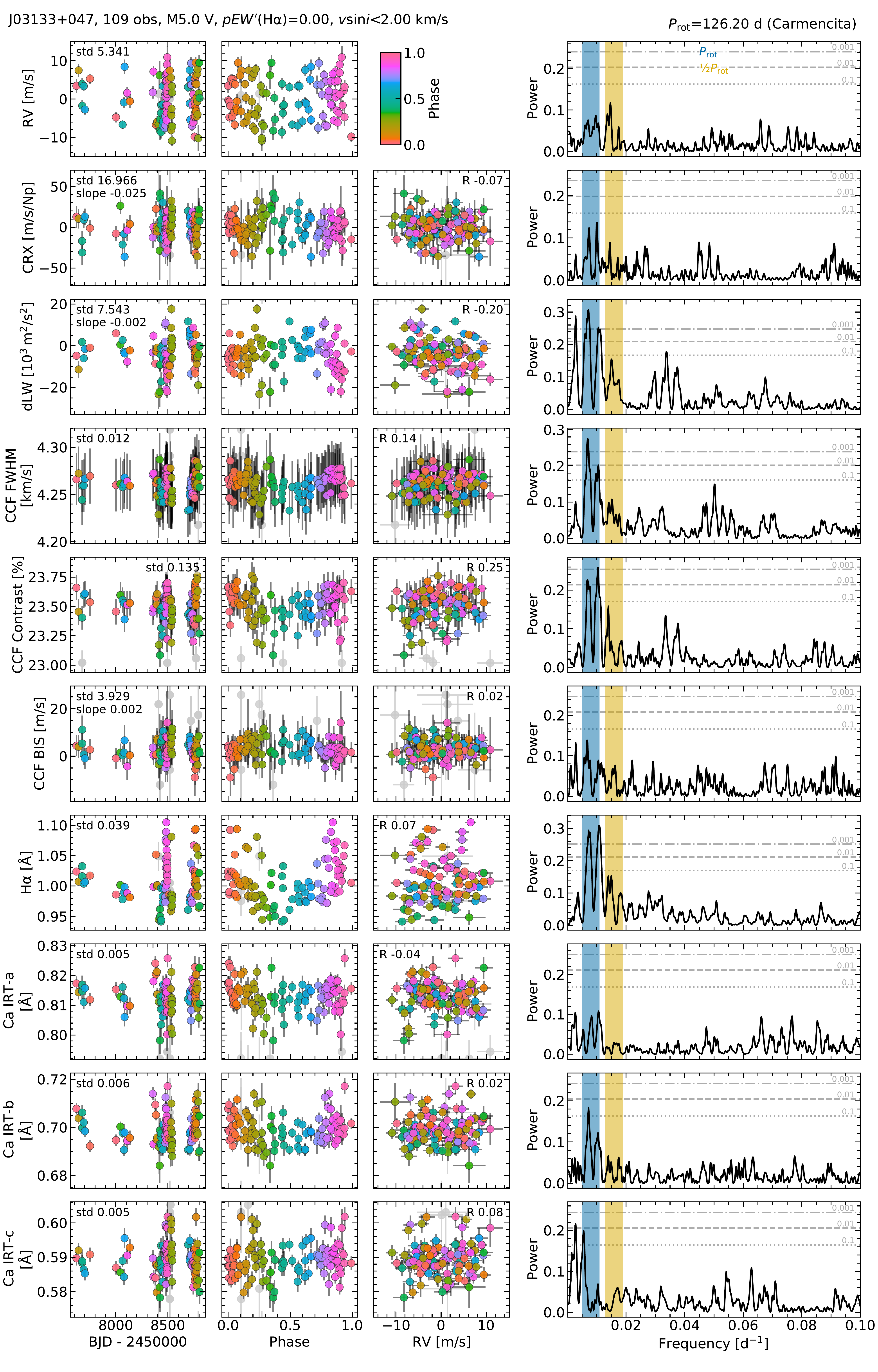}
\caption{Same as Fig. \ref{fig:actind_tsperiodogramJ07446+035}, but for J03133+047 (CD~Cet, GJ~1057). In this case, the periodogram shows a zoom in at the low frequency range, where we find activity-related peaks (the planetary companion shows a significant peak in the RVs at $2.29\,\days = 0.44\,\days^{-1}$, not shown here).}
\label{fig:actind_tsperiodogramJ03133+047}
\end{figure*}

\begin{figure*}
\centering
\includegraphics[width=0.82\linewidth]{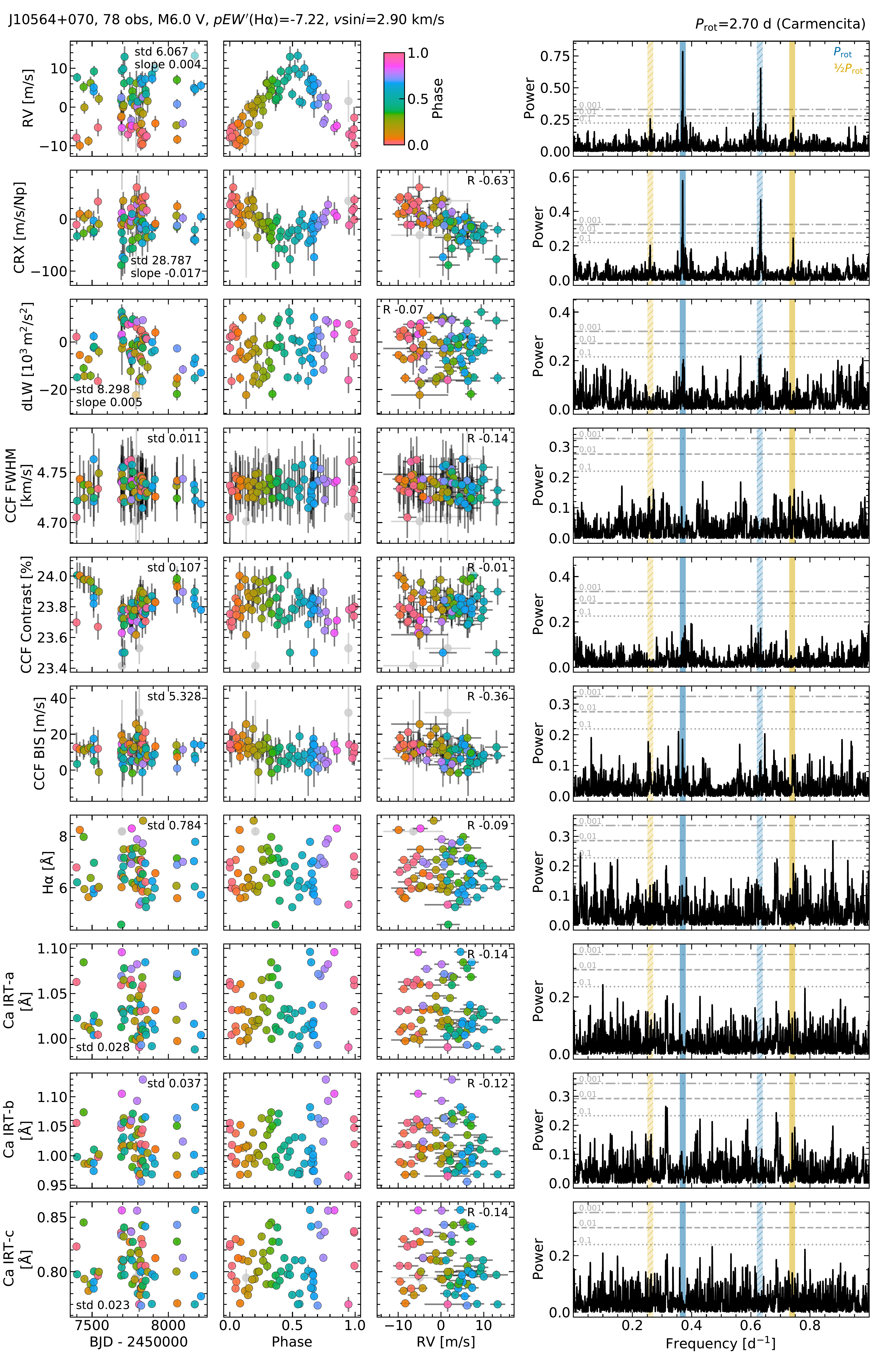}
\caption{Same as Fig. \ref{fig:actind_tsperiodogramJ07446+035}, but for J10564+070 (CN~Leo, GJ~406).}
\label{fig:actind_tsperiodogramJ10564+070}
\end{figure*}

\section{Periodogram analysis results}\label{sec:actindperiodogramresults}

\longtab[1]{
\begin{longtable}{lccccccccc}
\caption{\label{tab:starsactindselected_rv} Activity-related signals (periods and FAPs) at \Prot, \Prothalf and their 1 day aliases (1\,d alias) found in the periodograms of the \serval RVs. We also show literature periods compiled in Carmencita ($P_\mathrm{rot\,lit}$, same references as in Table \ref{tab:starsactindselected}). For stars with no known period, we take as \Prot the signal with the smallest FAP present in more than one indicator.}\\
\hline\hline
Karmn& $P_\mathrm{rot\,lit}$ & \multicolumn{2}{c}{\Prot} & \multicolumn{2}{c}{\Prothalf}&\multicolumn{2}{c}{\Prot 1\,d alias}&\multicolumn{2}{c}{\Prothalf 1\,d alias}\\
\cmidrule(lr){3-4}\cmidrule(lr){5-6}\cmidrule(lr){7-8}\cmidrule(lr){9-10}
& [d] &$P$ [d]&FAP [\%]&$P$ [d]&FAP [\%]&$P$ [d]&FAP [\%]&$P$ [d]&FAP [\%]\\
\hline
\noalign{\smallskip}
\endfirsthead
\caption{Continued.}\\
\hline\hline
Karmn& $P_\mathrm{rot\,lit}$ & \multicolumn{2}{c}{\Prot} & \multicolumn{2}{c}{\Prothalf}&\multicolumn{2}{c}{\Prot 1\,d alias}&\multicolumn{2}{c}{\Prothalf 1\,d alias}\\
\cmidrule(lr){3-4}\cmidrule(lr){5-6}\cmidrule(lr){7-8}\cmidrule(lr){9-10}
& [d] &$P$ [d]&FAP [\%]&$P$ [d]&FAP [\%]&$P$ [d]&FAP [\%]&$P$ [d]&FAP [\%]\\
\hline
\noalign{\smallskip}
\endhead
\hline
\endfoot
J00183+440  &          $45.00\pm 4.40^{*}$ &  40.71 &        0.69 &     ...&           ...&     ...&           ...&    ...&           ...\\
J01025+716  &              $52.00\pm 2.60$ &  50.99 &  $\leq$ 0.1 &     ...&           ...&  1.02 &  $\leq$ 0.1 &    ...&           ...\\
J01026+623  &              $20.00\pm 0.40$ &  18.89 &        1.03 &  9.43 &        1.26 &     ...&           ...& 1.12 &        6.12 \\
J01033+623  &             $1.020\pm 0.010$ &   1.08 &       45.38 &     ...&           ...& 13.67 &        5.20 &    ...&           ...\\
J01125-169  &              $69.00\pm 2.40$ &  80.67 &        9.56 &     ...&           ...&     ...&           ...&    ...&           ...\\
J02002+130  &                        ... &   1.95 &  $\leq$ 0.1 &     ...&           ...&  2.03 &  $\leq$ 0.1 &    ...&           ...\\
J02222+478  &              $30.00\pm 0.40$ &  28.20 &        2.21 &     ...&           ...&  1.03 &        9.51 &    ...&           ...\\
J04153-076  &                        ... &   1.80 &  $\leq$ 0.1 &     ...&           ...&     ...&           ...&    ...&           ...\\
\end{longtable}
}


\longtab[2]{
\begin{longtable}{lccccccccc}
\caption{Same as Table \ref{tab:starsactindselected_rv}, but for CRX.}\label{tab:starsactindselected_crx}\\
\hline\hline
Karmn& $P_\mathrm{rot\,lit}$ & \multicolumn{2}{c}{\Prot} & \multicolumn{2}{c}{\Prothalf}&\multicolumn{2}{c}{\Prot 1\,d alias}&\multicolumn{2}{c}{\Prothalf 1\,d alias}\\
\cmidrule(lr){3-4}\cmidrule(lr){5-6}\cmidrule(lr){7-8}\cmidrule(lr){9-10}
& [d] &$P$ [d]&FAP [\%]&$P$ [d]&FAP [\%]&$P$ [d]&FAP [\%]&$P$ [d]&FAP [\%]\\
\hline
\noalign{\smallskip}
\endfirsthead
\caption{Continued.}\\
\hline\hline
Karmn& $P_\mathrm{rot\,lit}$ & \multicolumn{2}{c}{\Prot} & \multicolumn{2}{c}{\Prothalf}&\multicolumn{2}{c}{\Prot 1\,d alias}&\multicolumn{2}{c}{\Prothalf 1\,d alias}\\
\cmidrule(lr){3-4}\cmidrule(lr){5-6}\cmidrule(lr){7-8}\cmidrule(lr){9-10}
& [d] &$P$ [d]&FAP [\%]&$P$ [d]&FAP [\%]&$P$ [d]&FAP [\%]&$P$ [d]&FAP [\%]\\
\hline
\noalign{\smallskip}
\endhead
\hline
\endfoot
J01025+716 &              $52.00\pm 2.60$ &  50.99 &       42.41 &    ...&           ...&  1.02 &        5.62 &    ...&           ...\\
J01033+623 &             $1.020\pm 0.010$ &   1.06 &        0.40 &    ...&           ...& 16.59 &  $\leq$ 0.1 &    ...&           ...\\
J01125-169 &              $69.00\pm 2.40$ &  80.67 &       12.43 &    ...&           ...&     ...&           ...&    ...&           ...\\
J02002+130 &                        ... &   1.96 &        0.79 &    ...&           ...&  2.03 &       19.09 &    ...&           ...\\
J02530+168 &                        ... &  96.20 &       73.37 &    ...&           ...&     ...&           ...&    ...&           ...\\
J03133+047 &               $130.0^{\dag}$ & 100.24 &       25.37 &    ...&           ...&     ...&           ...&    ...&           ...\\
J04153-076 &                        ... &   1.80 &  $\leq$ 0.1 &    ...&           ...&     ...&           ...&    ...&           ...\\
J05019+011 &             $2.120\pm 0.020$ &   2.09 &  $\leq$ 0.1 &    ...&           ...&  1.90 &  $\leq$ 0.1 &    ...&           ...\\
\end{longtable}
}


\longtab[3]{
\begin{longtable}{lccccccccc}
\caption{Same as Table \ref{tab:starsactindselected_rv}, but for dLW.}\label{tab:starsactindselected_dlw}\\
\hline\hline
Karmn& $P_\mathrm{rot\,lit}$ & \multicolumn{2}{c}{\Prot} & \multicolumn{2}{c}{\Prothalf}&\multicolumn{2}{c}{\Prot 1\,d alias}&\multicolumn{2}{c}{\Prothalf 1\,d alias}\\
\cmidrule(lr){3-4}\cmidrule(lr){5-6}\cmidrule(lr){7-8}\cmidrule(lr){9-10}
& [d] &$P$ [d]&FAP [\%]&$P$ [d]&FAP [\%]&$P$ [d]&FAP [\%]&$P$ [d]&FAP [\%]\\
\hline
\noalign{\smallskip}
\endfirsthead
\caption{Continued.}\\
\hline\hline
Karmn& $P_\mathrm{rot\,lit}$ & \multicolumn{2}{c}{\Prot} & \multicolumn{2}{c}{\Prothalf}&\multicolumn{2}{c}{\Prot 1\,d alias}&\multicolumn{2}{c}{\Prothalf 1\,d alias}\\
\cmidrule(lr){3-4}\cmidrule(lr){5-6}\cmidrule(lr){7-8}\cmidrule(lr){9-10}
& [d] &$P$ [d]&FAP [\%]&$P$ [d]&FAP [\%]&$P$ [d]&FAP [\%]&$P$ [d]&FAP [\%]\\
\hline
\noalign{\smallskip}
\endhead
\hline
\endfoot
J01025+716  &              $52.00\pm 2.60$ &  51.73 &  $\leq$ 0.1 &    ...&      ...&  1.02 &       33.48 &    ...&      ...\\
J01125-169  &              $69.00\pm 2.40$ &  80.67 &  $\leq$ 0.1 &    ...&      ...&     ...&           ...&    ...&      ...\\
J02002+130  &                        ... &   1.95 &        1.99 &    ...&      ...&  2.05 &        6.65 &    ...&      ...\\
J02222+478  &              $30.00\pm 0.40$ &  28.88 &  $\leq$ 0.1 &    ...&      ...&  1.03 &        1.15 &    ...&      ...\\
J02530+168  &                        ... &  97.56 &  $\leq$ 0.1 &    ...&      ...&     ...&           ...&    ...&      ...\\
J03133+047  &               $130.0^{\dag}$ & 140.81 &  $\leq$ 0.1 &    ...&      ...&     ...&           ...&    ...&      ...\\
J04290+219  &              $25.00\pm 0.30$ &  25.14 &  $\leq$ 0.1 &    ...&      ...&     ...&           ...&    ...&      ...\\
J04376+528  &                        ... &  16.30 &  $\leq$ 0.1 &    ...&      ...&     ...&           ...&    ...&      ...\\
\end{longtable}
}


\longtab[4]{
\begin{longtable}{lccccccccc}
\caption{Same as Table \ref{tab:starsactindselected_rv}, but for CCF FWHM.}\label{tab:starsactindselected_fwhm}\\
\hline\hline
Karmn& $P_\mathrm{rot\,lit}$ & \multicolumn{2}{c}{\Prot} & \multicolumn{2}{c}{\Prothalf}&\multicolumn{2}{c}{\Prot 1\,d alias}&\multicolumn{2}{c}{\Prothalf 1\,d alias}\\
\cmidrule(lr){3-4}\cmidrule(lr){5-6}\cmidrule(lr){7-8}\cmidrule(lr){9-10}
& [d] &$P$ [d]&FAP [\%]&$P$ [d]&FAP [\%]&$P$ [d]&FAP [\%]&$P$ [d]&FAP [\%]\\
\hline
\noalign{\smallskip}
\endfirsthead
\caption{Continued.}\\
\hline\hline
Karmn& $P_\mathrm{rot\,lit}$ & \multicolumn{2}{c}{\Prot} & \multicolumn{2}{c}{\Prothalf}&\multicolumn{2}{c}{\Prot 1\,d alias}&\multicolumn{2}{c}{\Prothalf 1\,d alias}\\
\cmidrule(lr){3-4}\cmidrule(lr){5-6}\cmidrule(lr){7-8}\cmidrule(lr){9-10}
& [d] &$P$ [d]&FAP [\%]&$P$ [d]&FAP [\%]&$P$ [d]&FAP [\%]&$P$ [d]&FAP [\%]\\
\hline
\noalign{\smallskip}
\endhead
\hline
\endfoot
J00183+440 &          $45.00\pm 4.40^{*}$ &  42.38 &       10.03 & 21.38 &       38.78 &     ...&           ...&     ...&           ...\\
J01025+716 &              $52.00\pm 2.60$ &  50.28 &  $\leq$ 0.1 &     ...&           ...&  1.02 &        0.55 &     ...&           ...\\
J01026+623 &              $20.00\pm 0.40$ &  18.80 &        1.70 &     ...&           ...&  1.05 &        0.58 &     ...&           ...\\
J01125-169 &              $69.00\pm 2.40$ &  79.48 &  $\leq$ 0.1 &     ...&           ...&     ...&           ...&     ...&           ...\\
J02002+130 &                        ... &   1.96 &        1.29 &     ...&           ...&  2.03 &       14.83 &     ...&           ...\\
J02530+168 &                        ... &  97.56 &  $\leq$ 0.1 &     ...&           ...&     ...&           ...&     ...&           ...\\
J03133+047 &               $130.0^{\dag}$ & 144.25 &  $\leq$ 0.1 &     ...&           ...&     ...&           ...&     ...&           ...\\
J04290+219 &              $25.00\pm 0.30$ &  25.14 &        3.20 &     ...&           ...&     ...&           ...&     ...&           ...\\
\end{longtable}
}


\longtab[5]{
\begin{longtable}{lccccccccc}
\caption{Same as Table \ref{tab:starsactindselected_rv}, but for CCF contrast.}\label{tab:starsactindselected_contrast}\\
\hline\hline
Karmn& $P_\mathrm{rot\,lit}$ & \multicolumn{2}{c}{\Prot} & \multicolumn{2}{c}{\Prothalf}&\multicolumn{2}{c}{\Prot 1\,d alias}&\multicolumn{2}{c}{\Prothalf 1\,d alias}\\
\cmidrule(lr){3-4}\cmidrule(lr){5-6}\cmidrule(lr){7-8}\cmidrule(lr){9-10}
& [d] &$P$ [d]&FAP [\%]&$P$ [d]&FAP [\%]&$P$ [d]&FAP [\%]&$P$ [d]&FAP [\%]\\
\hline
\noalign{\smallskip}
\endfirsthead
\caption{Continued.}\\
\hline\hline
Karmn& $P_\mathrm{rot\,lit}$ & \multicolumn{2}{c}{\Prot} & \multicolumn{2}{c}{\Prothalf}&\multicolumn{2}{c}{\Prot 1\,d alias}&\multicolumn{2}{c}{\Prothalf 1\,d alias}\\
\cmidrule(lr){3-4}\cmidrule(lr){5-6}\cmidrule(lr){7-8}\cmidrule(lr){9-10}
& [d] &$P$ [d]&FAP [\%]&$P$ [d]&FAP [\%]&$P$ [d]&FAP [\%]&$P$ [d]&FAP [\%]\\
\hline
\noalign{\smallskip}
\endhead
\hline
\endfoot
J01025+716  &   $52.00\pm 2.60$ &  57.52 &       19.03 &    ...&      ...&     ...&           ...&    ...&     ...\\
J01125-169  &   $69.00\pm 2.40$ &  70.19 &  $\leq$ 0.1 &    ...&      ...&     ...&           ...&    ...&     ...\\
J02002+130  &             ... &   1.96 &  $\leq$ 0.1 &    ...&      ...&  2.03 &  $\leq$ 0.1 &    ...&     ...\\
J02530+168  &             ... &  96.20 &  $\leq$ 0.1 &    ...&      ...&     ...&           ...&    ...&     ...\\
J03133+047  &    $130.0^{\dag}$ & 147.86 &        0.42 &    ...&      ...&     ...&           ...&    ...&     ...\\
J04376+528  &             ... &  16.52 &       30.43 &    ...&      ...&     ...&           ...&    ...&     ...\\
J05314-036  &   $34.00\pm 0.60$ &  34.07 &        0.51 &    ...&      ...&     ...&           ...&    ...&     ...\\
J07446+035  &  $2.780\pm 0.010$ &   2.77 &  $\leq$ 0.1 &    ...&      ...&  1.56 &        1.58 &    ...&     ...\\
\end{longtable}
}


\longtab[6]{
\begin{longtable}{lccccccccc}
\caption{Same as Table \ref{tab:starsactindselected_rv}, but for CCF BIS.}\label{tab:starsactindselected_bis}\\
\hline\hline
Karmn& $P_\mathrm{rot\,lit}$ & \multicolumn{2}{c}{\Prot} & \multicolumn{2}{c}{\Prothalf}&\multicolumn{2}{c}{\Prot 1\,d alias}&\multicolumn{2}{c}{\Prothalf 1\,d alias}\\
\cmidrule(lr){3-4}\cmidrule(lr){5-6}\cmidrule(lr){7-8}\cmidrule(lr){9-10}
& [d] &$P$ [d]&FAP [\%]&$P$ [d]&FAP [\%]&$P$ [d]&FAP [\%]&$P$ [d]&FAP [\%]\\
\hline
\noalign{\smallskip}
\endfirsthead
\caption{Continued.}\\
\hline\hline
Karmn& $P_\mathrm{rot\,lit}$ & \multicolumn{2}{c}{\Prot} & \multicolumn{2}{c}{\Prothalf}&\multicolumn{2}{c}{\Prot 1\,d alias}&\multicolumn{2}{c}{\Prothalf 1\,d alias}\\
\cmidrule(lr){3-4}\cmidrule(lr){5-6}\cmidrule(lr){7-8}\cmidrule(lr){9-10}
& [d] &$P$ [d]&FAP [\%]&$P$ [d]&FAP [\%]&$P$ [d]&FAP [\%]&$P$ [d]&FAP [\%]\\
\hline
\noalign{\smallskip}
\endhead
\hline
\endfoot
J01026+623  &              $20.00\pm 0.40$ &  18.27 &        1.65 &     ...&           ...&  1.05 &        0.38 &     ...&           ...\\
J02002+130  &                        ... &   1.96 &        0.16 &     ...&           ...&  2.03 &        0.76 &     ...&           ...\\
J03133+047  &               $130.0^{\dag}$ & 147.86 &       33.13 &     ...&           ...&     ...&           ...&     ...&           ...\\
J04153-076  &                        ... &   1.80 &  $\leq$ 0.1 &     ...&           ...&     ...&           ...&     ...&           ...\\
J04290+219  &              $25.00\pm 0.30$ &  24.97 &        0.43 &     ...&           ...&     ...&           ...&     ...&           ...\\
J04376+528  &                        ... &  16.30 &       32.79 &  7.90 &        1.42 &     ...&           ...&     ...&           ...\\
J04588+498  &                        ... &      ...&           ...&  8.97 &  $\leq$ 0.1 &     ...&           ...&  1.13 &        0.67 \\
J05019+011  &             $2.120\pm 0.020$ &   2.09 &  $\leq$ 0.1 &     ...&           ...&  1.90 &  $\leq$ 0.1 &     ...&           ...\\
\end{longtable}
}


\longtab[7]{
\begin{longtable}{lccccccccc}
\caption{Same as Table \ref{tab:starsactindselected_rv}, but for \Halpha.}\label{tab:starsactindselected_halpha}\\
\hline\hline
Karmn& $P_\mathrm{rot\,lit}$ & \multicolumn{2}{c}{\Prot} & \multicolumn{2}{c}{\Prothalf}&\multicolumn{2}{c}{\Prot 1\,d alias}&\multicolumn{2}{c}{\Prothalf 1\,d alias}\\
\cmidrule(lr){3-4}\cmidrule(lr){5-6}\cmidrule(lr){7-8}\cmidrule(lr){9-10}
& [d] &$P$ [d]&FAP [\%]&$P$ [d]&FAP [\%]&$P$ [d]&FAP [\%]&$P$ [d]&FAP [\%]\\
\hline
\noalign{\smallskip}
\endfirsthead
\caption{Continued.}\\
\hline\hline
Karmn& $P_\mathrm{rot\,lit}$ & \multicolumn{2}{c}{\Prot} & \multicolumn{2}{c}{\Prothalf}&\multicolumn{2}{c}{\Prot 1\,d alias}&\multicolumn{2}{c}{\Prothalf 1\,d alias}\\
\cmidrule(lr){3-4}\cmidrule(lr){5-6}\cmidrule(lr){7-8}\cmidrule(lr){9-10}
& [d] &$P$ [d]&FAP [\%]&$P$ [d]&FAP [\%]&$P$ [d]&FAP [\%]&$P$ [d]&FAP [\%]\\
\hline
\noalign{\smallskip}
\endhead
\hline
\endfoot
J00183+440  &          $45.00\pm 4.40^{*}$ &  47.40 &  $\leq$ 0.1 &     ...&           ...&    ...&           ...&    ...&           ...\\
J01025+716  &              $52.00\pm 2.60$ &  50.28 &  $\leq$ 0.1 &     ...&           ...&    ...&           ...& 1.02 &  $\leq$ 0.1 \\
J01026+623  &              $20.00\pm 0.40$ &  18.89 &  $\leq$ 0.1 &     ...&           ...& 1.05 &  $\leq$ 0.1 &    ...&           ...\\
J01125-169  &              $69.00\pm 2.40$ &  70.19 &       25.22 &     ...&           ...&    ...&           ...&    ...&           ...\\
J02002+130  &                        ... &   1.95 &        2.58 &     ...&           ...& 2.04 &       39.57 &    ...&           ...\\
J02222+478  &              $30.00\pm 0.40$ &  28.20 &        5.39 &     ...&           ...& 1.04 &        6.35 &    ...&           ...\\
J02530+168  &                        ... &  96.20 &       24.18 &     ...&           ...&    ...&           ...&    ...&           ...\\
J03133+047  &               $130.0^{\dag}$ & 137.54 &  $\leq$ 0.1 &     ...&           ...&    ...&           ...&    ...&           ...\\
\end{longtable}
}


\longtab[8]{
\begin{longtable}{lccccccccc}
\caption{Same as Table \ref{tab:starsactindselected_rv}, but for Ca IRT-a.}\label{tab:starsactindselected_cairt1}\\
\hline\hline
Karmn& $P_\mathrm{rot\,lit}$ & \multicolumn{2}{c}{\Prot} & \multicolumn{2}{c}{\Prothalf}&\multicolumn{2}{c}{\Prot 1\,d alias}&\multicolumn{2}{c}{\Prothalf 1\,d alias}\\
\cmidrule(lr){3-4}\cmidrule(lr){5-6}\cmidrule(lr){7-8}\cmidrule(lr){9-10}
& [d] &$P$ [d]&FAP [\%]&$P$ [d]&FAP [\%]&$P$ [d]&FAP [\%]&$P$ [d]&FAP [\%]\\
\hline
\noalign{\smallskip}
\endfirsthead
\caption{Continued.}\\
\hline\hline
Karmn& $P_\mathrm{rot\,lit}$ & \multicolumn{2}{c}{\Prot} & \multicolumn{2}{c}{\Prothalf}&\multicolumn{2}{c}{\Prot 1\,d alias}&\multicolumn{2}{c}{\Prothalf 1\,d alias}\\
\cmidrule(lr){3-4}\cmidrule(lr){5-6}\cmidrule(lr){7-8}\cmidrule(lr){9-10}
& [d] &$P$ [d]&FAP [\%]&$P$ [d]&FAP [\%]&$P$ [d]&FAP [\%]&$P$ [d]&FAP [\%]\\
\hline
\noalign{\smallskip}
\endhead
\hline
\endfoot
J01025+716  &              $52.00\pm 2.60$ &  59.42 &        4.07 &    ...&      ...&    ...&           ...&    ...&      ...\\
J01026+623  &              $20.00\pm 0.40$ &  18.89 &  $\leq$ 0.1 &    ...&      ...& 1.05 &  $\leq$ 0.1 &    ...&      ...\\
J02222+478  &              $30.00\pm 0.40$ &  31.13 &        2.25 &    ...&      ...&    ...&           ...&    ...&      ...\\
J03133+047  &               $130.0^{\dag}$ & 120.70 &       98.07 &    ...&      ...&    ...&           ...&    ...&      ...\\
J03463+262  &                        ... &  10.07 &        0.23 &    ...&      ...&    ...&           ...&    ...&      ...\\
J04290+219  &              $25.00\pm 0.30$ &  25.14 &  $\leq$ 0.1 &    ...&      ...&    ...&           ...&    ...&      ...\\
J04376+528  &                        ... &  15.47 &  $\leq$ 0.1 &    ...&      ...&    ...&           ...&    ...&      ...\\
J04429+189  &              $41.00\pm 0.40$ &  38.83 &        0.65 &    ...&      ...&    ...&           ...&    ...&      ...\\
\end{longtable}
}


\longtab[9]{
\begin{longtable}{lccccccccc}
\caption{Same as Table \ref{tab:starsactindselected_rv} but for Ca IRT-b.}\label{tab:starsactindselected_cairt2}\\
\hline\hline
Karmn& $P_\mathrm{rot\,lit}$ & \multicolumn{2}{c}{\Prot} & \multicolumn{2}{c}{\Prothalf}&\multicolumn{2}{c}{\Prot 1\,d alias}&\multicolumn{2}{c}{\Prothalf 1\,d alias}\\
\cmidrule(lr){3-4}\cmidrule(lr){5-6}\cmidrule(lr){7-8}\cmidrule(lr){9-10}
& [d] &$P$ [d]&FAP [\%]&$P$ [d]&FAP [\%]&$P$ [d]&FAP [\%]&$P$ [d]&FAP [\%]\\
\hline
\noalign{\smallskip}
\endfirsthead
\caption{Continued.}\\
\hline\hline
Karmn& $P_\mathrm{rot\,lit}$ & \multicolumn{2}{c}{\Prot} & \multicolumn{2}{c}{\Prothalf}&\multicolumn{2}{c}{\Prot 1\,d alias}&\multicolumn{2}{c}{\Prothalf 1\,d alias}\\
\cmidrule(lr){3-4}\cmidrule(lr){5-6}\cmidrule(lr){7-8}\cmidrule(lr){9-10}
& [d] &$P$ [d]&FAP [\%]&$P$ [d]&FAP [\%]&$P$ [d]&FAP [\%]&$P$ [d]&FAP [\%]\\
\hline
\noalign{\smallskip}
\endhead
\hline
\endfoot
J00183+440  &          $45.00\pm 4.40^{*}$ &  46.19 &  $\leq$ 0.1 &    ...&     ...&    ...&           ...&    ...&     ...\\
J01025+716  &              $52.00\pm 2.60$ &  51.73 &  $\leq$ 0.1 &    ...&     ...& 1.02 &  $\leq$ 0.1 &    ...&     ...\\
J01026+623  &              $20.00\pm 0.40$ &  18.89 &  $\leq$ 0.1 &    ...&     ...& 1.05 &  $\leq$ 0.1 &    ...&     ...\\
J02222+478  &              $30.00\pm 0.40$ &  31.13 &  $\leq$ 0.1 &    ...&     ...&    ...&           ...&    ...&     ...\\
J03133+047  &               $130.0^{\dag}$ & 140.81 &        4.01 &    ...&     ...&    ...&           ...&    ...&     ...\\
J03463+262  &                        ... &  10.07 &        0.11 &    ...&     ...&    ...&           ...&    ...&     ...\\
J04290+219  &              $25.00\pm 0.30$ &  25.14 &  $\leq$ 0.1 &    ...&     ...&    ...&           ...&    ...&     ...\\
J04376+528  &                        ... &  15.47 &  $\leq$ 0.1 &    ...&     ...&    ...&           ...&    ...&     ...\\
\end{longtable}
}


\longtab[10]{
\begin{longtable}{lccccccccc}
\caption{Same as Table \ref{tab:starsactindselected_rv}, but for Ca IRT-c.}\label{tab:starsactindselected_cairt3}\\
\hline\hline
Karmn& $P_\mathrm{rot\,lit}$ & \multicolumn{2}{c}{\Prot} & \multicolumn{2}{c}{\Prothalf}&\multicolumn{2}{c}{\Prot 1\,d alias}&\multicolumn{2}{c}{\Prothalf 1\,d alias}\\
\cmidrule(lr){3-4}\cmidrule(lr){5-6}\cmidrule(lr){7-8}\cmidrule(lr){9-10}
& [d] &$P$ [d]&FAP [\%]&$P$ [d]&FAP [\%]&$P$ [d]&FAP [\%]&$P$ [d]&FAP [\%]\\
\hline
\noalign{\smallskip}
\endfirsthead
\caption{Continued.}\\
\hline\hline
Karmn& $P_\mathrm{rot\,lit}$ & \multicolumn{2}{c}{\Prot} & \multicolumn{2}{c}{\Prothalf}&\multicolumn{2}{c}{\Prot 1\,d alias}&\multicolumn{2}{c}{\Prothalf 1\,d alias}\\
\cmidrule(lr){3-4}\cmidrule(lr){5-6}\cmidrule(lr){7-8}\cmidrule(lr){9-10}
& [d] &$P$ [d]&FAP [\%]&$P$ [d]&FAP [\%]&$P$ [d]&FAP [\%]&$P$ [d]&FAP [\%]\\
\hline
\noalign{\smallskip}
\endhead
\hline
\endfoot
J00183+440  &          $45.00\pm 4.40^{*}$ & 46.19 &        6.52 &    ...&     ...&    ...&           ...&    ...&     ...\\
J01025+716  &              $52.00\pm 2.60$ & 52.48 &  $\leq$ 0.1 &    ...&     ...& 1.02 &        9.53 &    ...&     ...\\
J01026+623  &              $20.00\pm 0.40$ & 18.98 &  $\leq$ 0.1 &    ...&     ...& 1.05 &  $\leq$ 0.1 &    ...&     ...\\
J02002+130  &                        ... &  1.95 &        4.54 &    ...&     ...&    ...&           ...&    ...&     ...\\
J02222+478  &              $30.00\pm 0.40$ & 28.65 &  $\leq$ 0.1 &    ...&     ...&    ...&           ...&    ...&     ...\\
J03463+262  &                        ... & 10.07 &        0.33 &    ...&     ...&    ...&           ...&    ...&     ...\\
J04290+219  &              $25.00\pm 0.30$ & 25.14 &  $\leq$ 0.1 &    ...&     ...&    ...&           ...&    ...&     ...\\
J04376+528  &                        ... & 15.47 &        0.39 &    ...&     ...&    ...&           ...&    ...&     ...\\
\end{longtable}
}

\end{appendix}

\end{document}